\documentclass[preprint,11pt]{aastex}
\usepackage{epsfig}
\usepackage{times}
\shorttitle{Star formation in Lynds dark clouds}
\shortauthors{Visser, Richer, \& Chandler}

\begin{document}

\title{Completion of a SCUBA survey of Lynds dark clouds and
implications for low-mass star formation}
\author{Anja E. Visser, John S. Richer}
\affil{Mullard Radio Astronomy Observatory, Cavendish Laboratory,
Madingley Road, Cambridge CB3 0HE, United Kingdom}
\email{anja.visser@wanadoo.nl, jsr@mrao.cam.ac.uk}
\and
\author{Claire J. Chandler}
\affil{National Radio Astronomy Observatory\footnote{The National Radio
Astronomy Observatory is a facility of the National Science Foundation
operated under cooperative agreement by Associated Universities, Inc.},
PO Box O, Socorro, NM 87801, USA}
\email{cchandle@nrao.edu}

\begin{abstract}

We have carried out a survey of optically-selected dark clouds using the
bolometer array SCUBA on the James Clerk Maxwell Telescope, at $\lambda
= 850$~$\mu$m.  The survey covers a total of 0.5 square degrees and is
unbiased with reference to cloud size, star formation activity, or the
presence of infrared emission.  Several new protostars and starless cores
have been discovered; the protostars are confirmed through the detection
of their accompanying outflows in CO(2--1) emission.  The survey is
believed to be complete for Class~0 and Class~I protostars, and yields
two important results regarding the lifetimes of these phases.  First,
the ratio of Class~0 to Class~I protostars in the sample is roughly
unity, very different from the 1:10 ratio that has previously been
observed for the $\rho$ Ophiuchi star-forming region.  Assuming star
formation to be a homogeneous process in the dark clouds, this implies
that the Class~0 lifetime is similar to the Class~I phase, which from
infrared surveys has been established to be $\sim 2 \times 10^5$~yr.
It also suggests there is no rapid initial accretion phase in Class~0
objects.  A burst of triggered star formation some $\sim 10^5$~yr
ago can explain the earlier results for $\rho$ Ophiuchus.  Second,
the number of starless cores is approximately twice that of the total
number of protostars, indicating a starless core lifetime of $\sim 8
\times 10^5$~yr.  These starless cores are therefore very short-lived,
surviving only two or three free-fall times.  This result suggests that,
on size scales of $\sim 10^4$~AU at least, the dynamical evolution of
starless cores is probably not controlled by magnetic processes.

\end{abstract}


\section{Introduction}

Systematic surveys of the earliest stages of low-mass star formation
provide vital information on the physics of cloud collapse.
In particular, if complete samples of protostars and pre-stellar
cores can be identified, their relative lifetimes can be estimated
if it is assumed that the clouds are not being observed at a special
time in their evolution.  Such studies are vital for differentiating
between models of star formation, which can predict very different
protostellar accretion histories.  In some models, the evolution of
molecular cloud cores is controlled entirely by strong magnetic fields,
leading to lifetimes determined by the ambipolar diffusion timescale,
which is typically $10^7$~yr (Ciolek \& Mouschovias 1994).  However,
recent 3-dimensional numerical simulations of turbulent, gravitationally
unstable, clouds show that cores can form and evolve over only a few
dynamical timescales, even if significant magnetic fields are present
(Li et al.\ 2000; Heitsch, Mac Low, \& Klessen 2001).  For typical cloud
conditions, this implies evolution on timescales of about $10^6$~yr.

Once a gravitationally unstable core has formed and started to collapse,
the accretion rate of the protostar is governed by the initial conditions
at the onset of collapse.  The idealized collapse of a singular isothermal
sphere (Shu 1977) proceeds with a uniform accretion rate.  Other accretion
models starting from different initial conditions predict a short phase
of rapid accretion, succeeded by a more or less constant accretion
rate (e.g., Foster \& Chevalier 1993; McLaughlin \& Pudritz 1997).
If evolutionary phases of different accretion rates can be identified
observationally then their relative lifetimes can then be estimated,
possibly enabling the discrimination between these different models.

Complete surveys of infrared protostars have been used to obtain good
estimates for the lifetime of infrared protostars, in particular the
Class I, II, and III phases of protostellar evolution.  From such
surveys Wilking et al.\ (1989) and Kenyon et al.\ (1990) conclude
that the Class I phase lasts $\sim 2 \times 10^5$~yr in the low-mass,
star-forming regions of $\rho$ Ophiuchus and Taurus, and this lifetime
is consistent with typical cloud collapse models (e.g., Adams, Lada, \&
Shu 1987).  However, the timescales associated with the earlier phases
of star formation, specifically the Class~0 phase (Andr\'e et al.\ 1993)
and the pre-collapse phase (Ward-Thompson et al.\ 1994), are much more
uncertain.  Most samples of dense cores derive originally from catalogues
of optically-selected dark clouds, which were subsequently searched
for associated IRAS emission (e.g., Beichman et al.\ 1986; Clemens \&
Barvainis 1988; Benson \& Myers 1989; Lee \& Myers 1999; Jijina, Myers, \&
Adams 1999).  Since Class~0 protostars are typically too cold and faint to
have been detected by IRAS, complete samples of both Class~0 protostars
and truly ``starless'' cores (as opposed to cores lacking an associated
IRAS source or 2~$\mu$m emission) have been difficult to obtain.  In one
of the few regions where complete samples of Class~0 and Class~I objects
exist --- the $\rho$ Ophiuchi cloud --- the results imply that the Class~0
phase lasts only one tenth of the Class~I phase in this region, i.e.,
$2 \times 10^4$~yr (Andr\'e \& Montmerle 1994).  However, there are few,
if any, good statistical constraints in other regions of star formation.

The advent of large bolometer arrays on millimeter and submillimeter
telescopes has made the first systematic surveys for the earliest
phases of star formation possible.  Typical observing wavelengths,
$\lambda \sim 0.5$ to 1~mm, lie towards the Rayleigh-Jeans side of the
Planck function for all reasonable dust temperatures $T \ga 7$~K\@.
Millimeter and submillimeter dust emission is therefore an excellent
tracer of the youngest embedded protostars and starless dense cores,
which are regions of high dust column density that can be too cold for
detection by far-infrared instruments.  In this paper we present the
second and concluding part of a survey, using the submillimeter camera
SCUBA on the James Clerk Maxwell Telescope (JCMT), of dust continuum
emission from typical, nearby, molecular clouds forming low-mass stars.
The observations are sufficiently sensitive to detect all the embedded
protostars and starless cores above a certain mass limit, and so allow
us to estimate the lifetimes of these earliest phases of star formation.
In an earlier paper (Visser, Richer, \& Chandler 2001, hereafter Paper I)
we presented initial results derived from images of the smaller clouds
in the sample.  Here we include the images of the larger clouds, and
using the full survey present a detailed analysis of the structure
of the cores, their star forming properties, and the properties of
their molecular outflows.  We also estimate the relative and absolute
lifetimes of the various pre-stellar and protostellar phases based on
their detection rates.

\section{Source sample}

The clouds observed were selected from those listed as opacity class 6 in
the Lynds catalogue of dark nebulae, which contains 1802 optically-dark
clouds selected from the POSS plates (Lynds 1962).  Lynds catalogued the
clouds by eye based on their apparent opacity, using an arbitrary scale
of 1 to 6, with the class 6 clouds being the most opaque ($A_V \ga 10$
magnitudes).  These estimates were based on a comparison of the cloud
with a neighboring field on the POSS plate, and the classification
is therefore somewhat subjective.  The Lynds catalogue contains a
total of 147 opacity class 6 clouds covering a total area of $\sim 2$
square degrees, with individual clouds having a mean minor axis of 4$'$
and major axis of 11$'$.  Some of these clouds are small and round,
like globules, but others are extended and filamentary.  The optical
appearance of most clouds is quiescent, with only a small fraction of
the clouds looking shocked.

Star formation in Lynds class 6 clouds has been studied by Parker
(1991) following a re-analysis of their positions (Parker 1988) and
an investigation of their association with IRAS sources.  However,
the selection criteria he used based on IRAS detections failed to find
the well-known Class~0 source B335 (L663) (Parker 1989), demonstrating
the importance of understanding the selection effects introduced by
such wavelength-dependent criteria.  Our optical selection criterion
corresponds to a limiting column density, and is biased towards finding
nearby clouds.

Forty two opacity class 6 clouds were selected from the Lynds catalogue
with sizes and positions determined by the observing time allocated
and observing mode available at the JCMT (see Section~\ref{obs_cont}
below).  This resulted in an essentially random selection of clouds.
The first to be observed were the small or filamentary clouds that could
be mapped using the single SCUBA field of view, before the commissioning
of large-area (scan) mapping.  These results are presented in Paper I\@.
The larger clouds were mapped when scan mapping became available.
Because of the LST ranges allocated to the observations we were able to
map several areas in $\rho$~Ophiuchus, and more than one third of all the
clouds are part of the $\rho$~Ophiuchi cloud complex.  The full sample
contains no biases relating to size, star formation activity, or the
presence of infrared emission, and so is statistically representative of
nearby, dark clouds.  The names and positions of all the clouds surveyed
are given in Table~\ref{tab_sample}, along with identifications from
the Barnard (1927) catalogue (B) and the Clemens \& Barvainis (1988)
catalogue (CB).

\subsection{Distances}

The distances to nearby clouds can be difficult to determine because
neither star counts nor kinematic determinations are very reliable.
The use of foreground star counts requires a precise knowledge of
the distribution of the stars in the direction of the cloud, and
the cloud needs to be large to make the count statistically relevant.
The LSR velocity of the cloud is usually influenced by local motions and
cannot, on its own, be used to determine the distance towards the cloud.
Nevertheless, to be able to obtain physical parameters from the data
the correct distances need to be known.  Here we use the method of
associating the dark clouds with larger molecular cloud complexes (see,
e.g., Launhardt \& Henning 1997).  Fifteen of our clouds have distances
previously reported in the literature (Hilton \& Lahulla 1995) with which
our distance estimates agree very well.  A discussion of the distances for
the clouds presented in Paper I is given in an appendix to that paper.
For the larger clouds that were scan mapped details are given by Visser
(2000).  The assumed distances and associated cloud complexes are
summarized in Table~\ref{tab_sample}.

\subsection{IRAS sources}

Table~\ref{tab_iras} lists the IRAS sources in the Point Source Catalogue
(PSC) associated with the Lynds clouds in our sample.  Association is
defined to be all IRAS PSC sources within the SCUBA maps, and the IRAS
sources that were associated with the clouds by Parker (1988).  Most of
the IRAS sources in Table~\ref{tab_iras} had already been associated
with the Lynds clouds by Parker.  Five IRAS sources covered by the SCUBA
maps were not associated by Parker (even though they sometimes satisfy
Parker's selection criteria, for example IRAS~19186+2325 in L771), and
these are marked with an ``a''.  Sources associated with the cloud by
Parker, but not covered by the maps, are marked with a ``b''.  The eight
candidate protostars in common with those selected and studied by Parker
(1991) are indicated with a ``c''.

Information about the nature of some of these IRAS sources is
available in the literature, and is included in Table~\ref{tab_iras}.
IRAS~16445$-$1352 is probably cirrus of the type very common in
$\rho$~Ophiuchus, since it is only detected at 60 and 100~$\mu$m
and is extended at both wavelengths (Beichman et al.\ 1986).
IRAS~16455$-$1405\footnote{mistyped as 16445$-$1405 by Bontemps et al.\
(1996)} is listed as a Class~I source by Bontemps et al.\ (1996).
An outflow was not detected, however, and the source was identified
by Parker (1991) as a T~Tauri star with an optical counterpart.  This
source is not covered by the SCUBA map, and is not discussed further.
T~Tauri stars covered by this survey are IRAS~16459$-$1411 (Andr\'e \&
Montmerle 1994), and IRAS~19181+1056, which is associated with HH~32
(Curiel et al.\ 1997).  IRAS sources 16442$-$0930, 18148$-$0440,
19345+0727, 23238+7401, and 16285$-$2355 are all known protostars with
outflows partly mapped by Bontemps et al.\ (1996).  IRAS~19184+1055 is
identified as an OH/IR star by Chengalur et al.\ (1993), and 16285$-$2358
is an optically visible star (Ichikawa \& Nishida 1989).

IRAS~22051+5848 is a protostar with an outflow presented by Parker,
Padman, \& Scott (1991).  Launhardt, Ward-Thompson, \& Henning (1997)
suggest that IRAS~23228+6320 is a protostar, but an outflow is not
detected (Parker et al.\ 1991).  Massive outflows are detected from
IRAS~21017+6742 (Myers et al.\ 1988) and from the vicinity of 19180+1116
and 19180+1114 (Armstrong \& Winnewisser 1989).  Anglada, Sepulveda,
\& Gomez (1997) argue that of these latter two it is more likely that
19180+1114 drives the outflow, taking into account ammonia emission,
their SEDs, and the location of the sources with respect to the CO
flow.  Identifications from the literature are listed in column 7 of
Table~\ref{tab_iras}.

In general, field stars and T~Tauri stars are detected in the 12
and 25~$\mu$m IRAS bands with their spectral energy distributions
(SEDs) rising towards shorter wavelengths, while embedded protostars
(and cirrus clouds) are detected at 60 and 100~$\mu$m with SEDs rising
towards longer wavelengths.  This trend is seen in the sample presented
in Table~\ref{tab_iras}, although many of the IRAS fluxes are high upper
limits due to confusion.

\section{Submillimeter continuum observations and data reduction}
\label{obs_cont}

Observations were made using the submillimeter camera, SCUBA, on the 15-m
James Clerk Maxwell Telescope (JCMT) on Mauna Kea, Hawaii (Holland et
al.\ 1999).  SCUBA comprises two arrays of bolometer detectors, which can
be used simultaneously by means of a dichroic beam splitter.  The arrays
are cooled to a temperature of 0.1~K to achieve high sensitivity.  The 37
bolometers of the long-wavelength array are optimised for observations
at 850~$\mu$m, and the 91 bolometers of the short-wavelength array are
optimised for operation at 450~$\mu$m.  The diffraction-limited FWHM
beam sizes at these two wavelengths are $14''$ and $8''$ respectively.
At 850~$\mu$m the beam is well approximated by a single Gaussian, but at
450~$\mu$m there is an extended error beam containing a roughly equal
amount of power to the main diffraction beam, spread over a diameter
of approximately 40$''$.  Most of the conclusions in this paper are
therefore based on the 850~$\mu$m data.

The bolometers form a hexagonal pattern, and are spaced approximately two
beamwidths apart.  To produce fully-sampled images, either the secondary
mirror is moved in a so-called ``jiggle'' pattern, or the arrays are
scanned across the sky in ``scan map'' mode.  For both of these methods
the secondary mirror also chops at a rate of 7.8~Hz to remove most of
the sky emission.  The jiggle mode is ideal for sources smaller than
the field of view of the arrays (2.3$'$), while scan mapping is used
for more extended sources.  Further technical details of observing with
SCUBA on the JCMT are described by Holland et al.\ (1999).

A sample of 24 compact Lynds clouds was observed in 1997 September using
jiggle mode, before scan mapping was fully commissioned (Paper I)\@.
Three larger clouds were also scan-mapped at this time as part of the
commissioning of this mode (L663, L1165, and L1262).  Weather conditions
were average and stable (zenith atmospheric opacity at 850~$\mu$m,
$\tau_{850}$ = 0.2$-$0.7), but not good enough to obtain 450-$\mu$m
data.  A further 15 clouds were scan mapped under extremely good weather
conditions ($\tau_{850}$ = 0.12) in 1998 July, providing good data at both
850 and 450~$\mu$m.  Measurements at 450~$\mu$m were also made of some of
the clouds observed in 1997 during this second observing run.  The total
survey of 42 clouds covered an area of $\sim 0.5$ square degrees, or
$\sim 7$~pc$^2$ for the distances listed in Table~\ref{tab_sample}.
Table~\ref{tab_fluxes} indicates whether a cloud was observed in jiggle
mode or scan mode.

The sky transmission was monitored by performing skydips throughout each
night, and the telescope pointing was checked regularly using nearby
bright sources (usually a blazar or a planet).  The pointing accuracy
was 2--3 arcsec.  The focus and alignment of the secondary mirror were
also checked two or three times every night.  The absolute calibration
is obtained from observations of Mars and Uranus, observed using the
same mode and chop throw as the observations of the dark clouds.

The chop throw for the jiggle maps was 150$''$, with the chop direction
dependent on the structure of the cloud.  To remove slowly varying
background emission, instrumental offsets in SCUBA, and telescope
asymmetries, the telescope is also nodded every 16 seconds, to place
the source in the opposite beam.

When mapping extended sources in scan-map mode, smaller chop throws must
be used to avoid significant degredation of the beam efficiency at the
edges of the arrays.  In addition the telescope is not nodded, so that
instrumental and other offsets have to be established during the data
reduction using baselining algorithms.  The resulting maps are images
of the sky convolved with a dual beam function, which is the single
telescope beam convolved with positive and negative delta functions
separated by the chop throw (the chop function).  The chop function must
be deconvolved from the image to obtain a representation of the sky.
Conceptually this deconvolution is equivalent to a division of the
Fourier Transform (FT) of the image by the FT of the chop function.
The FT of the chop function, however, is a sine function, and clearly
problems will arise at spatial frequencies close to nulls in the sine
function, where the noise will be greatly amplified by the division.

There are two ways of solving the problem of noise amplification
near nulls in the sine function, and both have been used here.
The conventional observing method is to scan the telescope in the same
direction as the chop throw (usually azimuth).  The resulting dual-beam
map can be restored with the Emerson-Klein-Haslam (EKH) algorithm
(Emerson, Klein, \& Haslam 1979).  The deconvolution of the chop function
is essentially carried out by dividing the FT of the map with a weighted
FT of the chop function, where the weighting excludes spatial frequencies
which have been attenuated to less than 0.5 of their original amplitude.
The scan maps obtained during commissioning of this mode were observed
in this way.

The other method is the ``Emerson2'' technique (Emerson 1995; Jenness,
Lightfoot, \& Holland 1998), used during the 1998 July observations.
Here six maps are made, three of which are chopped in right ascension, and
three of which are chopped in declination.  Chop throws of 20$''$, 30$''$,
and 65$''$ were used for each map, and the resulting spatial frequency
coverage at 850 and 450~$\mu$m is shown in Figure~\ref{coverage}.
With the exception of the zero spatial frequency, these chop throws
ensure that the nulls of the FT of each do not coincide, up to the
spatial frequency limit of the telescope beam.  The disadvantages of
not chopping in the azimuth direction (which gives the best subtraction
of sky emission and ground spillover) is outweighed by the improved
signal-to-noise ratio due to the more efficient deconvolution of the
chop function (Jenness et al.\ 1998).

The data were reduced using the SCUBA User Reduction Facility, SURF
(Jenness \& Lightfoot 1998).  Data from noisy or bad bolometers have
been removed.  Details of the reduction of the jiggle maps are given
by Visser (2000) and Paper I\@.  For the scan maps a baseline must
be removed for each scan and for each bolometer.  Various methods for
baseline removal are provided in SURF\@.  Some work better than others
for a given map, but none is entirely satisfactory and baseline removal
is still an unsolved problem.  The chop function is then deconvolved from
the data, adjoining maps are mosaicked together, and are interpolated
from the coordinates of the SCUBA arrays to equatorial coordinates.
This interpolation can result in a slight degradation of the resolution,
compared with the diffraction beams given above, for both jiggle maps
and scan maps.  The effective resolution at 850~$\mu$m is $\sim 15''$.

The incomplete spatial frequency coverage for scan maps
(Figure~\ref{coverage}) results in the low spatial frequencies
(corresponding to large-scale structure in the image) being measured
with low sensitivity, and the zero spatial frequency (i.e., the total
power in the map) not being measured at all.  The deconvolution of the
chop function is therefore similar in many respects to the deconvolution
of the dirty beam from an image made by an interferometer, where an
attempt is made to extrapolate the measured visibility function to zero
spatial frequency while maintaining consistency with the measured data.
In the Emerson2 algorithm as implemented in SURF the effect is to amplify
the noise on low spatial frequencies, so that the final images have the
equivalent of `1/f' noise in the image plane: spatial scales corresponding
to a few chop throws or more are measured with low sensitivity and thus
have high noise values, whereas structures on scales between the maximum
chop throw and the beam size are measured with dynamic ranges limited
only by the thermal noise.

\section{Results of the submillimeter continuum imaging}

\subsection{Cloud structure}
\label{sec_images}

The 24 compact Lynds clouds observed using SCUBA in jiggle mode have
been presented in Paper I and will not be reproduced here.  The 15
clouds detected in scan map mode are shown in Figures~\ref{fig_L57} to
\ref{fig_L1709}.  Contour plots of the 850-$\mu$m or 450-$\mu$m continuum
emission are overlaid on optical images from the STScI Digitized Sky
Survey, with contour levels indicated in the figure captions.  All the
images have been smoothed with a 20$''$ Gaussian (FWHM), apart from
those of L158 and L1172 (Figures~\ref{fig_L158} and \ref{fig_L1172}),
for which a 25$''$ Gaussian was used.  This smoothing improves the
signal-to-noise ratio in the maps, and brings out the faint, extended
structure.  The smoothing also suppresses high spatial frequency noise
introduced by the data reduction process.  The area of each cloud surveyed
is outlined by a dashed line.  Eight clouds from the total of 42 were
not detected, 5 in jiggle mode and 3 in scan map mode.  Further details
of these non-detections, including the areas covered by the SCUBA maps
of these clouds, are given by Visser (2000).  The positions of sources
from the IRAS PSC are indicated by triangles in each image.

The optical shapes of the clouds in this sample are varied.  The clouds
L57, L543, L663, and L675 are round and globule-like.  The clouds L860,
L951, and L1103 appear to have been swept up by a shock or wind, and
are cometary-like.  The structure of L673, on the other hand, is very
filamentary.  The submillimeter continuum emission also encompasses a
wide range of structures.  Of the clouds that were detected, some exhibit
purely diffuse, extended emission, some contain compact submillimeter
sources, and others, such as L63, L694, and L944, have a combination of
both.  The extended submillimeter emission from L31 and L953 correlates
very well with extinction in the optical images.

The extended submillimeter emission from a few clouds does not
correlate well with the dust extinction: see, for instance, L162, and
the southern part of L673.  In the case of L673 this might be caused by
poor atmospheric conditions during observation of the southern part.
This would increase the noise level towards the south, and may also
explain why 450-$\mu$m emission is only detected towards the northern
parts of the cloud.  The lowest contour levels in the south may therefore
trace correlated noise and is not very reliable.  L162 was also observed
under less than optimum weather conditions.  In general, scan maps
can sometimes exhibit not only suspicious extended positive features,
but also negative features -- artifacts introduced by the deconvolution
process and the lack of information at low spatial frequencies.  Several
techniques have been used in the literature to solve this problem,
such as fitting two-dimensional polynomials or surfaces to background
emission in the image plane (e.g., Johnstone \& Bally 1999; Bianchi et
al.\ 2000).  Such methods are, however, ad hoc, and no such subtraction
has been performed on the images presented in Figures~\ref{fig_L57}
to \ref{fig_L1709}.  Unsatisfactory as this may be, uncertainty in
the structure of extended features in these images will not affect the
identification of compact objects such as protostars and starless cores.

\subsection{General properties of the clouds}

The peak flux densities in the unsmoothed maps are listed in
Table~\ref{tab_fluxes} for both observed wavelengths.  The peak
position in the 850-$\mu$m map of a cloud does not necessarily coincide
with the peak position in the 450-$\mu$m image (see, e.g., L63 in
Figure~\ref{fig_L63}).  For those scan maps exhibiting negative features
the peak fluxes have been measured from a local zero level, determined
from cuts through the peak of the emission.  This increased the peak
flux densities of 10 sources by $\sim 10$\%.  The total systematic error
in the 450 and 850-$\mu$m peak flux densities is about 50\% and 20\%
respectively, and is due to negative features and fluctuations in the
optical depth of the atmosphere.  Also presented in Table~\ref{tab_fluxes}
are the statistical uncertainties and integrated flux densities for
extended emission.  The typical noise level in a 850-$\mu$m map is
25~mJy~beam$^{-1}$.  The noise levels on the 850-$\mu$m EKH maps of L663,
L1165, and L1262 are considerably higher, however, due to the poor weather
conditions.  The integrated flux densities were obtained by summing the
flux density within a box around the extended emission of the cloud.
The box sizes varied depending on the size of the clouds, but the boxes
always covered the extended emission down to the 1$\sigma$ detection
limit, staying clear of the noisy edges of the maps.  No correction has
been made for the (potentially spurious) negative or positive features
in the scan maps when measuring the integrated flux densities.

The 850-$\mu$m peak flux densities have been used to calculate masses,
column densities, space densities and extinction, $A_V$, and are
presented in Table~\ref{tab_fluxes}.  Assuming low optical depth at
850~$\mu$m the total mass (gas plus dust) traced by the submillimeter
dust emission is found from $M = D^2 F_\nu / B_\nu(T) \kappa_\nu$,
where $D$ is the distance to the source, $F_\nu$ is the flux density at
frequency $\nu$, $B_\nu(T)$ is the Planck function for dust temperature
$T$, and $\kappa_\nu$ is the dust opacity per unit mass of gas plus dust.
For $\kappa_{850\mu{\rm m}} = 0.0012$~m$^2$~kg$^{-1}$, and a temperature
$T = 12$~K, this gives $(M/M_\odot) = 1.9 \times 10^{-8} (D/{\rm pc})^2
(F_{850\mu{\rm m}}/$(mJy~beam$^{-1}$)).  The peak masses obtained in this
way range from 0.04~$M_\odot$ to 4.07~$M_\odot$.  The column density has
been calculated using $N({\rm H}_2) = F_\nu/B_\nu(T) \kappa_\nu m_{{\rm
H}_2} \mu \Omega_b$, where $m_{{\rm H}_2}$ is the mass of an H$_2$
molecule, $\mu$ is the mean molecular weight of the particles, assumed
to be 1.36, and $\Omega_b$ is the solid angle of the beam.  For the same
values of $\kappa_{850\mu{\rm m}}$ and $T$ used above, and $\Omega_b =
15''$, this reduces to $(N({\rm H}_2)/{\rm m}^{-2}) = 1.33 \times 10^{24}
(F_{850\mu{\rm m}}/$(mJy~beam$^{-1}$)).  An estimate of the mean space
density is obtained by assuming the depth of the cloud is similar to the
beamsize at the distance of the cloud.  The mean peak column density
is $3.9 \times 10^{26}$~m$^{-2}$, and the mean peak density is $7.6
\times 10^{11}$~m$^{-3}$, but the values for individual clouds cover a
broad range.  The peak extinction is calculated using $(A_V/{\rm mag})
= N({\rm H}_2)/(0.95 \times 10^{25}~{\rm m}^{-2})$ (Bohlin, Savage,
\& Drake 1978), and for the detected clouds all have $A_V \geq 8$~mag,
agreeing well with that measured for dark clouds using isotopomers of
CO (Myers, Linke, \& Benson 1983).  The total gas mass and mean column
densities are also presented in Table~\ref{tab_fluxes}, calculated using
the integrated flux densities.  The total mass observed in the survey
is $\sim 400~M_\odot$, with the most mass ($\sim 90~M_\odot$) associated
with L673.  The mean cloud column density varies between $\sim 1$ and $10
\times 10^{25}$~m$^{-2}$ with three exceptions: L663, L1165, and L1262.
These three clouds were those observed with the EKH technique, and the
dynamic range on these maps is too small to be able to detect the lower
column densities.

It is clear that the above calculations of mass, column, and space
densities depend on the assumptions made about the dust opacity, dust
temperature, and distance of the cloud.  The dust opacity used is that
of Hildebrand (1983), and is probably uncertain by a factor of 3 or so
(Henning, Michel, \& Stognienko 1995).  The assumed dust temperature
of 12~K is typical for quiescent, dark, molecular clouds (Myers et al.\
1983), but is likely to be an underestimate for those clouds containing
young stellar objects.  Radiation from protostars heat the surrounding
material out to $\sim 10^4$~AU (Chandler \& Richer 2000), and protostellar
outflows can also heat the cloud through shocks.  However, temperatures
in the range 8~K to 20~K will not change the derived cloud masses by more
than a factor of two.  Uncertainties in the cloud distances potentially
have a stronger influence on the cloud mass.  The numbers presented in
Table~\ref{tab_fluxes} should therefore be regarded just as indicative
of general cloud conditions.

\section{A catalogue of compact dust cores}
\label{sec_cores}

The submillimeter images of the 42 Lynds clouds presented
in Section~\ref{sec_images} and Paper I show many compact objects,
possibly pre-protostellar or protostellar in nature.  A compact object
not containing a protostar is not necessarily pre-protostellar, however.
A dense core must be gravitationally bound before it can collapse and
form a star, and it is not possible to determine whether a clump is
gravitationally bound from dust emission alone.  The term ``starless
core'' is therefore used here to denote a compact source not associated
with an outflow.

Various definitions of a ``core'' can also be found in the literature.
Myers et al.\ (1983) were the first to study cores in dark molecular
clouds systematically, and defined a core as a small (0.1~pc), dense ($3
\times 10^{10}$~m$^{-3}$) and cold (10~K) condensation which is close to
stable equilibrium and {\it may} evolve into a low-mass star.  Williams et
al.\ (2000) define a core as a region in which a single star (or multiple
system such as a binary) forms and which is gravitationally bound.
Here we use the term core to refer to the compact submillimeter objects
in the SCUBA survey, which are not necessarily gravitationally bound.

\subsection{Identification of the cores}
\label{sec_core_id}

Compact submillimeter objects in the Lynds clouds (SMM) have been
identified by spatially filtering the SCUBA images.  Each map was
smoothed with a 2$'$ Gaussian, and the smoothed map was then subtracted
from the original image.  This removed all large cloud structures, as
well as the unreliable positive and negative features resulting from the
beam deconvolution of the scan maps.  The remaining image contains only
compact structures, and all structures at a level $\geq 3 \sigma$ were
identified as cores.  Although this method is somewhat subjective, it does
select in a quantitative way all the features seen by eye, and all known
protostars and pre-protostellar cores.  Ideally, one would like to use a
fixed physical scale (such as a limiting $A_V$) to select the dense cores.
However, this fixed value would be dominated by the noisiest map.

Forty SMM cores were identified in the manner described above, and are
listed in Table~\ref{tab_cores}.  The positions of the cores have been
obtained from Gaussian fits to the emission, and so do not necessarily
coincide exactly with the emission peak.  Four cores were rejected from
the list (in L31, L55, and two in L917), since they appear near the
edge of the map where spiky negative and positive features can occur, a
known artifact resulting from the regridding algorithm.  L158$-$SMM2 is
suspiciously spiky too, but since it is not near the edge of the map it
is included in the list.

There are 10 cores that lie within 12$''$ of sources in the IRAS PSC,
and so can be directly associated with those sources.  These associations
are listed in Table~\ref{tab_cores}.  All 10 cores are known protostars.
Both L673$-$SMM1 and L1709$-$SMM5 lie within 40$''$ of an IRAS source
(19180+1116 and 16285$-$2356 respectively), and are possible associations.
In order to verify these possible assocations, HIRES images of the IRAS
data for L673 and L1709 were obtained from IPAC\footnote{The infrared
Processing and Analysis Center (IPAC) is funded by NASA as part of the
Infrared Astronomical Satellite (IRAS) extended mission under contract
to the Jet Propulsion Laboratory (JPL).}.  Figures~\ref{L673_hires}
and \ref{L1709_hires} demonstrate that IRAS~19180+1116 can confidently
be associated with L673$-$SMM1, and IRAS~16285$-$2356 with L1709$-$SMM5.

Figure~\ref{fig_cores} shows 850~$\mu$m emission from all the cores in
the sample on the same angular scale, with the central positions from
Table~\ref{tab_cores} marked as stars.  All associated IRAS sources are
very near the core positions, apart from IRAS~19180+1116 and 16285$-$2356.
Some of the cores are very compact and round (see, e.g., L260$-$SMM1,
L673$-$SMM1, and L1262$-$SMM1), while others are more extended (e.g.,
L63$-$SMM1, L158$-$SMM1, and L673$-$SMM8).  It has been suggested
that emission from T~Tauri stars and Class~I sources is more centrally
condensed than emission from Class~0 sources (Andr\'e \& Montmerle 1994).
Certainly, the only T~Tauri star detected, L162$-$SMM1, is very compact,
as expected.  L63$-$SMM1, a starless core, appears extended, and
the known protostars vary from compact (L483$-$SMM1) to very compact
(L1709$-$SMM1).  Figures~\ref{fig673_cores} and \ref{fig1709_cores}
illustrate the locations of the submillimeter cores in L673 and L1709,
the two clouds containing multiple sources.

\subsection{Completeness}

The detection limit in terms of mass is different for each cloud, due
to varying noise levels and distances.  In order to establish that we
have complete (or at least, representative) samples of starless cores,
Class~0, and Class~I sources down to a certain mass limit, we must
consider each of these effects.  Starless cores and Class~0 sources
are easier to detect than Class~I sources at submillimeter wavelengths.
Class~0 sources (for which the envelope mass, $M_{\rm env}$, is greater
than the mass of the forming protostar, $M_*$) have more material in
their envelopes than Class~I sources (for which $M_{\rm env} < M_*$),
and starless, pre-protostellar objects still have {\it all} their
material in an ``envelope'', since they have not yet formed a hydrostatic
protostellar core.  Our main problem is therefore potentially missing
Class~I sources associated with the observed Lynds clouds.

In Section~\ref{stats} we compare the detection rates of Class~0
and Class~I sources in our sample with those previously observed
for $\rho$~Ophiuchus by Andr\'e \& Montmerle (1994), for which the
detected ratio of Class~0 sources to Class~I was 1:10.  In order for
such a comparison to be meaningful, it is important to establish whether
our samples of Class~I protostars meet the same completeness criteria.
Andr\'e \& Montmerle found that, for Class~I sources in $\rho$~Ophiuchus,
$M_{\rm env}$ ranges from $\sim 0.015~M_\odot$ to $\sim 0.15~M_\odot$
with a median value of $\sim 0.06~M_\odot$, assuming a dust temperature
of 30~K, and $\kappa_{1.3{\rm mm}} = 0.001$~m$^{2}$~kg$^{-1}$.
The detection mass limit for the Lynds clouds mapped using SCUBA can
be calculated using these same parameters, adopting a $3 \sigma$ flux
detection limit, the distances given in Table~\ref{tab_sample}, and
scaling $\kappa_{1.3{\rm mm}}$ to a wavelength of 850~$\mu$m assuming
a frequency dependent dust opacity, $\kappa_\nu \propto \nu^\beta$,
with $\beta = 1.5$.  The detection limit for most clouds is then $<
0.015~M_\odot$ in our survey.  Only the furthest clouds (L543, L771,
L860, L917, L944, L951, L953, L1172, L1246), and the clouds for which the
maps are noisy (L663, L1165, L1262), have higher mass detection limits,
ranging from $\sim 0.02~M_\odot$ to $\sim 0.09~M_\odot$.  We could,
therefore, have missed some Class~I sources in these clouds.  However,
we would still have expected these sources to have been detected by IRAS:
even for the most distant cloud its 60~$\mu$m detection limit of 0.25~Jy
is equivalent to $M_{\rm env} \sim 0.003~M_\odot$.

In total there are nine IRAS sources associated with the clouds
mentioned above (Table~\ref{tab_iras}), and five of these sources are
known protostars which we have detected.  We do, therefore, need to
investigate the nature of the remaining four IRAS sources, in order
to exclude the possibility that these are low-mass Class~I protostars,
undetected in the SCUBA maps.  The IRAS colors of 22051+5849 in L1165
do not correspond to those of a protostar, and this is certainly not a
Class~I source.  Two IRAS sources not detected as submillimeter cores,
19343+0727 in L663 and 19186+2325 in L771, do have the colors consistent
with being protostellar, but these sources can be identified with
bright, optically-visible stars.  The remaining undetected IRAS source is
21186+4320, a source with an entry in the PSC at only 100~$\mu$m.  A HIRES
60-$\mu$m image of this cloud showed far-infrared emission displaced from
the PSC position, and the 60~$\mu$m peak in the HIRES map was not covered
by the SCUBA image.  None of the undetected IRAS sources is therefore a
Class~I source, and our sample of submillimeter cores should be complete
for Class~I sources down to $M_{\rm env} = 0.015~M_\odot$.

\section{A $^{12}$CO J=2--1 survey for outflows from the compact cores}

\subsection{Observations}

To determine the nature of the newly-identified compact cores a search
for outflows was carried out in the J=2--1 transition of $^{12}$CO
(230.538~GHz) using receiver A2 on the JCMT\@.  Observations were
made in 1998 July, and by JCMT staff in service mode throughout 1999.
Spectra were obtained in a five- or nine-point pattern, with the central
spectrum on the dust peak from the SCUBA 850-$\mu$m map.  The system
temperature was typically 440~K and the integration time of each spectrum
was 2 minutes, to give an rms level of 0.2~K in a bandwidth of 0.16~MHz
(0.2~km~s$^{-1}$).

Table~\ref{tab_cores} shows which submillimeter cores have been searched
for outflows, and whether or not high-velocity gas was detected.  Ten
cores were not observed in $^{12}$CO(2--1): L483$-$SMM1 is a well known
protostar whose outflow has already been extensively studied; L162$-$SMM1
is a T~Tauri star, and L158$-$SMM2 is the unreliable ``spiky'' source.
The other seven cores were not searched due to limited observing time,
but seem extended and are unlikely to contain embedded protostars
(Figure~\ref{fig_cores}).  They have therefore been classified as
``probably starless'' cores since, strictly speaking, their nature is
still unidentified.  Of the 40 cores identified in Section~\ref{sec_cores}
ten are known protostars, 8 of which were already known to drive outflows.
The two protostars for which this was not known are L162$-$SMM1 (a
Class~II object), and L1246$-$SMM1 (which does show high-velocity
line wings in these new data).  Of the 30 remaining cores, 22 were
searched for high-velocity gas, and the $^{12}$CO(2--1) line emission
from three of these cores (L673$-$SMM1, L944$-$SMM1, and L1709$-$SMM5)
shows high-velocity emission, indicative of an outflow.

Larger maps of the three new outflows, and seven of the known 8 flows
(all but L483$-$SMM1), were subsequently observed in $^{12}$CO(2--1)
in 1998 July and throughout 1999 in service mode with receiver A2 on
the JCMT\@.  Many of the old maps of the known outflows are incomplete,
of poor resolution, or not in $^{12}$CO(2--1), and our new observations
provide a homogeneous sample.  All these maps were made using the
on-the-fly mapping mode with a cell size of 5$''$ and an integration
time of 5~s per point.  The spectra were taken by position-switching to
eliminate instrumental and atmospheric effects.  Pointing was checked
regularly and the system was calibrated using standard chopper wheel
techniques to obtain spectra on the $T_{A}^{*}$ scale.  The main beam
efficiency at this wavelength is 0.66.

\subsection{Outflow maps}

The large-scale maps of the outflows have been slightly smoothed, and
are shown in Figures~\ref{fig_260outflow} to \ref{fig_1709smm5outflow}.
Previous observations and results from these new maps are described below.

\subsubsection{L260}

L260$-$SMM1 drives a very compact bipolar outflow, which was classed
only as an uncertain detection by Bontemps et al.\ (1996).  The outflow
has been detected again here.  This source has been classified as Class~I
(Bontemps et al.\ 1996), and also has been identified optically (Ichikawa
\& Nishida 1989).  Optical identifications are rare for Class~I sources,
and this fact, together with the detection of only a very compact and
weak outflow, suggest that this protostar is either fairly evolved and
accreting the last of its material, or is being observed close to pole-on
(Figure~\ref{fig_260outflow}).

\subsubsection{L483}
\label{sec_l483outflow}

The outflow from L483$-$SMM1 has been extensively studied so no new map
was obtained for this flow.  The high-velocity gas exhibits a distinct
east-west bipolar structure, with a high degree of symmetry about the
IRAS source (Parker 1989; Fuller et al.\ 1995).  The low-level, extended
emission in the SCUBA images of this source (Figure~\ref{fig_L483})
reflects the shape of the outflow cavities, as has been observed for
the Class~0 protostar L1527 (Chandler \& Richer 2000).  Indeed, the
detection of the outflow from L483$-$SMM1 in $^{12}$CO $J$=4--3 by
Hatchell, Fuller, \& Ladd (1999) confirms the presence of heated gas,
and the dust emission probably traces this same material.

\subsubsection{L663}

L663$-$SMM1 (B335) drives an outflow very similar in structure to that
of L483$-$SMM1, and has also been studied extensively (e.g., Cabrit,
Goldsmith, \& Snell 1988; Moriarty-Schieven \& Snell 1989).  The map
presented here (Figure~\ref{fig_663outflow}) is of limited extent and
does not cover the entire outflow, but is included for completeness.

\subsubsection{L673}
\label{sec_l673outflow}

The full extent of high-velocity gas in L673 mapped by Armstrong \&
Winnewisser (1989) in $^{12}$CO(1--0) emission is 12$'$.  The resolution
of their map was poor (HPBW = 3.9$'$), however, and it is not clear from
their data whether IRAS~19180+1114 or 19180+1116 drives the outflow.
Anglada et al.\ (1997) argue that it is more likely for IRAS~19180+1114 to
be the origin of the flow, but our new map (Figure~\ref{fig_673outflow})
suggests that both L673$-$SMM1 (associated with IRAS~19180+1116)
and L673$-$SMM2 (IRAS~19180+1114) have outflows.  At least, it is
clear that both sources have blue-shifted outflow, but the structure
of the red-shifted gas is more complicated.  The situation near the
southern cores in L673 is even more confused.  A small map ($120''
\times 80''$) was made to cover L673$-$SMM3 and L673$-$SMM5, and some
high velocity gas was detected (the $^{12}$CO line is 6.5~km~s$^{-1}$
wide), so it is possible that one of these two cores is a protostar.
The emission from L673$-$SMM3 is more centrally condensed than L673$-$SMM5
(Figure~\ref{fig673_cores}) and L673$-$SMM3 has therefore been classified
as a protostar candidate, and L673$-$SMM5 as a starless core, but
more observations are needed to confirm this.  Both L673$-$SMM6 and
L673$-$SMM8 also show some red-shifted high velocity gas, but this
is difficult to interpret from the limited $^{12}$CO data available.
Both cores have therefore been classified as starless.

\subsubsection{L944}

This protostar is newly detected by this survey, with both the continuum
emission and outflow previously unknown.  Its outflow is presented in
Paper I.

\subsubsection{L1165}

The bipolar outflow of L1165$-$SMM1 was partly observed by
Parker (1989), but is now fully covered by the new $^{12}$CO map
(Figure~\ref{fig_1165outflow}).

\subsubsection{L1172}

The outflow emanating from L1172 is very extended and was detected
in $^{12}$CO(1--0) by Myers et al.\ (1988).  Unfortunately the
$^{12}$CO(2--1) map presented here covers only a part of the outflow due
to limited observing time, and includes L1172$-$SMM2 but not L1172$-$SMM1
(see Figure~\ref{fig_1172outflow}).  It is, however, more likely that
L1172$-$SMM1 drives the outflow, since this submillimeter source is
closer to the IRAS source.  L1172$-$SMM1 has therefore been classified
as a protostar, and L1172$-$SMM2 as starless.  The properties of this
outflow have not been studied further, due to the limited map.

\subsubsection{L1246}

L1246$-$SMM1 is a previously known protostar (Launhardt et al.\ 1997), but
an outflow had not been detected, despite sensitive searches (Parker et
al.\ 1991).  The CO outflow has now been detected and mapped, confirming
the protostellar nature of this source, and is presented in Paper I.

\subsubsection{L1262}

The bipolar outflow from L1262$-$SMM1 has been observed by Bontemps
et al.\ (1996).  The new map made with the JCMT is larger in extent,
and is shown in Figure~\ref{fig_1262outflow}.

\subsubsection{L1709}

A blue-shifted outflow driven by L1709$-$SMM1 has been presented by
Bontemps et al.\ (1996).  Parker (1989) noted, however, that a red-shifted
outflow is very likely also present, and this red-shifted outflow has now
been detected (Figure~\ref{fig_1709smm1outflow}).  A new outflow emanating
from L1709$-$SMM5 has also been mapped (Figure~\ref{fig_1709smm5outflow}).
Spectra obtained in a five-point pattern centered on L1709$-$SMM3
showed some red-shifted, high-velocity gas, but this is probably due
to the outflow emanating from L1709$-$SMM1, and L1709$-$SMM3 has been
classified as starless.

\subsection{Outflow dynamics}
\label{sec_flowdyn}

The dynamical properties of the observed outflows are given in
Table~\ref{tab_outflows}.  The mass of high-velocity molecular gas in
the outflows has been calculated assuming low optical depth and local
thermodynamic equilibrium with an excitation temperature $T_{\rm ex}
= 20$~K\@.  The abundance of $^{12}$CO relative to H$_2$ is assumed to
be $5 \times 10^{-5}$, and the distances used in the calculations are
those in Table~\ref{tab_sample}.  The kinetic energy in the flow, $\Sigma
MV^2/2$, has been calculated for velocity intervals of 0.5~km~s$^{-1}$
or 1.0~km~s$^{-1}$ and summed.  For example, the energy in the L1262
outflow is summed between 6 and 10.5~km~s$^{-1}$ for the red-shifted
gas, and between $-$3 and 2~km~s$^{-1}$ for the blue-shifted gas.
The velocity boundaries for each outflow were established by comparison
with an ``ambient'' spectrum obtained away from the outflow, using the
same criteria as those of Bontemps et al.\ (1996): the high-velocity bound
of the blue-shifted wing (and low-velocity bound of the red-shifted wing)
was chosen to be the velocity at which the ambient spectrum had $T^*_A =
T_{\rm peak}/10$ on the blue side (red side); the low-velocity bound of
the blue-shifted wing (and high-velocity bound of the red-shifted wing)
correspond to the velocity at which emission is no longer positively
detected above the noise.  The extent of the outflow is the distance
from the core to the edge of the red-shifted outflow, plus the core to
the edge of the blue-shifted outflow.  The outflow in L663 has not been
completely mapped and the extent measured is therefore a lower limit.
The velocity width is the total width of the outflow at the one sigma
noise level.  The dynamical age, $\tau_{\rm d}$, is the outflow extent
divided by the velocity width, and is an indication of the lifetime
of the outflow.  No correction has been made for flow inclination in
calculating the above quantities.

The momentum flux (or force) of an outflow is the ratio of the flow
momentum to the dynamical age.  We have applied a correction factor
of 10 in calculating the values of the momentum flux displayed in
Table~\ref{tab_outflows} to account for a combination of the optical
depth of the CO emission and a mean inclination angle of the flow,
thereby enabling a direct comparison with the momentum fluxes presented
by Bontemps et al.\ (1996).  Bontemps et al.\ argue that, if outflows
are momentum driven and momentum is conserved along the flow direction,
full maps of an outflow are not needed in order to obtain an estimate of
the momentum flux.  In the case of L663$-$SMM1 (B335) we have investigated
this by comparison with values from the literature.  Based on a map even
more limited than that shown in Figure~\ref{fig_663outflow}, Bontemps et
al.\ derive a momentum flux (corrected using the above factor of 10 for
line opacity and flow inclination) for the L663 outflow of $1.2 \times
10^{-5}~M_\odot$~km~s$^{-1}$~yr$^{-1}$.  These authors do not specify
the assumed $^{12}$CO to H$_2$ abundance used in their calculations,
although much of their analysis is compared directly with the work
of Cabrit \& Bertout (1992), who use a value of $1 \times 10^{-4}$, a
factor of 2 higher than that assumed here.  Converting the Bontemps et
al.\ value for the momentum flux to an abundance of $5 \times 10^{-5}$
therefore gives $2.4 \times 10^{-5}~M_\odot$~km~s$^{-1}$~yr$^{-1}$.
Cabrit et al.\ (1988) mapped the entire flow, and using a $^{12}$CO
to H$_2$ abundance of $1.8 \times 10^{-4}$ obtain a momentum flux
of $2.4 \times 10^{-5}~M_\odot$~km~s$^{-1}$~yr$^{-1}$, including
measured corrections for optical depth and flow inclination.
Converting this to the same abundance used here gives $8.6 \times
10^{-5}~M_\odot$~km~s$^{-1}$~yr$^{-1}$.  Our value for the momentum flux
of $4.7 \times 10^{-5}~M_\odot$~km~s$^{-1}$~yr$^{-1}$ lies in the same
range as those previously reported, when the different abundances assumed
are taken into account.  Thus to within a factor of a few the estimates
of the outflow momentum flux using the Bontemps et al.\ assumptions seem
reasonable, and can be used for qualitative comparisons within the sample
since they are calculated for all sources in a consistent manner.

Most of the outflows are clearly bipolar.  The outflows emanating from
L260$-$SMM1 and L1709$-$SMM5 have their red- and blue-shifted lobes
overlapping to a large degree, and we may be observing these flows
close to pole-on.  The outflows in L673 are the most complex.  The two
submillimeter cores clearly drive blue-shifted outflows, but it is less
obvious which source drives the red-shifted gas.  The outflow properties
of both L673$-$SMM1 and L673$-$SMM2 in Table~\ref{tab_outflows} are
therefore based only on the blue-shifted gas.  The red-shifted outflow
is very massive (0.07~$M_\odot$) and energetic ($2.2 \times 10^{36}$~J)
compared with the other outflows in the sample, and may be a combination
of flows associated with both cores.

\section{Classification of the protostars}
\label{sec_protostars}

A statistical analysis of source properties and relative lifetimes
requires that all the protostars in our sample be classified according
to some scheme approximating age or evolutionary status.  Several such
classifications have been used in the past, from the rather coarse
division according to the near- to mid-infrared SED, to the more
continuous bolometric temperature ($T_{\rm bol}$).  Other source
properties have also been found to correlate with either the SED
classification or $T_{\rm bol}$, the subsequent interpretation of which
can be somewhat model dependent.  Here we place the protostars observed
in our sample into these classification schemes and compare the results
with those reported in the literature.

\subsection{Spectral energy distributions}

The flux densities of the protostars at 450 and 850~$\mu$m in a
50$''$ aperture have been measured after having subtracted background
emission, and are presented in Table~\ref{tab_proto}.  For those sources
with previously published submillimeter flux densities (L483$-$SMM1,
L663$-$SMM1, L1172$-$SMM1, L1246$-$SMM1, and L1262$-$SMM1: see Launhardt
et al.\ 1997; Huard, Sandell, \& Weintraub 1999; Shirley et al.\ 2000)
our measurements are in good agreement with those earlier results.
Our 450 and 850~$\mu$m data are also plotted in Figure~\ref{fig_seds}
along with the far-infrared IRAS fluxes from Table~\ref{tab_iras}.
The protostar L944$-$SMM1 is not associated with an IRAS source from the
PSC, but its far-infrared flux densities were obtained from HIRES data,
as described in Paper~I\@.  Far-infrared flux densities for the protostars
in L673 and L1709, where inspection of Table~\ref{tab_iras} indicates
considerable confusion in the PSC, were also obtained from HIRES images
(Figures~\ref{L673_hires} and \ref{L1709_hires}) for all IRAS bands except
100~$\mu$m, for which the resolution is insufficient to separate emission
from the multiple sources.  Upper limits only are therefore given for
100~$\mu$m flux densities.  At 60, 25, and 12~$\mu$m the emission was
integrated in a box around the protostar after subtraction of a background
level (a circular aperture does not work here because of the elliptical
IRAS beam).  The HIRES flux densities are listed in Table~\ref{tab_hires};
the absolute uncertainty in these values is $\sim 30$\%.

The SEDs have been fitted with a single-temperature greybody
spectrum, $F_\nu \propto \nu^\beta B_\nu(T_{\rm dust})$.  The dust
emission is assumed to be optically thin, and $\beta$ is fixed at 1.5.
The fitted values of the dust temperature, $T_{\rm dust}$, are given in
Table~\ref{tab_proto}, and spectra are plotted in Figure~\ref{fig_seds}.
Where available, the greybody spectrum is fitted to 850, 450, and
100~$\mu$m data.  For the protostars with no 450~$\mu$m flux density
measurement only the 850 and 100~$\mu$m flux densities are used (i.e.,
for L162$-$SMM1, L633$-$SMM1, L1165$-$SMM1, and L1262$-$SMM1).  For the
four protostars with 100~$\mu$m upper limits the 850, 450, and 60~$\mu$m
flux densities are used (i.e., L673$-$SMM1, L673$-$SMM2, L1709$-$SMM1
and L1709$-$SMM5).  All fits have reduced chi squared $\chi^2_\nu < 1$
apart from L1709$-$SMM5, for which $\chi^2_\nu = 1.3$.

In order to classify the protostars according to the scheme proposed
by Lada \& Wilking (1984) based on the near- to mid-infrared spectral
index, and extended to sources not detected in the infrared by Andr\'e,
Ward-Thompson, \& Barsony (1993), infrared measurements are needed.
L162$-$SMM1 is an optically-identified T~Tauri star (Ichikawa \& Nishida
1989) and is classified as Class~II by Andr\'e \& Montmerle (1994).
The protostars detected in the near-infrared are L260$-$SMM1 (Myers et
al.\ 1987), L1165$-$SMM1 (Tapia et al.\ 1997), and L1709$-$SMM1 (Myers et
al.\ 1987), and so are Class~I protostars.  The Class~0 protostars not
detected in the near infrared are L483$-$SMM1 (Fuller et al.\ 1995),
and L663$-$SMM1 (Hodapp 1998).  L1246$-$SMM1 and L1262$-$SMM1 have
been classified as Class~0 sources by Launhardt et al.\ (1997) based
on unpublished near-infrared photometry.  The remaining five protostars
have to be classified without near-infrared measurments.

\subsubsection{Ratio of $L_{\rm bol}/L_{\rm submm}$}
\label{sec_lbol_lsubmm}

A method proposed for classifying protostars by Andr\'e et al.\ (1993)
that does not depend on near- and mid-infrared detections is to evaluate
the ratio $L_{\rm bol}/L_{\rm submm}$, which for a constant mass accretion
rate should be approximately proportional to $M_*/M_{\rm env}$.  The ratio
$L_{\rm bol}/L_{\rm submm}$ will increase with age, as material from
the envelope is accreted onto the protostar.  Andr\'e et al.\ classify
protostars as Class~0 when $L_{\rm bol}/L_{\rm submm} < 200$, where
$L_{\rm bol}$ is measured from 1 to 1300~$\mu$m, and $L_{\rm submm}$
is the submillimeter luminosity from 350 to 1300~$\mu$m.  Unfortunately,
$L_{\rm bol}$ cannot be determined without near-infrared measurements,
and the best we can do here is to calculate the luminosity from 12 to
1300~$\mu$m.  This will underestimate $L_{\rm bol}$, particularly for
the older sources.  Lower limits to $L_{\rm bol}$ have therefore been
calculated by integrating under the greybody fit from 1300 to 150~$\mu$m,
and by regarding the IRAS PSC or HIRES upper limits as real detections.
This latter assumption may tend to overestimate $L_{\rm bol}$, and
Figure~\ref{fig_seds} shows it will affect mainly those protostars not
detected by IRAS at 12~$\mu$m.  For both L663$-$SMM1 and L944$-$SMM1,
which are not detected at either 12 or 25~$\mu$m, the bolometric
luminosity has been calculated by integrating under the greybody fit only.
Estimates of $L_{\rm bol}$, $L_{\rm submm}$, and $L_{\rm bol}/L_{\rm
submm}$ are listed in Table~\ref{tab_proto}.  Applying the criterion
that a protostar is a Class~0 source when $L_{\rm bol}/L_{\rm submm} <
200$ results in only one Class~I source in this sample, suggesting that
a different limit should be used for a sample that does not include
near-infrared emission in determining $L_{\rm bol}$.  Indeed, a limit
of $\sim 50$ is needed to reproduce most of the classifications of those
protostars with previous measurements or good upper limits in the near-
and mid-infrared.  The only source which then does not match its previous
classification is L483$-$SMM1, which has recently been described as a
transition Class~0/Class~I object by Tafalla et al.\ (2000) based on a
study of the emission from various molecular species.

\subsubsection{Bolometric temperature}

A more continuous indicator of age, the bolometric temperature ($T_{\rm
bol}$), was introduced by Myers \& Ladd (1993).  $T_{\rm bol}$ is
the temperature of a blackbody having the same mean frequency as
the observed continuum spectrum, and is a measure of circumstellar
obscuration.  Chen et al.\ (1995) quantify the Class~0/Class~I boundary
as $T_{\rm bol} = 70$~K, and Class~I/Class~II as $T_{\rm bol} = 650$~K\@.
Table~\ref{tab_proto} lists values of $T_{\rm bol}$ derived from the
SEDs in Figure~\ref{fig_seds} in a manner similar to the measurement of
$L_{\rm bol}$ described above.  Again, the lack of near- and mid-infrared
data for these sources will tend to lower the derived $T_{\rm bol}$.
Nevertheless, the values in Table~\ref{tab_proto} should be a good
indicator of evolutionary state, albeit with different class boundaries
from those obtained by Chen et al.  L162$-$SMM1 has a $T_{\rm bol}$
considerably lower than the 650~K limit assigned by Chen et al., but
this source is also likely to be the most seriously affected by the
lack of near-infrared data included in its SED\@.  A Class~0/Class~I
boundary at $T_{\rm bol} = 65$~K gives reasonably good agreement with
previous classifications for the other protostars.  The class implied
for L1709$-$SMM5 obtained from its $T_{\rm bol}$ is not consistent with
that implied by its $L_{\rm bol}/L_{\rm submm}$, and this source may be
a transition object similar to L483$-$SMM1.

\subsection{Envelope masses}

The mass of circumstellar material associated with each protostar has
been estimated from its 850~$\mu$m flux density and the dust temperature
derived from fitting its SED, and is also given in Table~\ref{tab_proto}.
A dust opacity $\kappa_{850\mu{\rm m}} = 0.002$~m$^2$~kg$^{-1}$ has
been assumed, consistent with the 1.3~mm dust opacity recommended by
Ossenkopf \& Henning (1994) for very dense regions, for $\beta = 1.5$.
The envelope masses of the protostars obtained in this way range from 0.01
to 2.15~$M_\odot$, and agree well with masses derived by other authors
for previously-known sources (Bontemps et al.\ 1996; Launhardt et al.\
1997; Andr\'e, Ward-Thompson, \& Barsony 2000).

\subsubsection{Ratio of $M_{\rm env}/L_{\rm bol}$}

Under the assumption that for Class~0 and Class~I protostars derive
their luminosity primarily from accretion such that $L_{\rm bol} = G
M_* \dot{M}/R_*$ , and that the accretion rate, $\dot{M}$, and stellar
radius, $R_*$, are constant, the ratio $M_{\rm env}/L_{\rm bol}$ is
proportional to $M_{\rm env}/M_*$.  If the Class~0/Class~I division is set
at $M_{\rm env}/M_* = 1$ (such that Class~0 sources have $M_{\rm env}/M_*
> 1$), then $\dot{M} = 10^{-6} M_\odot$~yr$^{-1}$ and $R_* = 3 R_\odot$
results in $M_{\rm env}/L_{\rm bol} > 0.1 M_\odot/L_\odot$ for Class~0
sources (Andr\'e \& Montmerle 1994).  This criterion for classification
identifies those same objects as Class~0 sources as does the $L_{\rm
bol}/L_{\rm submm} > 50$ criterion above (Section~\ref{sec_lbol_lsubmm};
see also Table~\ref{tab_proto}).

\subsection{Outflow properties}

A correlation between outflow momentum flux, $F_{\rm CO}$, with
envelope mass was found by Bontemps et al.\ (1996), and has been
confirmed for other source samples (e.g., Henning \& Launhardt 1998).
Class~0 sources drive more powerful, and better-collimated outflows
than Class~I protostars.  Figure~\ref{fig_correlatie} plots $F_{\rm CO}$
versus $M_{\rm env}$ for the ten protostars for which we have outflow data
(Figures~\ref{fig_260outflow} to \ref{fig_1709smm5outflow}).  The outflow
momentum flux for both L673$-$SMM1 and L673$-$SMM2 are lower limits due
to the confusion concerning the red outflow.  The dashed line plotted
is the best linear fit between $\log(F_{\rm CO})$ and $\log(M_{\rm
env})$ found by Bontemps et al.\ (1996).  It is clear that the momentum
fluxes of outflows in our sample agree very well with those observed by
Bontemps et al., with only two sources lying at some distance from the
relationship found by those authors.  These are L1246$-$SMM1 (the Class~0
source, to the right of the dashed line in Figure~\ref{fig_correlatie}),
and L1709$-$SMM5 (the Class~I source to the left of the dashed line).
The displacement of L1246$-$SMM1 might be explained by a wrongly-assigned
distance.  It is listed in Table~\ref{tab_sample} as having $D = 730$~pc,
and as such is the most distant in this sample.  If this source were
actually closer its position in Figure~\ref{fig_correlatie} would shift
to the left and down, since $M_{\rm env} \propto D^2$, and $F_{\rm
CO} \propto D$.  The overlapping red- and blue-shifted lobes of the
L1709$-$SMM5 outflow was suggested above (Section~\ref{sec_flowdyn})
to indicate we may be observing this source pole-on.  Were this the
case, the correction factor applied to $F_{\rm CO}$ that accounts for
a mean inclination angle and CO line optical depth may be severely
overestimated, and could explain the location of this source in
Figure~\ref{fig_correlatie}.

Bontemps et al.\ (1996) also find that outflows from Class~0 protostars
are more efficient than those from Class~I protostars, when $F_{\rm CO}$
is compared to the maximum momentum flux available in stellar photons,
$L_{\rm bol}/c$.  The outflow efficiency, given by the ratio $F_{\rm CO}
c/L_{\rm bol}$, is listed in Table~\ref{tab_proto} for the ten outflow
sources mapped and this trend is clearly evident: the Class~0 sources
typically have $F_{\rm CO} c/L_{\rm bol} \ga 300$.  Table~\ref{tab_proto}
also illustrates the nature of L1709$-$SMM5 as an object in transition
between Class~0 and Class~I\@.  L1246$-$SMM1 again stands out as having
an outflow considerably weaker than other Class~0 sources.  The outflow
efficiency scales as $1/D$, so an overestimated distance might again help
bring this source into line with the others in this sample.  The distance
of L1246 has been estimated by associating it with Cep OB3 (Paper I).
If it were close enough to raise its outflow efficiency by the factor
of 4 or 5 needed to bring it in line with to the Bontemps et al.\
correlation it would have to be rather isolated.

\subsection{Sizes}

The emission from the Class~0 sources is extended, and some objects,
such as L483$-$SMM1, clearly exhibit structure related to the outflow.
All the sources identified as Class~I protostars in Table~\ref{tab_proto}
are unresolved apart from L1709$-$SMM5.  L1709$-$SMM5 stands out among
all the cores identified as protostellar in nature as being the most
extended, and in many respects resembles the starless cores.  This raises
the possibility that L1709$-$SMM5 in fact comprises several sources in
superposition along the line of sight.  Higher resolution observations
are needed to confirm this idea.

\section{Physical properties of the starless cores}

The masses, column densities and inferred space densities of
those cores not containing embedded protostars are listed in
Table~\ref{tab_starless}, along with the protostar candidate L673$-$SMM3.
The integrated flux densities at 850~$\mu$m have been measured within
an aperture after the subtraction of background emission as described in
Section~\ref{sec_core_id}.  For some of the more extended cores, such as
L260$-$SMM2, the subtraction of background emission may subtract extended
emission from the core itself.  However, in most cases the background
subtraction is clearly desirable, in order to obtain integrated flux
densities representing only the compact submillimeter core.  The size
of the aperture used was determined from radially-averaged intensity
profiles, after background subtraction, to include all the observed
emission.  Table~\ref{tab_starless} gives the diameter of the aperture
used for each core.  The masses were calculated assuming $T_{\rm
dust} = 12$~K and $\kappa_{850\mu{\rm m}} = 0.001$~m$^2$~kg$^{-1}$.
The space densities assume that the core has a depth along the line of
sight the same as the aperture diameter.  The mean mass of the cores is
1.1~$M_\odot$, the mean column density $3.4 \times 10^{25}$~m$^{-2}$,
and the mean space density $1.0 \times 10^{10}$~m$^{-3}$.

The masses of the starless cores are systematically higher than those
of the protostellar envelopes, by approximately 0.6~$M_\odot$.  If the
starless cores are indeed the progenitors of the protostars in this
sample, we might therefore expect that the starless cores will, in time,
produce stars of $\sim 0.5$--1~$M_\odot$.  This result must be taken with
some caution, since different dust opacities and dust temperatures have
been used for calculating the masses for each population, in such a way
that these two factors both serve to increase the apparent mass of the
starless cores.  However, the assumption that the temperatures in the
protostellar envelopes are higher than those in the starless cores is
probably a good one, and even if the same dust opacity were used for
all sources the starless cores would still be more massive than the
protostellar envelopes.

The 26 starless cores in Table~\ref{tab_starless} represent the largest
sample of these objects for which submillimeter continuum images have
been made to date, and therefore provide a unique dataset with which to
study the properties of these objects.  In particular, if the cores are
isothermal, the radial intensity profile of the submillimeter emission
can be directly inverted to determine the radial density profile in
the cores.  Analytic and numerical studies show that the mass infall
rate during the protostellar phase depends on the radial density profile
at the onset of collapse, and on the equation of state of the material.
Singular models have an $n \propto r^{-2}$ density profile resulting in
a constant accretion rate (Shu 1977).  Density profiles with a flatter
inner region, however, will result in a high mass-infall rate initially,
possibly corresponding to the Class~0 phase, followed by a phase with a
low accretion rate, possibly corresponding to the Class~I phase (Foster
\& Chevalier 1993; Henriksen, Andr\'e, \& Bontemps 1997).  The evolution
of a protostar will also depend on whether the core from which it is
forming has an ``edge.''  The Shu picture has star formation taking
place in an essentially infinite reservoir of material, in which case
the final mass of the star being formed will be determined by a local
process, such as the interaction between its outflow and accretion.
If, however, the core has a well-defined edge, the final stellar mass
may determined by the structure of the core in the pre-collapse phase.

The starless cores that have been studied to date in
millimeter/submillimeter continuum emission do not appear to show the
$r^{-2}$ density profile predicted by the singular isothermal model of
Shu (1977).  Instead, they have shallow density profiles in their inner
regions, in qualitative agreement with models of magnetically-supported
cloud cores (Ward-Thompson et al.\ 1994; Andr\'e, Ward-Thompson, \&
Motte 1996; Motte, Andr\'e, \& Neri 1998; Ward-Thompson et al.\ 1999;
Shirley et al.\ 2000).  The density profiles of a number of starless
cores have also been studied in absorption in the mid-infrared (Bacmann
et al.\ 2000).  These also exhibit density profiles that flatten toward
the center, and several also have sharp edges.  Such profiles need very
strong background magnetic fields to be supported, but they are also
well-fitted by Bonnor-Ebert-type hydrostatic equilibrium models for
isothermal, self-gravitating, cores confined by external pressure.

Radial intensity profiles for all the starless cores in our sample
were obtained from the SCUBA 850-$\mu$m images by averaging data
points in a 3$''$ wide ring at radius $r$ from the core center
(Table~\ref{tab_cores}), from $r = 0''$ to 450$''$ or the edge of the map.
The mean ellipticity of the starless cores is 1.5, and the assumption of
circular symmetry should therefore result in a representative intensity
profile (see also Andr\'e et al.\ 1996), although there is clearly
scope for more detailed modelling.  Eighteen of the cores have had DC
offsets subtracted to obtain sensible zero levels for the background.
Emission from nearby cores or protostars were not removed before the
profile fitting.  A single power-law model for the density distribution
did not fit most of the starless cores, so the intensity profiles have
been modeled with a density distribution consisting of two power laws.
The density is assumed to have the form $n \propto r^{-\gamma}$ from
$r = 100$~AU to a radius $r_0$, beyond which $n \propto r^{-\delta}$.
The density at $r < 100$~AU was assumed to be zero to avoid any
singularity.  An outer cutoff radius $r_1$ has also been implemented to
model the observed edge of the profile.  A temperature $T_{\rm dust}
= 12$~K has been assumed, and dust opacity $\kappa_{850\mu{\rm m}} =
0.0012$~m$^2$~kg$^{-1}$.  The resulting intensity profile was calculated
from the model by integrating the density distribution along the line of
sight, and convolving it with a 15$''$ Gaussian to simulate the JCMT beam
at 850~$\mu$m.  The four parameters $r_0$, $r_1$, $\gamma$, and $\delta$
were then adjusted to produce the best fit.  The density at $r_0$, $n_0$,
was normalized to reproduce the observed flux density.

Profiles of all the starless cores were fitted except for
L158$-$SMM2, which is a suspected artifact of the regridding algorithm
(Secton~\ref{sec_core_id}), and L1709$-$SMM4, which is an extended
ridge and clearly should not be modelled by a spherically-symmetric core
(Figure~\ref{fig_cores}).  All the fitted cores apart from L673$-$SMM6
and L951$-$SMM2 show a flat inner part in the intensity profile,
with a steepening in the radial power-law exponent beyond $r_0$,
out to $r_1$ where the intensity profile merges into the background.
Excluding these two cores we find mean values of $\gamma = 0.9 \pm 0.3$
and $\delta = 1.9 \pm 0.3$.  Figure~\ref{fig_profiles} shows one of
the best fits, L694$-$SMM1, an average fit, L1262$-$SMM2, and a very
poor fit, L673$-$SMM7.  The parameters of these fits are also given in
Table~\ref{tab_starless}.  For none of our cores is the radial density
profile steeper than $r^{-3}$, as was found for three starless cores
by Bacmann et al.\ (2000).  However, the best-fit values of $\delta$
are positively correlated with values of $r_1$ in our fitting, so that
a larger value of $\delta$ can, to some extent, be compensated for by
a larger value of $r_1$.

Both the results presented here and previous work seem to rule out the
singular isothermal spheres developed by Shu and coworkers (but see also
the recent, more detailed, modeling of starless cores by Evans et al.\
2001), and the logotropic model having $n \propto r^{-1}$ at large radii
proposed by McLaughlin \& Pudritz (1997), as the pre-collapse density
profile of starless cores.  Instead, the radial density profiles resemble
more closely those expected for pressure-bounded-Ebert spheres (e.g.,
Foster \& Chevalier 1993) or for which magnetic pressure is important
(e.g., Ciolek \& Mouschovias 1994; Basu \& Mouschovias 1995).

\section{Statistics, timescales, and implications for low-mass star
formation}
\label{stats}

Assuming star formation is a homogeneous process in our sample of dark
clouds, and that the starless cores $\longrightarrow$ Class~0 sources
$\longrightarrow$ Class~I protostars represent an evolutionary sequence,
statistical lifetimes of these different stages can be derived from the
relative numbers of objects in these different phases.  We find a total
of 7 Class~0 sources, 5 Class~I sources, one candidate protostar, and
26 starless cores.  All these cores and protostars are observed in 21
clouds, i.e., in half of the sample of optically-dark molecular clouds.
Fifty percent of the clouds are therefore quiescent.

\subsection{Ratio of Class~0 to Class~I sources}

One of the most remarkable results of this survey is that the ratio of
Class~0 sources to Class~I sources is not 1 to 10, as has been observed
in previous surveys of the $\rho$ Ophiuchi main cloud (Andr\'e \&
Montmerle 1994; Motte et al.\ 1998), which are also believed to be
complete.  Furthermore, this result comes in spite of the fact that
a significant fraction of the cores in our sample are associated with
the $\rho$ Ophiuchi complex.  Separating our sample into those clouds
that are associated with $\rho$ Ophiuchus and those that are not
results in rather small numbers, but the comparison is nevertheless
worth investigating.  For the Lynds clouds associated with $\rho$
Ophiuchus (see Table~\ref{tab_sample}) we find 3 Class~I protostars, 1
Class~II source, and 10 starless cores, giving a Class~0 to Class~I ratio
consistent with earlier results.  For the non-$\rho$ Ophiuchi clouds
we find 7 Class~0 sources, 2 Class~I sources, and 17 starless cores.
Excluding the clouds in $\rho$ Ophiuchus therefore makes the potential
ratio of Class~0 to Class~I sources even higher.

We have considered whether the discrepancy between the ratio of Class~0
to Class~I sources observed in Lynds dark clouds compared with that
observed for $\rho$ Ophiuchus is the result of the completeness limit of
our sample ($M_{\rm env} = 0.015~M_\odot$).  If, for example, the Lynds
clouds form predominantly lower-mass stars than does $\rho$ Ophiuchus,
then our mass completeness limit would be insufficient to detect all the
Class~I sources associated with the Lynds clouds.  In order for such an
effect to account for the observed discrepancy, approximately 90\% of
the Class~I sources in the Lynds clouds would have to have $M_{\rm env} <
0.015~M_\odot$.  In a recent study of embedded protostars in the low-mass,
star-forming region of Taurus, Motte \& Andr\'e (2001) found that in a
total of 23 Class~I sources only 5 have $M_{\rm env} < 0.015~M_\odot$.
It therefore seems unlikely that our completeness limit can account for
the discrepancy, unless star formation in the Lynds clouds proceeds very
differently from star formation in both Taurus and $\rho$ Ophiuchus.

While there are difficulties in the classification of some of our objects
(Section~\ref{sec_protostars}), we can fairly easily rule out a 1:10
ratio for the full sample, and the statistics are more consistent with
a 1:1 ratio.  This suggests that the lifetime of a Class~0 source in a
Lynds dark cloud is as long as the lifetime of a Class~I source, a result
that is supported by the dynamical lifetimes of the outflows.  The mean
dynamical age of the Class~0 outflows presented here is $2.7 \times
10^4$~yr, compared with the Class~I outflows, of $1.8 \times 10^4$~yr.
Even if the dynamical lifetime of the outflow in L1246 is overestimated
due to the uncertainty in its distance, the mean age of the remaining
Class~0 outflows is $2.2 \times 10^4$~yr, similar to that measured for
the Class~I sources.

The lifetime of the Class~I phase is believed to be reasonably well
established due to complete infrared surveys of the $\rho$ Ophiuchus
and Taurus star forming regions (Wilking, Lada, \& Young 1989; Kenyon
et al.\ 1990).  If a lifetime of $\sim 2 \times 10^5$~yr is adopted
for Class~I protostars, the statistical lifetime of Class~0 sources in
this sample will be similar.  Of course, this assumes that stars form in
molecular clouds continuously and in a homogeneous fashion.  However, the
different ratio of the numbers of Class~0 to Class~I sources found here,
compared with other surveys, might indicate that stars do not form in such
a homogeneous way, demonstrating that the star formation process is highly
dependent on the local environment.  There are three ways this might work:

\begin{itemize}

\item Class~0 sources may represent a different branch of star formation,
and may not be precursors of Class~I sources.  Indeed, Jayawardhana,
Hartmann, \& Calvet (2001) have suggested that some Class~0 sources
may be protostars forming in very high density regions, and are in fact
at the same evolutionary stage as the Class~I sources forming in lower
density environments.  If this were true, however, it would imply that
the Lynds dark clouds are of higher density than the cores in $\rho$
Ophiuchus.  A comparison of our results with those of Motte et al.\
(1998) shows the opposite to be the case.

\item If Class~0 sources are precursors to Class~I sources, it may be that
the environment, with its local initial conditions, determines how long
the Class~0 phase lasts.  For example, if the Lynds clouds represent a
more quiescent environment than the clouds in $\rho$ Ophiuchus, there
may be more time for the starless cores in the Lynds clouds to evolve
towards a singular isothermal sphere before the onset of collapse.
In this case the collapse might progress with a constant accretion rate,
and the Class~0 phase could then last longer in these clouds.  In $\rho$
Ophiuchus the star formation may be triggered before the starless cores
reach a singular state, in which case a more dramatic Class~0 phase
(higher accretion rates) would precede the Class~I phase.  However,
none of the starless cores detected here has the characteristics of a
singular isothermal sphere on small scales.

\item Perhaps a more likely possibility is that star formation is not
a steady process.  If star formation relies heavily on triggering, the
sample of embedded protostars in $\rho$ Ophiuchus may be dominated by
an older, more evolved population due to a burst of star formation some
$\sim 10^5$~yr ago.  Indeed, on large scales it is thought that star
formation in $\rho$ Ophuichus may have been triggered by a supernova
in the Scorpio-Centaurus OB association (de Geus 1992).  Motte et al.\
(1998) conclude that the small-scale structure of the $\rho$ Ophiuchi
main cloud is also consistent with the passage of a shock, and that the
youth of the protostars over a large region indicate star formation must
have been synchronized by an external mechanism.

\end{itemize}

The previous, short, lifetimes derived for Class~0 protostars in $\rho$
Ophiuchus also had another important implication for accretion during
early protostellar evolution.  A distinguishing feature of Class~0
protostars is that they are surrounded by a considerably larger
reservoir of circumstellar material than Class~I protostars (see, e.g.,
Table~\ref{tab_proto}), and a short time spent in this phase implied
a considerably higher accretion rate for Class~0 protostars compared
with the Class~I phase.  If, instead, protostars spend a similar
amount of time in the Class~0 and Class~I phases, there may be no such
initial period of rapid accretion.

\subsection{Ratio of starless cores to protostars}

There are 26 starless cores compared with a total of 12 or 13 embedded
protostars in our sample of Lynds clouds, so the ratio is approximately
2:1.  This suggests that the lifetime of a starless core as detected in
the SCUBA survey is only twice as long as the lifetime of an embedded
protostar.  The lifetime of the Class~0 and Class~I phase combined is
$\sim 4 \times 10^5$~yr, and from this the estimated lifetime of starless
cores is $\sim 8 \times 10^5$~yr.  This is only 2--3 times the free-fall
time for an average core in the sample, $\sim 3 \times 10^5$~yr.  Using
the power-law density profiles given in Table~\ref{tab_starless} we have
examined whether the thermal pressure of gas at a temperature of 10--12~K
is sufficient to provide support against gravitational collapse for the
starless cores on scales $r \sim 10^4$~AU\@.  In all cases a velocity
dispersion of $\sim 0.1$--0.2~km~s$^{-1}$ is required, comparable to the
isothermal sound speed for gas at 10~K\@.  It therefore seems entirely
feasible that the starless cores are in hydrostatic equilibrium and
do not need the support of magnetic fields {\it on these size scales}.
The cores are, therefore, on the verge of collapse, and it is unlikely
that strong magnetic fields are important for the collapse dynamics of
starless cores once they reach densities of $\sim 10^{10}$~m$^{-3}$.
Of course, this does not rule out the possibility that magnetic fields
are important for providing support in molecular clouds on larger size
scales, nor does it suggest that magnetic fields may not be important
dynamically for the {\it formation} of the starless cores.

\subsection{Ratio of quiescent clouds to star-forming clouds}

Since all the compact cores are observed in half the sample of dark clouds
the other half of the sample are quiescent clouds.  The combined duration
of the starless core phase and embedded phase is $\sim 1 \times 10^6$~yr,
and the statistical lifetime of a quiescent Lynds Class 6 cloud is thus
also $\sim 1 \times 10^6$~yr.  The comparison made here is, however,
difficult, since all the clouds have different sizes, and some of the
clouds are complexes of smaller clouds (e.g., L673).  Nevertheless,
these clouds are known to have reasonably long lifetimes and should at
least be stable for long enough to form stars.  Some T~Tauri stars are
also still found in this type of cloud, suggesting that the clouds have
lifetimes $\ga 10^6$~yr.

\section{Summary and conclusions}

We have surveyed a sample of optically-selected dark clouds for the
submillimeter dust emission associated with embedded protostars and
starless dense cores.  The optical selection criterion is equivalent to
a limiting column density ($A_V \ga 10$ mag), and avoids biases relating
to the infrared properties of older protostars and other indicators
of star formation potential or activity.  Furthermore, the clouds are
predominantly nearby, giving a sample for which good spatial resolution
can be obtained.  All clouds were imaged at $\lambda = 850$~$\mu$m,
with a spatial resolution of $\sim 2,000$--10,000~AU, depending on the
distance.  A total of 42 dark clouds, covering an area of 0.5 square
degrees, are included in the survey.

Compact submillimeter cores have been identified in the clouds, and we
have established whether the cores contain embedded protostars through
a combination of association with IRAS emission and/or the presence of
high-velocity CO(2--1) emission from protostellar outflows.  The survey
is complete for starless cores, and for Class~0 and Class~I protostars,
to a mass limit of 0.015~$M_\odot$.  Half of the clouds, 21 in total, do
not contain any compact submillimeter cores, and are therefore quiescent
as far as star formation activity is concerned.  The other half contain a
total of 7 Class~0 protostars, 5 Class~I protostars, one Class~II source,
a candidate protostar (L673$-$SMM3), and 26 starless cores.  The ratio of
Class~0 to Class~I protostars found in this survey is therefore close to
unity, and is not consistent with the previous result found for the $\rho$
Ophiuchi cloud, where a ratio of 1 to 10 has been observed (Andr\'e \&
Montmerle 1994; Motte et al.\ 1998).  The ratio of starless cores to
cores containing embedded protostars is approximately 2 to 1.  The implied
lifetimes of the Class~0 and starless core phases are therefore $\sim 2
\times 10^5$~yr and $\sim 8 \times 10^5$~yr respectively.

The different ratio of Class~0 to Class~I protostars detected in our
survey compared with previous work suggests star formation is highly
dependent on the local environment.  Our new results provide several
possibilities for the nature of low-mass star formation: {\it (i)}
Class~0 sources are at the same evolutionary stage as Class~I sources,
but represent a different branch of star formation, e.g., star formation
in high density environments.  However, a comparison between our study
and that of Motte et al.\ (1998) suggests the Lynds clouds are actually
less dense than those in $\rho$ Ophiuchus; {\it (ii)} if Class~0 sources
are precursors to Class~I sources, the local environment may determine
the lifetime of the Class~0 phase.  For example, if the Lynds clouds are
more quiescent than those in $\rho$ Ophiuchus, they may have more time
to evolve towards the singular $n \propto r^{-2}$, which predicts a more
uniform accretion rate.  However, none of the starless cores detected
here has the characteristics of a singular isothermal sphere; {\it
(iii)} star formation is not steady, and relies heavily on triggering.
The ratio of Class~0 to Class~I protostars in $\rho$ Ophiuchus can then
be explained by being dominated by an older, more evolved population
caused by a burst of star formation some $\sim 10^5$~yr ago.  {\it We
regard this as the most likely possibility.} Furthermore, if the lifetime
of the Class~0 phase is actually similar to that of the Class~I phase,
Class~0 protostars may not have such dramatically high accretion rates
compared with Class~I protostars, as has previously been assumed.

The lifetime of the starless cores found in this survey is similar to the
$\sim 10^6$~yr derived from surveys using the lack of an IRAS association
to define ``starless'' (e.g., Lee \& Myers 1999).  The cores therefore
last only 2--3 free-fall times before collapsing.  Temperatures of only
$\sim 10$--12~K are needed for the dominant support mechanism of the cores
to be thermal pressure.  The starless cores are therefore on the verge of
collapse, and it is unlikely that strong magnetic fields are important
for the collapse dynamics of the cores on scales $r \sim 10^4$~AU\@.
Further work is needed to determine the velocity structure of these
cores through molecular line spectroscopy.

\acknowledgments

The authors thank the staff of the JCMT for carrying out some of the
$^{12}$CO observations of the outflows in service observing mode.  We also
thank the anonymous referee for his/her careful reading of the text, and
comments that have helped to clarify the paper.  A.E.V. was supported by
a Marie Curie Research Training Grant; J.S.R. and C.J.C. acknowledge the
support of a Royal Society Fellowship and a PPARC Advanced Fellowship,
respectively, during part of this work.  The James Clerk Maxwell
Telescope is operated by the Joint Astronomy Centre on behalf of the
Particle Physics and Astronomy Research Council of the United Kingdom,
the Netherlands Organization for Scientific Research, and the National
Research Council of Canada.  The Digitized Sky Surveys were produced
at the Space Telescope Science Institute under US Government grant
NAG W-2166.  The images of these surveys are based on photographic data
obtained using the Oschin Schmidt Telescope on Palomar Mountain and
the UK Schmidt Telescope.  The plates were processed into the present
compressed digital form with the permission of these institutions.

\newpage

$$\psfig{file=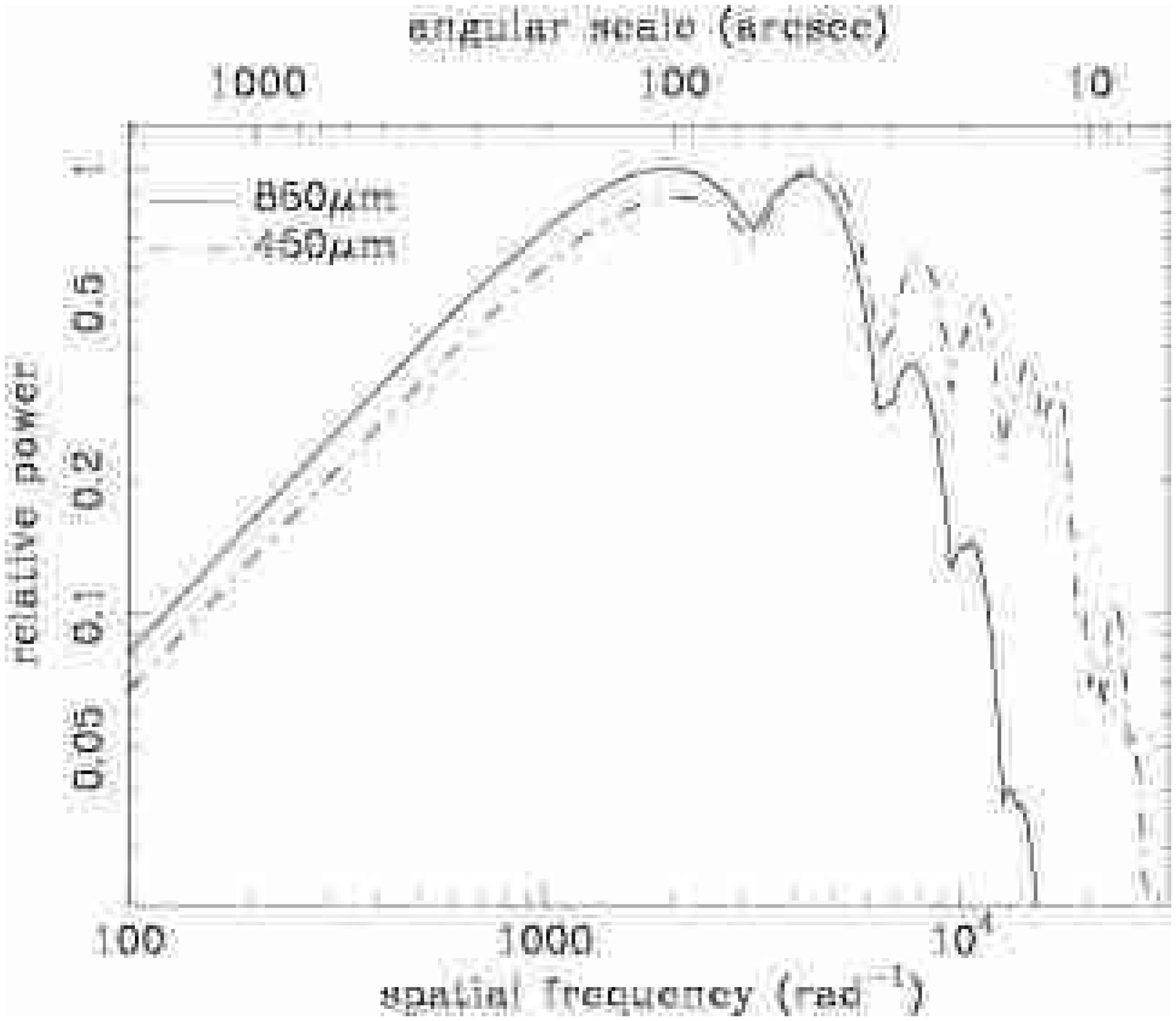,width=4in,angle=0}$$

\figcaption{\label{coverage}Spatial frequency coverage of scan maps
made using the Emerson2 technique, with chop throws of 20$''$, 30$''$,
and $65''$.  It has been calculated by adding the moduli of the FT of
the three dual beam functions.  The solid line shows the 850~$\mu$m
coverage, and the dot-dash line the 450~$\mu$m coverage.  Low spatial
frequencies are measured (apart from zero frequency) but with less
sensitivity than the higher spatial frequencies, and can result in
spatially extended noise features in the scan maps.}

$$\psfig{file=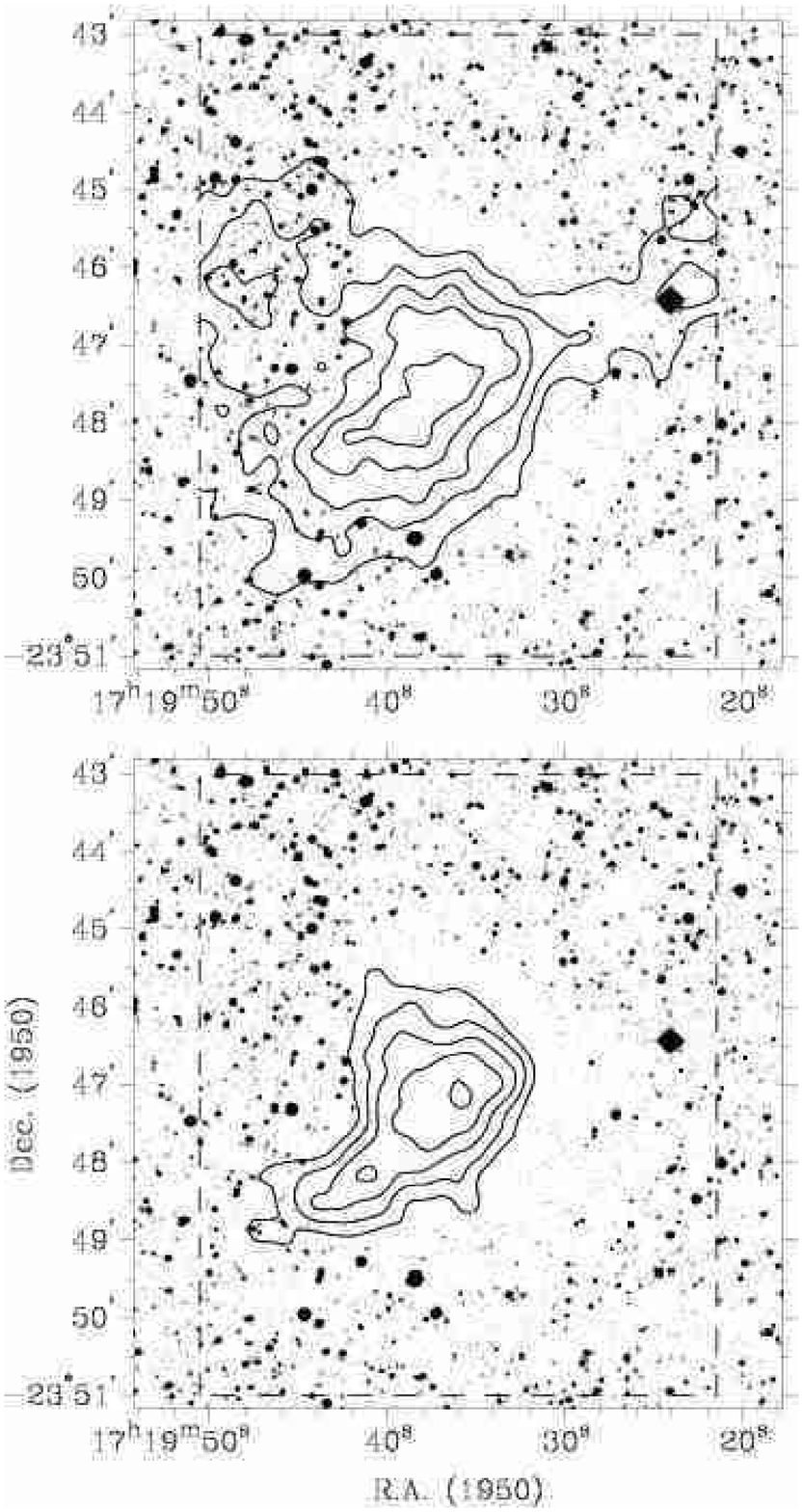,height=7in,angle=0}$$

\figcaption{\label{fig_L57}Deep SCUBA images of L57 (B68) at 450~$\mu$m
(top) and 850~$\mu$m (bottom).  The IRAS source associated with
this cloud by Launhardt \& Henning (1997) is 2$'$ south of the cloud
and is not covered by the SCUBA map.  The 450~$\mu$m contour levels
are at 3$\sigma$, 4.5$\sigma$, 6$\sigma$, 7.5$\sigma$, and 9$\sigma$
($\sigma = 67$~mJy~beam$^{-1}$).  The 850~$\mu$m contour levels are at
1$\sigma$, 2$\sigma$, 3$\sigma$, 4$\sigma$, and 5$\sigma$ with $\sigma =
14$~mJy~beam$^{-1}$.}

$$\psfig{file=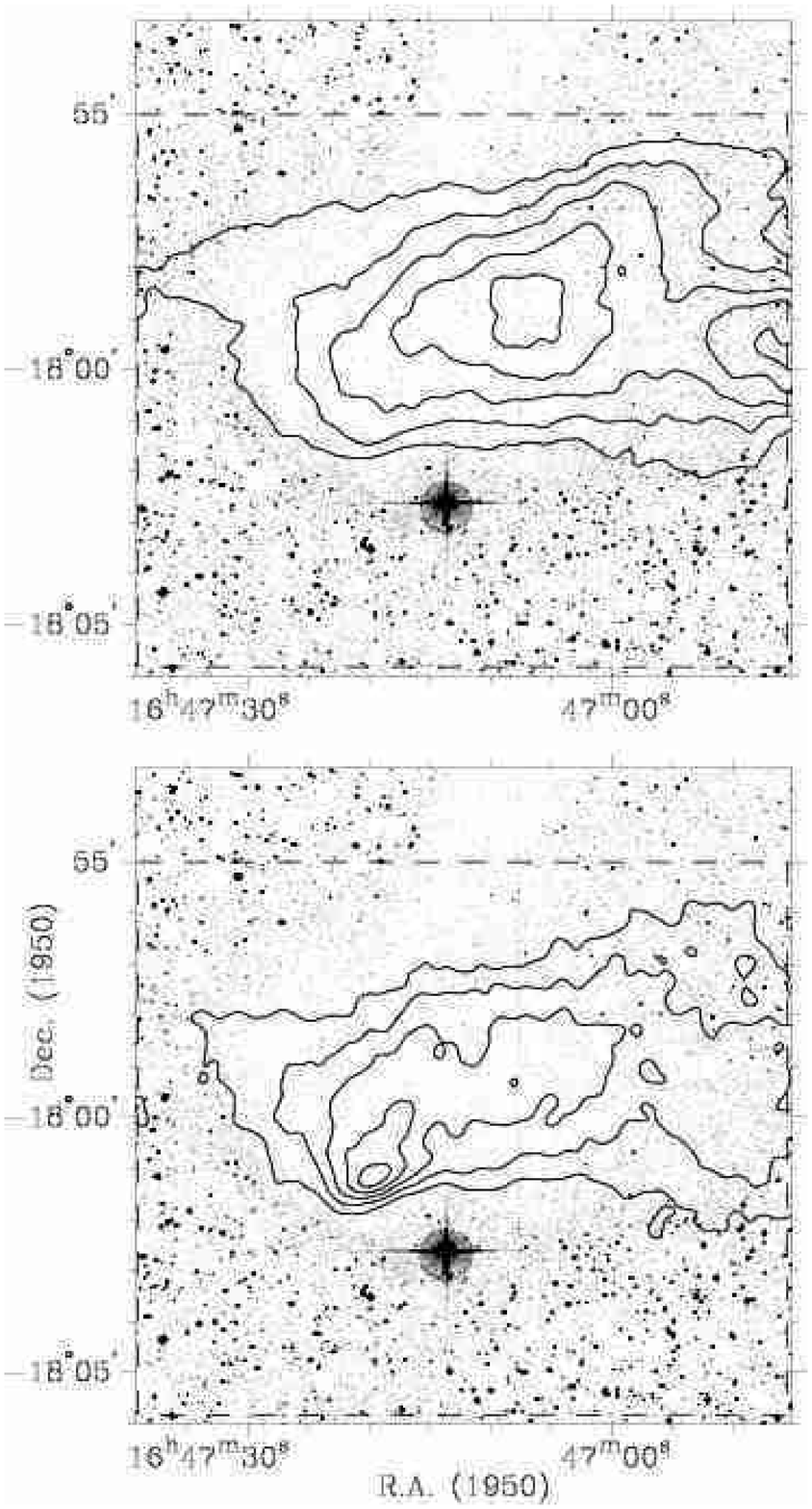,height=7in,angle=0}$$

\figcaption{\label{fig_L63}L63 contains a well-known starless core
(Ward-Thompson et al.\ 1994; Ward-Thompson, Motte, \& Andr\'e 1999)
in the south-east part of the cloud, detected here at 850~$\mu$m
(bottom) but not at 450~$\mu$m (top).  Extended emission is detected
at both wavelengths.  No IRAS sources are associated with this cloud.
Both the 450~$\mu$m and 850~$\mu$m contour levels are at 1$\sigma$,
2$\sigma$, 3$\sigma$, 4$\sigma$, and 5$\sigma$, with $\sigma_{450} =
335$~mJy~beam$^{-1}$, and $\sigma_{850} = 38$~mJy~beam$^{-1}$.}

$$\psfig{file=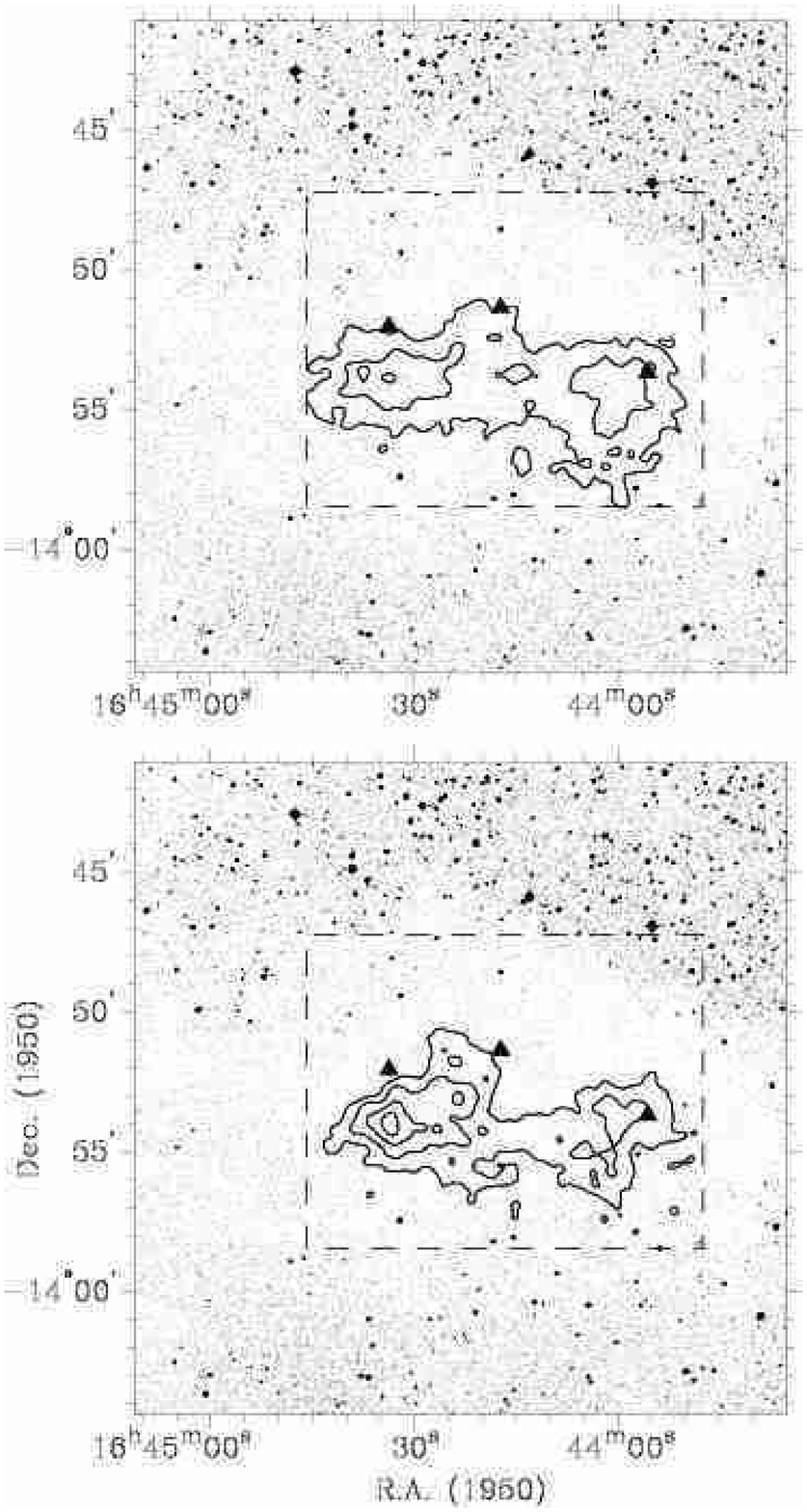,height=6.8in,angle=0}$$

\figcaption{\label{fig_L158}Extended emission is detected from L158 at
both 450~$\mu$m (top) and 850~$\mu$m (bottom).  Three IRAS sources are
covered by the SCUBA map (triangles), but none of these can be associated
with submillimeter emission.  IRAS~16445$-$1352 is not a Class~I object
as sometimes quoted, but cirrus (Beichman et al.\ 1986; Bontemps et al.\
1996).  The other two IRAS sources are both only detected at 60~$\mu$m,
and could also be cirrus.  The 450~$\mu$m contour levels are at 1$\sigma$,
2$\sigma$, and 3$\sigma$ ($\sigma = 268$~mJy~beam$^{-1}$).  The 850~$\mu$m
contour levels are at 1$\sigma$, 2$\sigma$, 3$\sigma$, and 4$\sigma$
($\sigma = 24$~mJy~beam$^{-1}$).}

$$\psfig{file=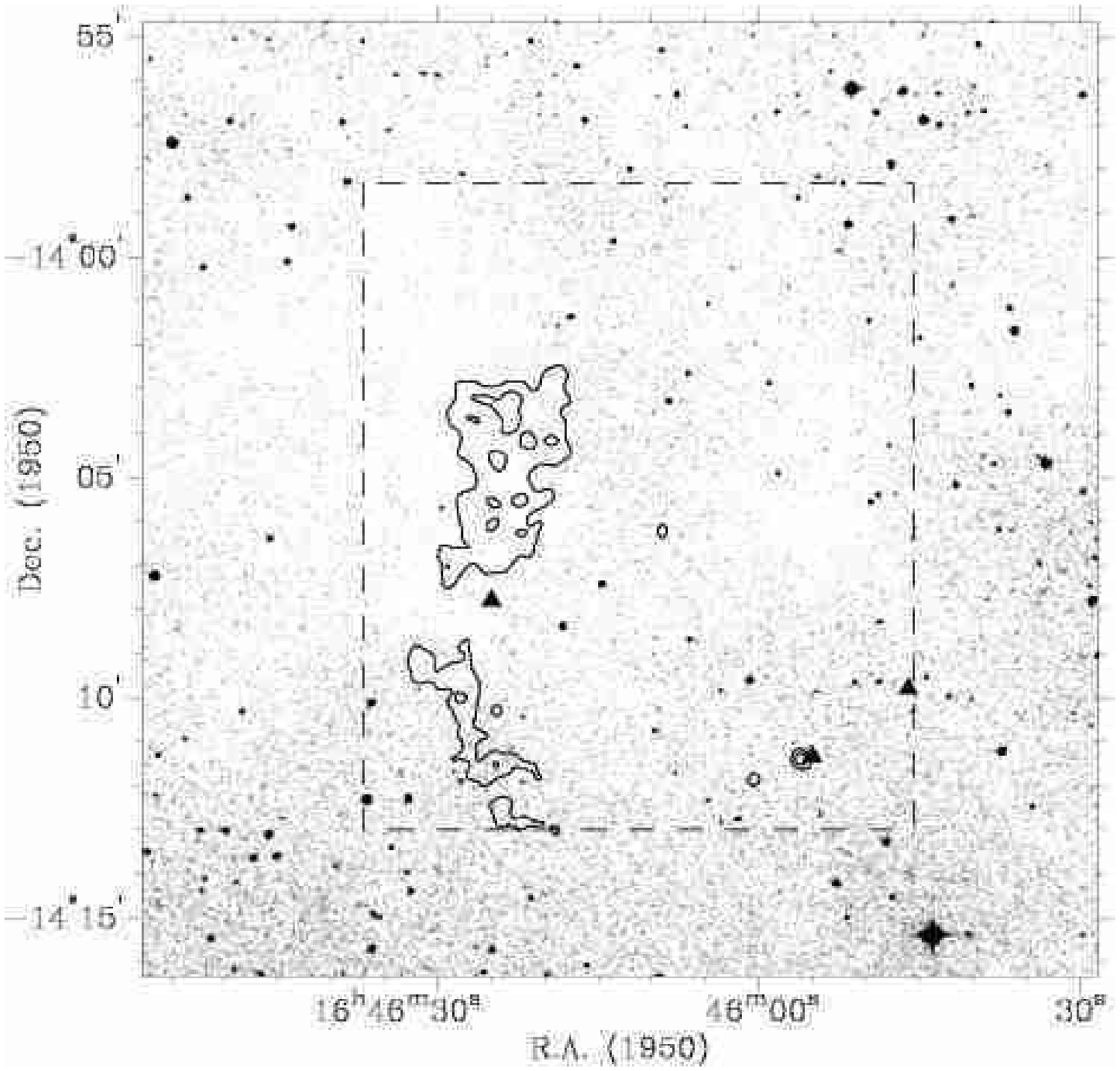,width=5in,angle=0}$$

\figcaption{\label{fig_L162}The central region of L162 was mapped
during weather conditions suitable only for observing at 850~$\mu$m.
Three IRAS sources were covered by the SCUBA map (triangles).  Emission
from the T~Tauri star IRAS~16459$-$1411 has been detected (lower right),
but the other two IRAS sources are not detected, and may be cirrus.
The contour levels are at 5$\sigma$ and 7$\sigma$ (with $\sigma =
20$~mJy~beam$^{-1}$).  Although the extended emission at the top of
the map correlates well with the dark cloud, the bottom part does not,
which may indicate problems with the baseline subtraction.}

$$\psfig{file=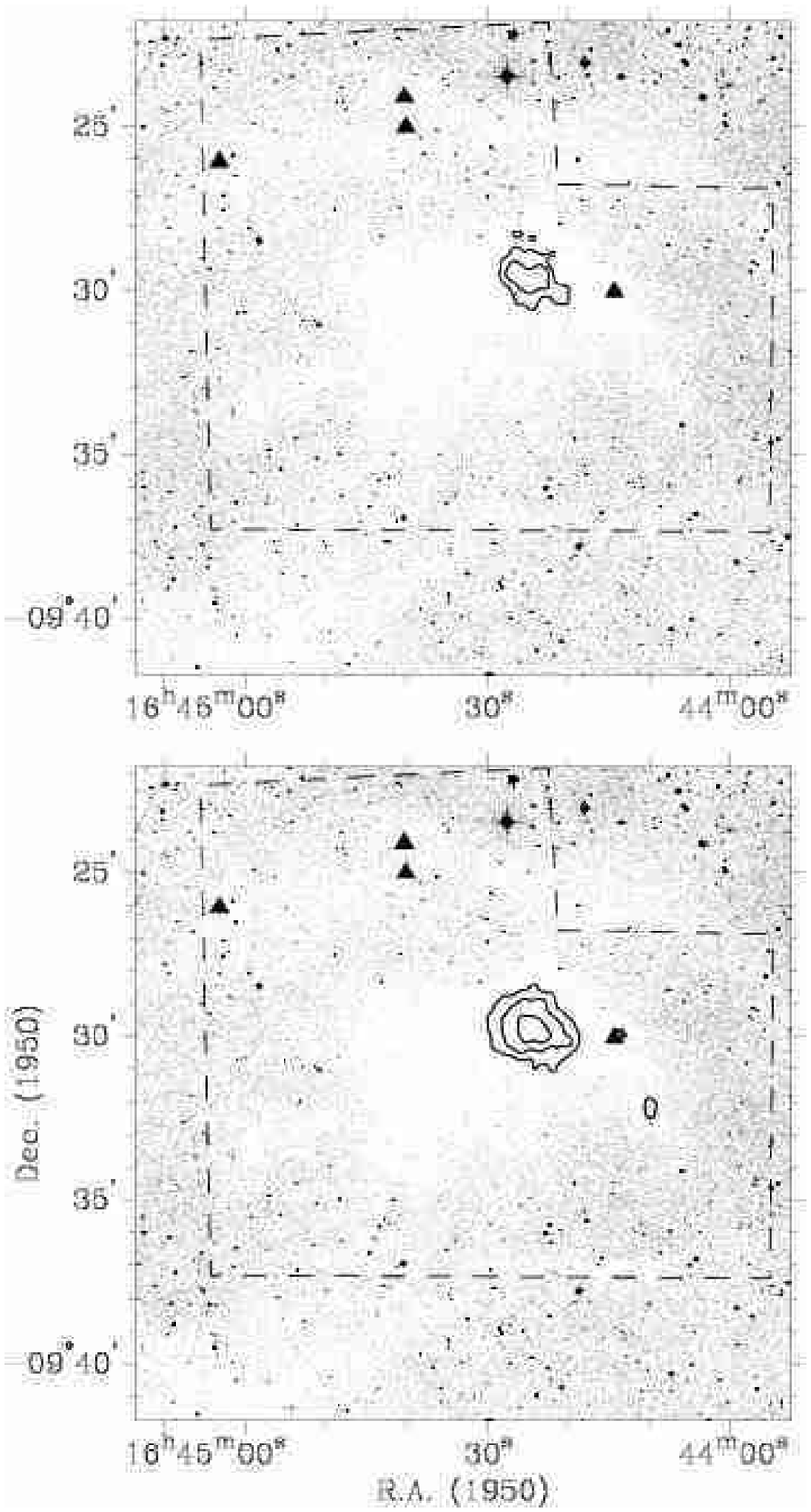,height=7in,angle=0}$$

\figcaption{\label{fig_L260}L260 contains a well known Class~I protostar
detected at 850~$\mu$m (bottom) but not at 450~$\mu$m (top).  A further
three IRAS sources (triangles) are covered by the SCUBA map, but none was
detected.  There is also extended emission east of the protostar at both
wavelengths.  The 450~$\mu$m contour levels are at 1$\sigma$ and 2$\sigma$
($\sigma = 134$~mJy~beam$^{-1}$).  The 850~$\mu$m contour levels are at
1$\sigma$, 2$\sigma$ and 3$\sigma$ ($\sigma = 24$~mJy~beam$^{-1}$).}

$$\psfig{file=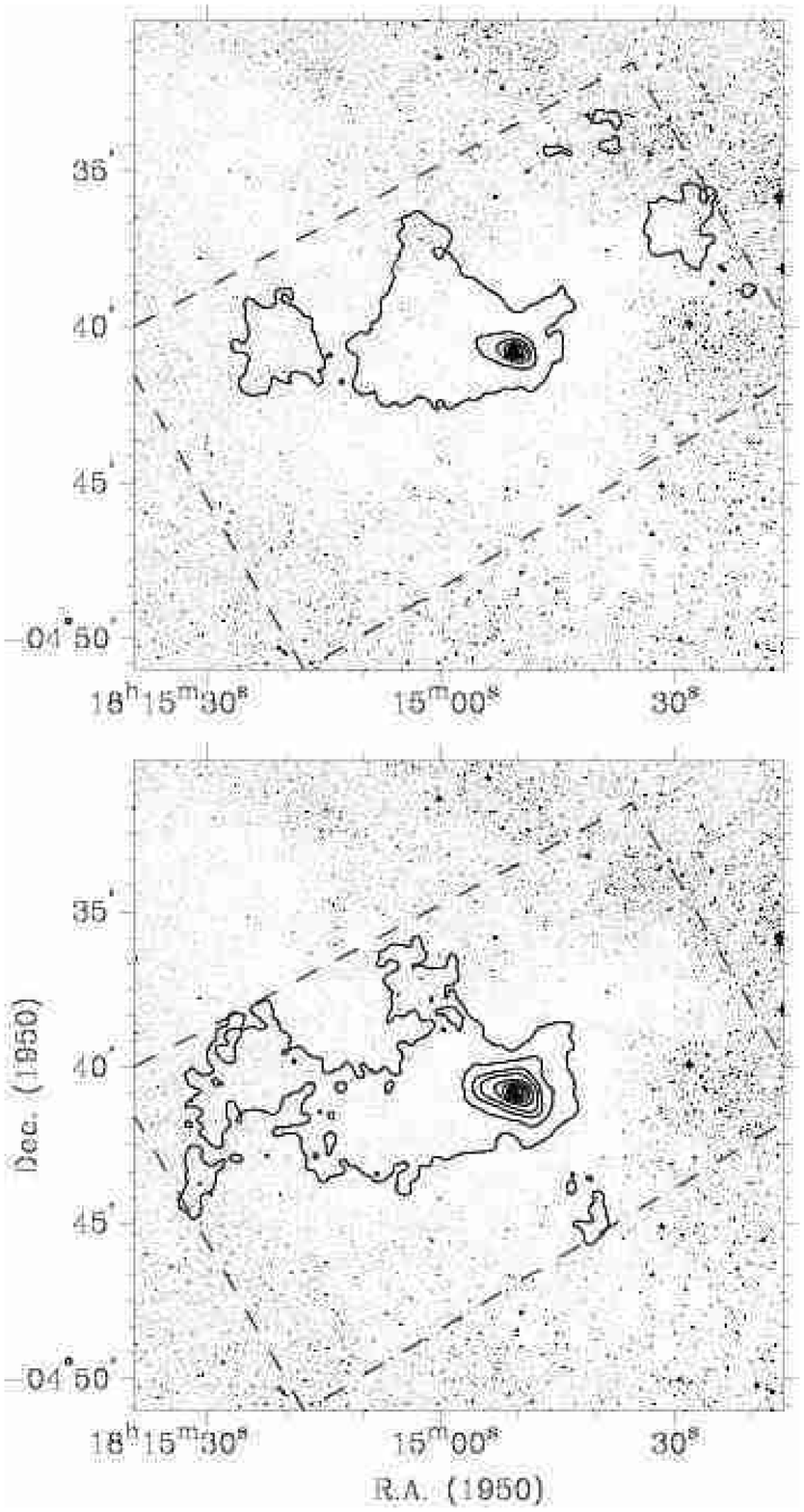,height=7in,angle=0}$$

\figcaption{\label{fig_L483}L483 contains a well-known protostar clearly
detected at both 450~$\mu$m (top) and 850~$\mu$m (bottom).  Extended
dust emission is also detected and appears to outline the outflow from
the central source.  The 450~$\mu$m contour levels are at 1$\sigma$,
3$\sigma$, 5$\sigma$, 7$\sigma$, 9$\sigma$, and 11$\sigma$ ($\sigma =
355$~mJy~beam$^{-1}$), and the 850~$\mu$m contour levels are at 1$\sigma$,
4$\sigma$, 7$\sigma$, 10$\sigma$, 16$\sigma$, 22$\sigma$, 28$\sigma$,
34$\sigma$, and 40$\sigma$, with $\sigma = 22$~mJy~beam$^{-1}$.}

$$\psfig{file=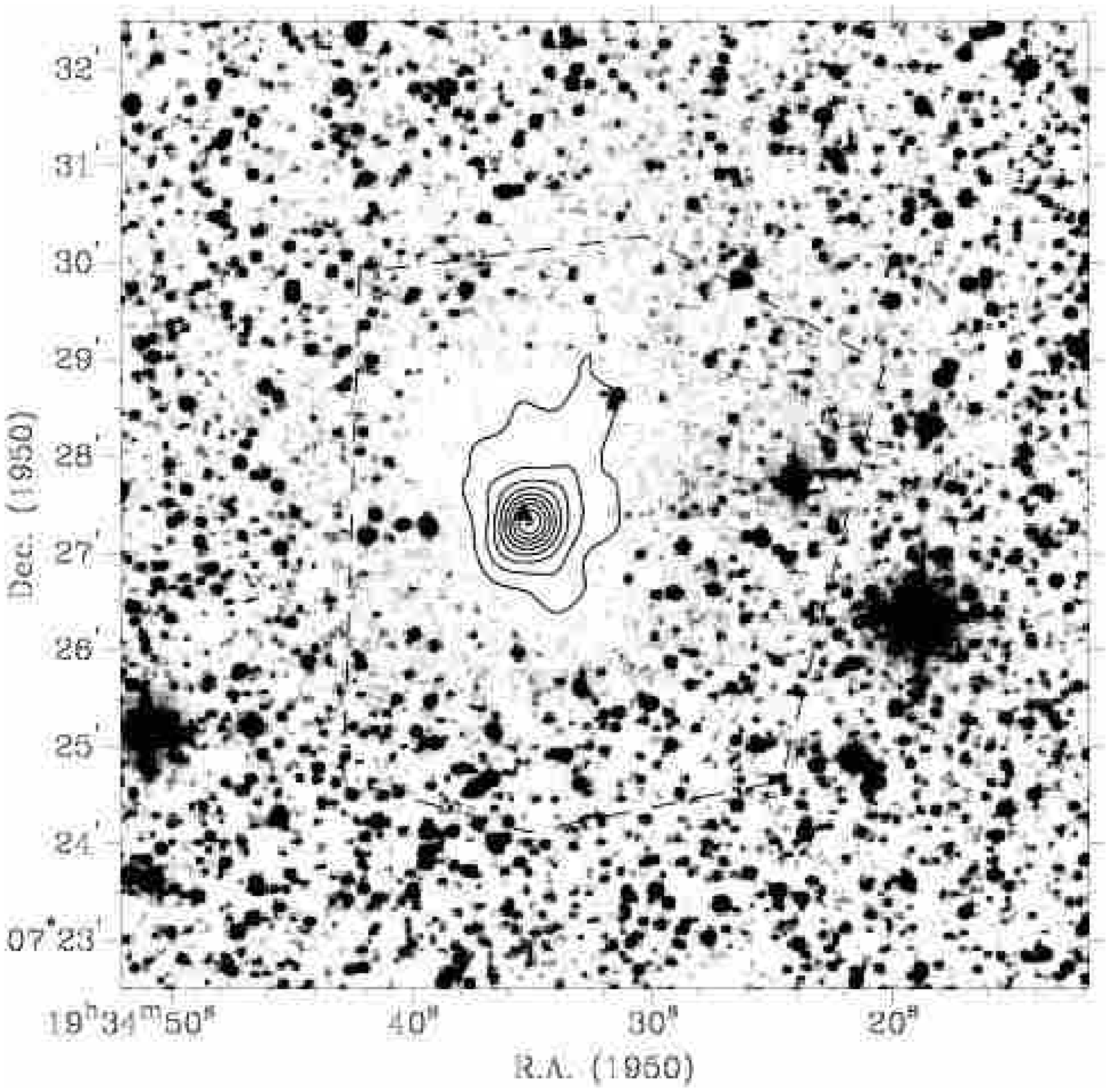,width=5in,angle=0}$$

\figcaption{\label{fig_L663}L663 (B335) is the most famous cloud in
this sample.  It has been imaged using the EKH technique at 850~$\mu$m.
Strong submillimeter emission originates from a Class~0 protostar
that has been identified with the IRAS source 19345+0727 (triangle).
A second IRAS source to the west, also covered by the SCUBA map but
not detected in the submillimeter, is associated with a bright star.
The 850~$\mu$m contour levels are at 1$\sigma$, 2$\sigma$, 3$\sigma$,
4$\sigma$, 5$\sigma$, 6$\sigma$, 7$\sigma$, 8$\sigma$, and 9$\sigma$
($\sigma = 74$~mJy~beam$^{-1}$).}

$$\psfig{file=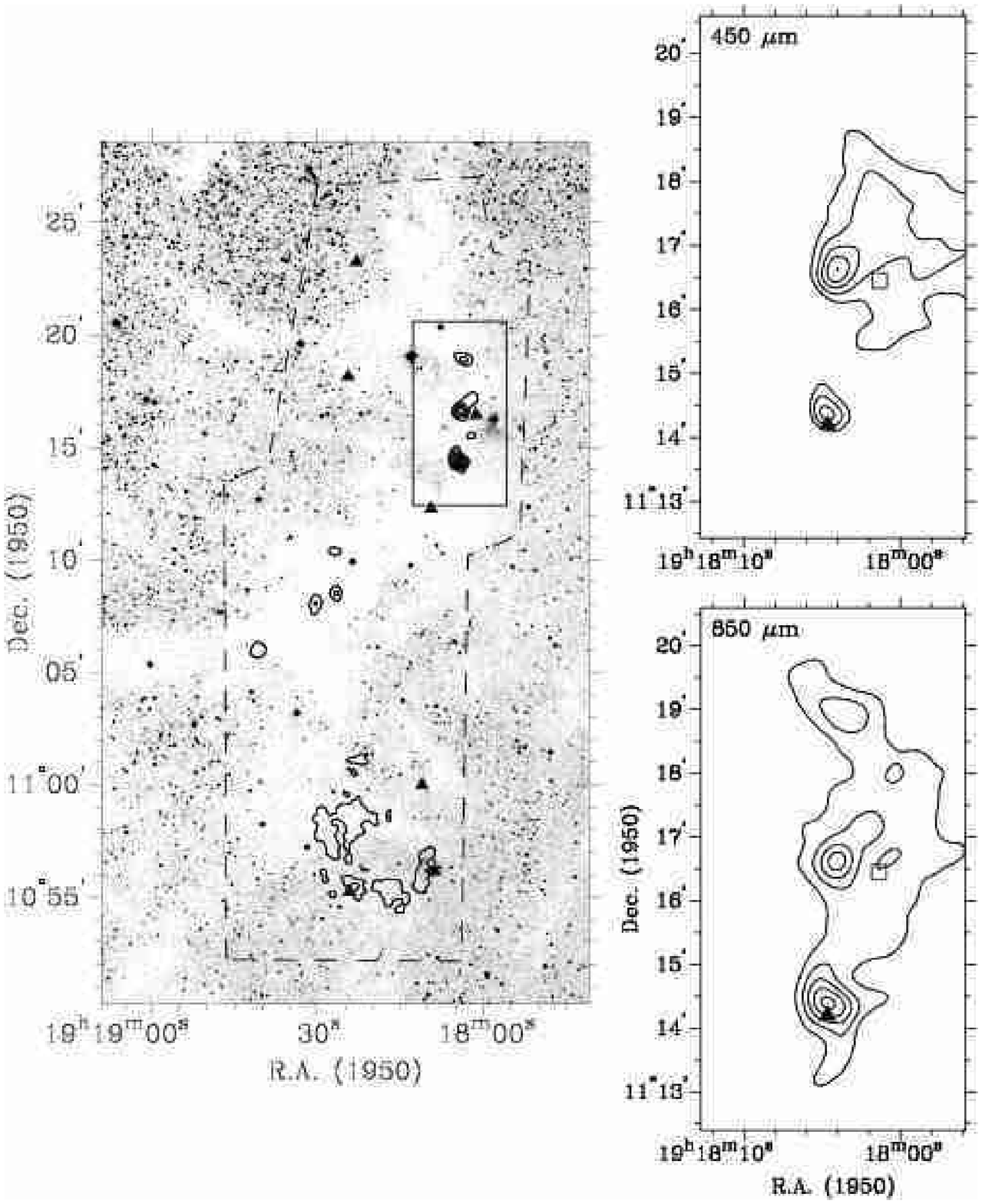,height=5.3in,angle=0}$$

\figcaption{\label{fig_L673}{\it Left:} Overlay of 850~$\mu$m contours
on the optical image of L673.  Some extended emission is detected towards
the southern part of the cloud, although it does not correlate very well
with the dust extinction.  Eight IRAS sources are covered by this SCUBA
map (triangles), but most have not been detected (e.g., the T~Tauri star
IRAS~19181+1056).  The box indicates the area displayed in more detail
on the right.  This northern region is the only part mapped in both 450
and 850~$\mu$m.  The 850~$\mu$m contours are at 2$\sigma$, 3$\sigma$,
4$\sigma$, 5$\sigma$, and 6$\sigma$ ($\sigma = 40$~mJy~beam$^{-1}$).
{\it Right:} The submillimeter cores in the northern part of L673.
IRAS~19180+1114 can be associated with the continuum source furthest south
in this ridge (triangle).  The submillimeter source north of 19180+1114
may be associated with 19180+1116 (square, see Section~\ref{sec_core_id})
suggesting both could be protostars.  The 450~$\mu$m contour levels
are at 2$\sigma$, 3$\sigma$, 4$\sigma$, 5$\sigma$, and 6$\sigma$
($\sigma = 147$~mJy~beam$^{-1}$).  The 850~$\mu$m contours are at
1$\sigma$, 3$\sigma$, 5$\sigma$, 7$\sigma$, and 9$\sigma$ ($\sigma =
28$~mJy~beam$^{-1}$).}

$$\psfig{file=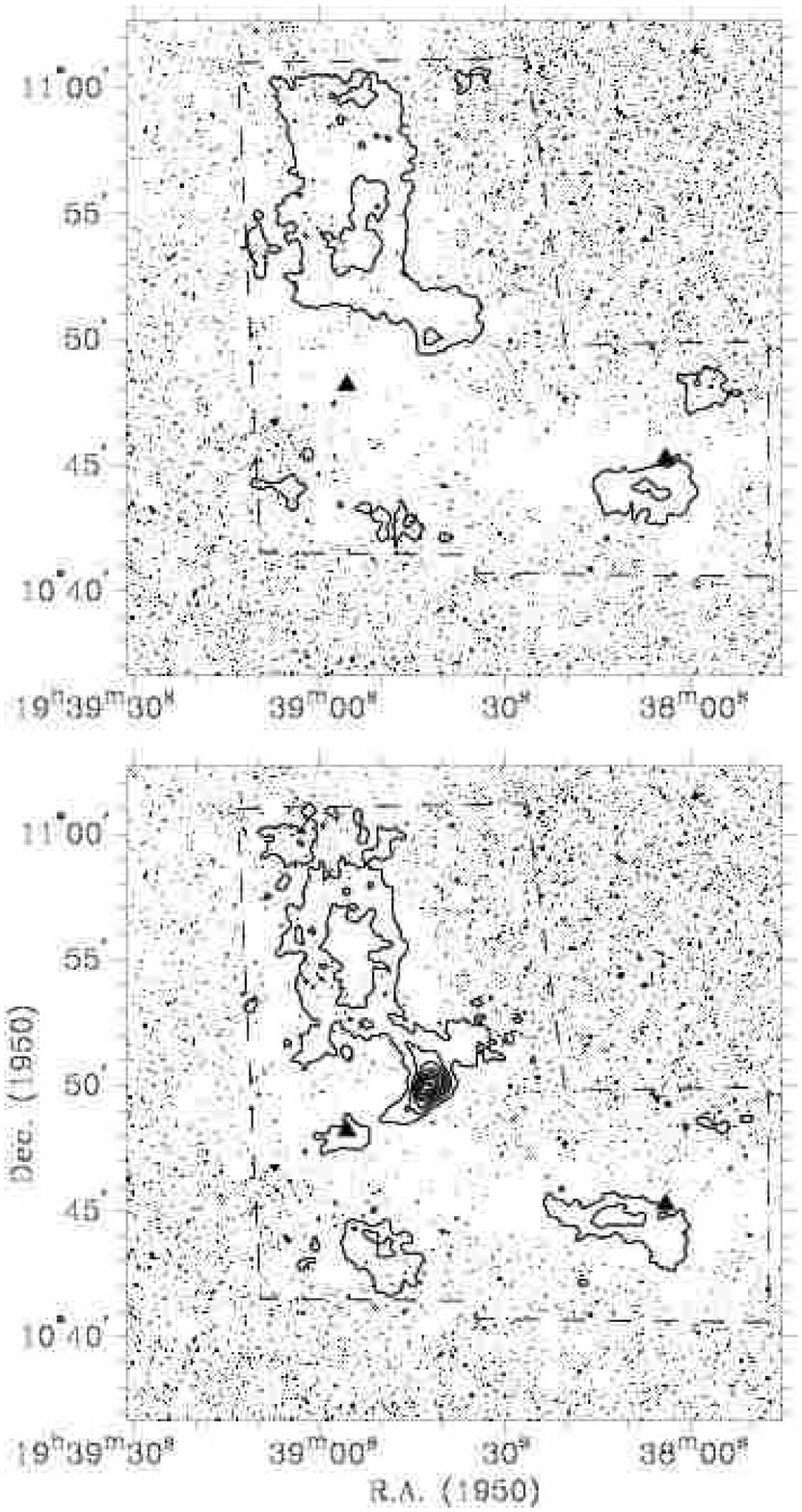,height=6.8in,angle=0}$$

\figcaption{\label{fig_L694}L694 contains a bright embedded submillimeter
continuum source, clearly detected at 850~$\mu$m (bottom) but not at
450~$\mu$m (top).  Extended emission is detected at both wavelengths
and correlates well with the optical dust extinction.  The maps cover
two IRAS sources, but neither has been detected.  IRAS~19389+1048 is
probably cirrus, and IRAS~19380+1045 is very likely a (young) field star.
The 450~$\mu$m contour levels are at 1$\sigma$ and 2$\sigma$ ($\sigma =
201$~mJy~beam$^{-1}$).  The 850~$\mu$m contour levels are at 1$\sigma$,
2$\sigma$, 3$\sigma$, 4$\sigma$, 5$\sigma$, 6$\sigma$, 7$\sigma$, and
8$\sigma$ ($\sigma = 25$~mJy~beam$^{-1}$).}

$$\psfig{file=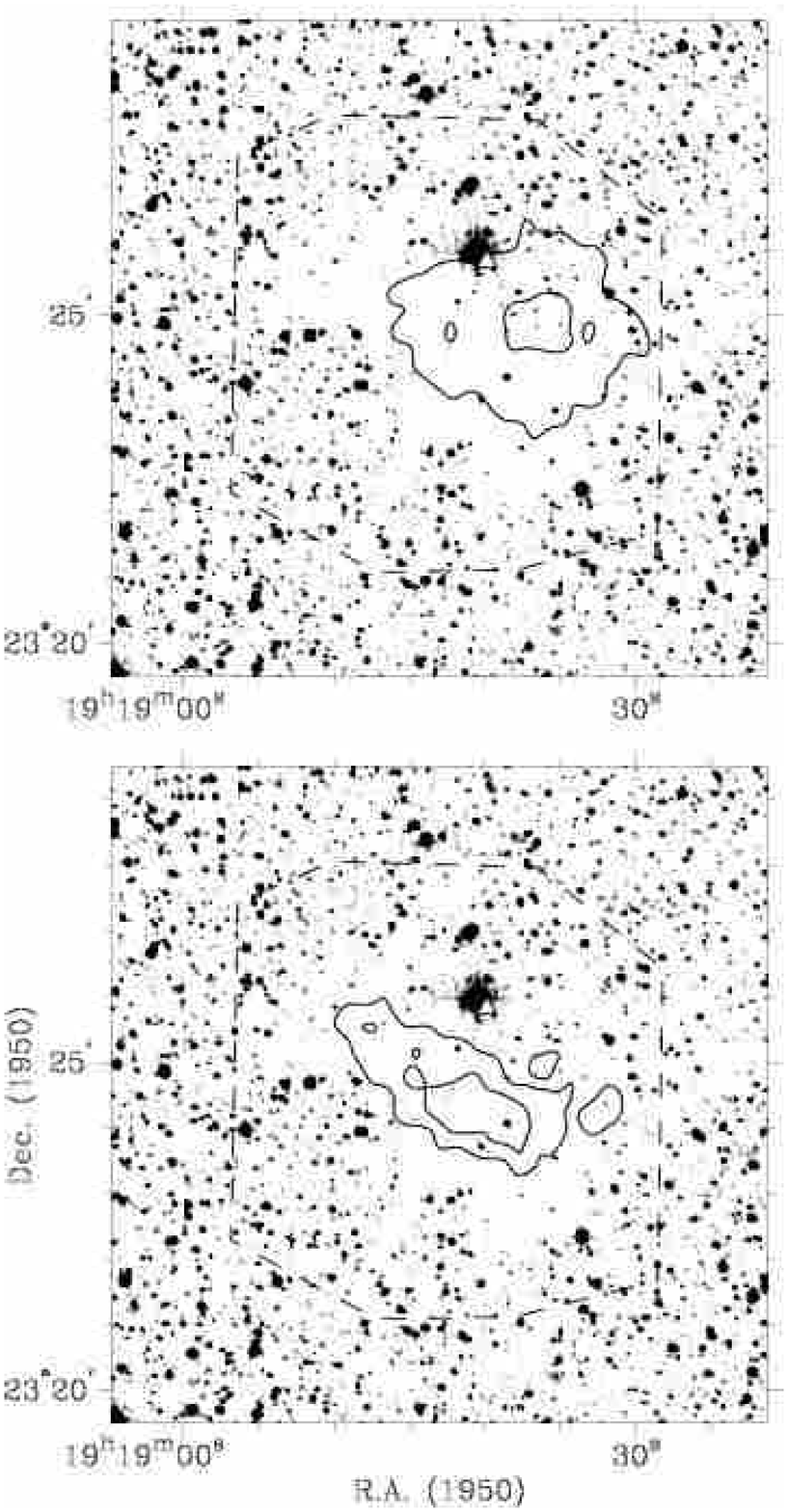,height=7in,angle=0}$$

\figcaption{\label{fig_L771}Extended emission has been detected from L771
at both 450~$\mu$m (top) and 850~$\mu$m (bottom), and appears different
at the two wavelengths.  The IRAS source in this field can probably
associated with the bright star (triangle).  The 450~$\mu$m contours
are at 3$\sigma$ and 6$\sigma$ ($\sigma = 100$~mJy~beam$^{-1}$).
The 850~$\mu$m contours are at 3$\sigma$ and 4$\sigma$ ($\sigma =
20$~mJy~beam$^{-1}$).}

$$\psfig{file=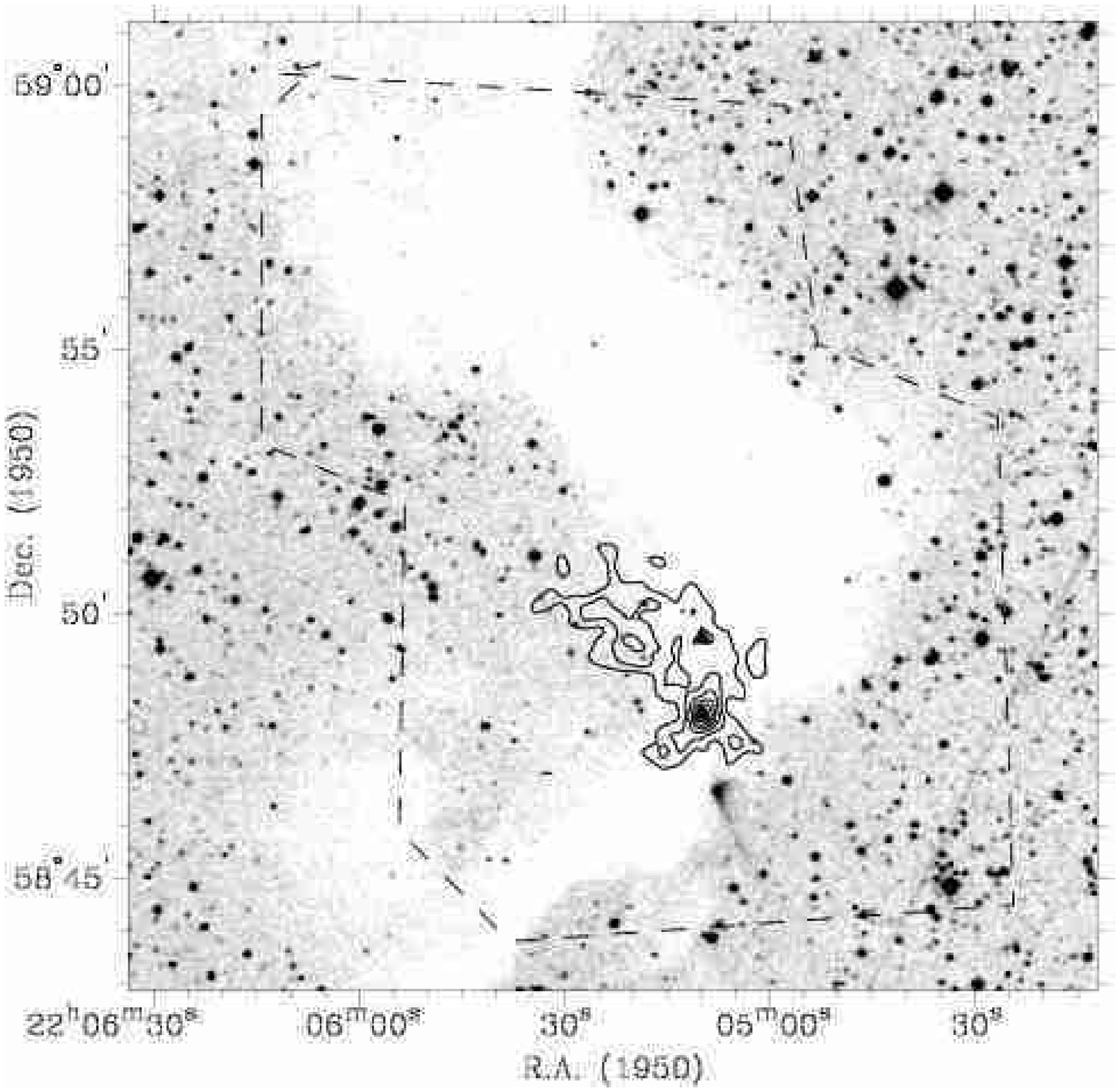,width=5in,angle=0}$$

\figcaption{\label{fig_L1165}The 850~$\mu$m map of L1165 shows clearly
the known protostar IRAS~22051+5848.  The other IRAS source covered by
this map is not detected, and is probably a field star or T~Tauri star.
The contour levels are at 1$\sigma$, 2$\sigma$, 3$\sigma$, 4$\sigma$,
5$\sigma$, and 6$\sigma$ ($\sigma = 67$~mJy~beam$^{-1}$).}

$$\psfig{file=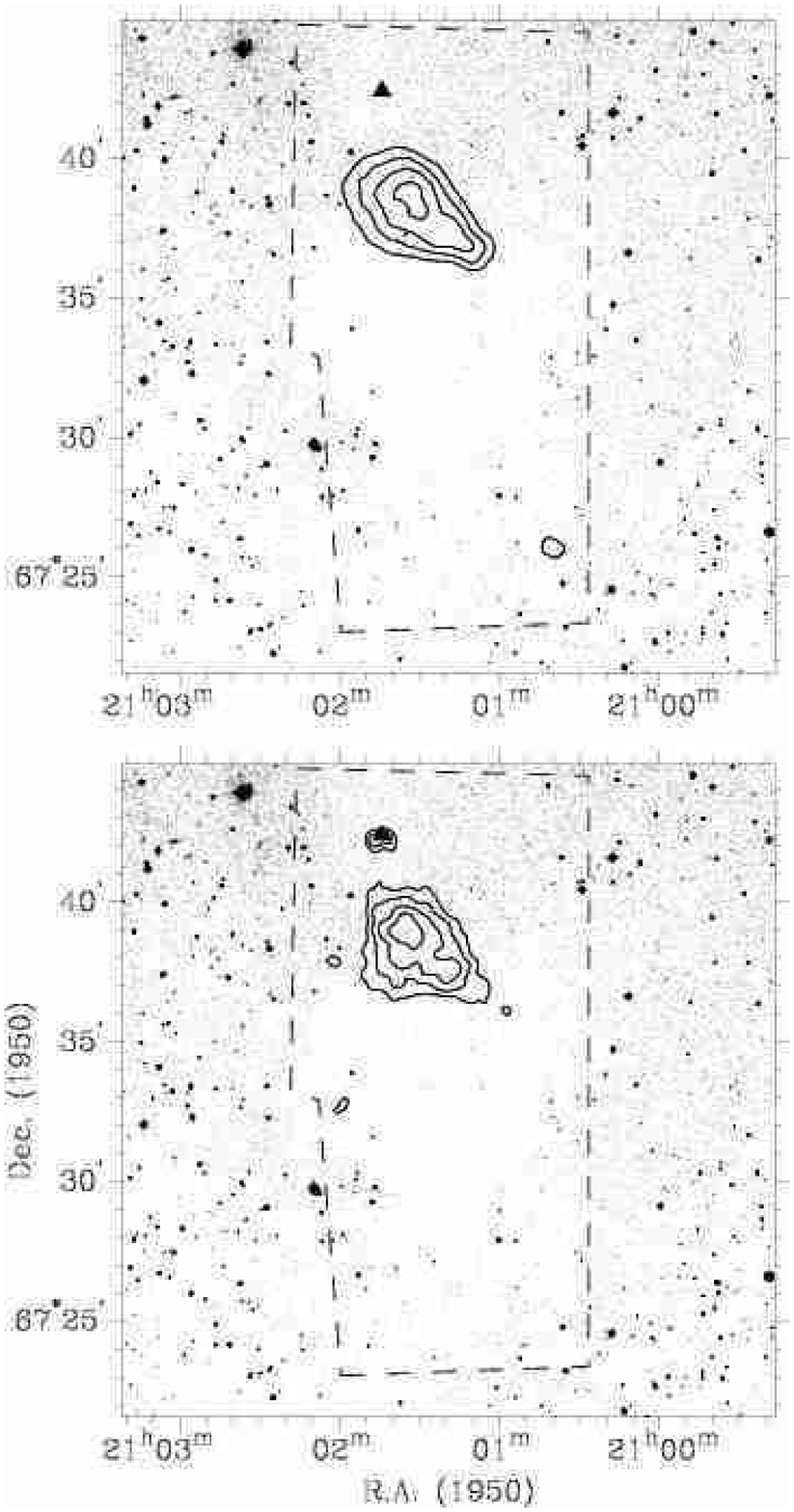,height=7in,angle=0}$$

\figcaption{\label{fig_L1172}L1172 shows a compact continuum source at
850~$\mu$m (bottom) which coincides with IRAS~21017+6742 (triangle), and
some extended emission detected at both 450~$\mu$m (top) and 850~$\mu$m.
The 450~$\mu$m contour levels are at 2$\sigma$, 3$\sigma$, 4$\sigma$, and
5$\sigma$ ($\sigma = 369$~mJy~beam$^{-1}$, top image).  The 850~$\mu$m
contour levels are at also at 2$\sigma$, 3$\sigma$, 4$\sigma$, and
5$\sigma$ ($\sigma = 22$~mJy~beam$^{-1}$).}

$$\psfig{file=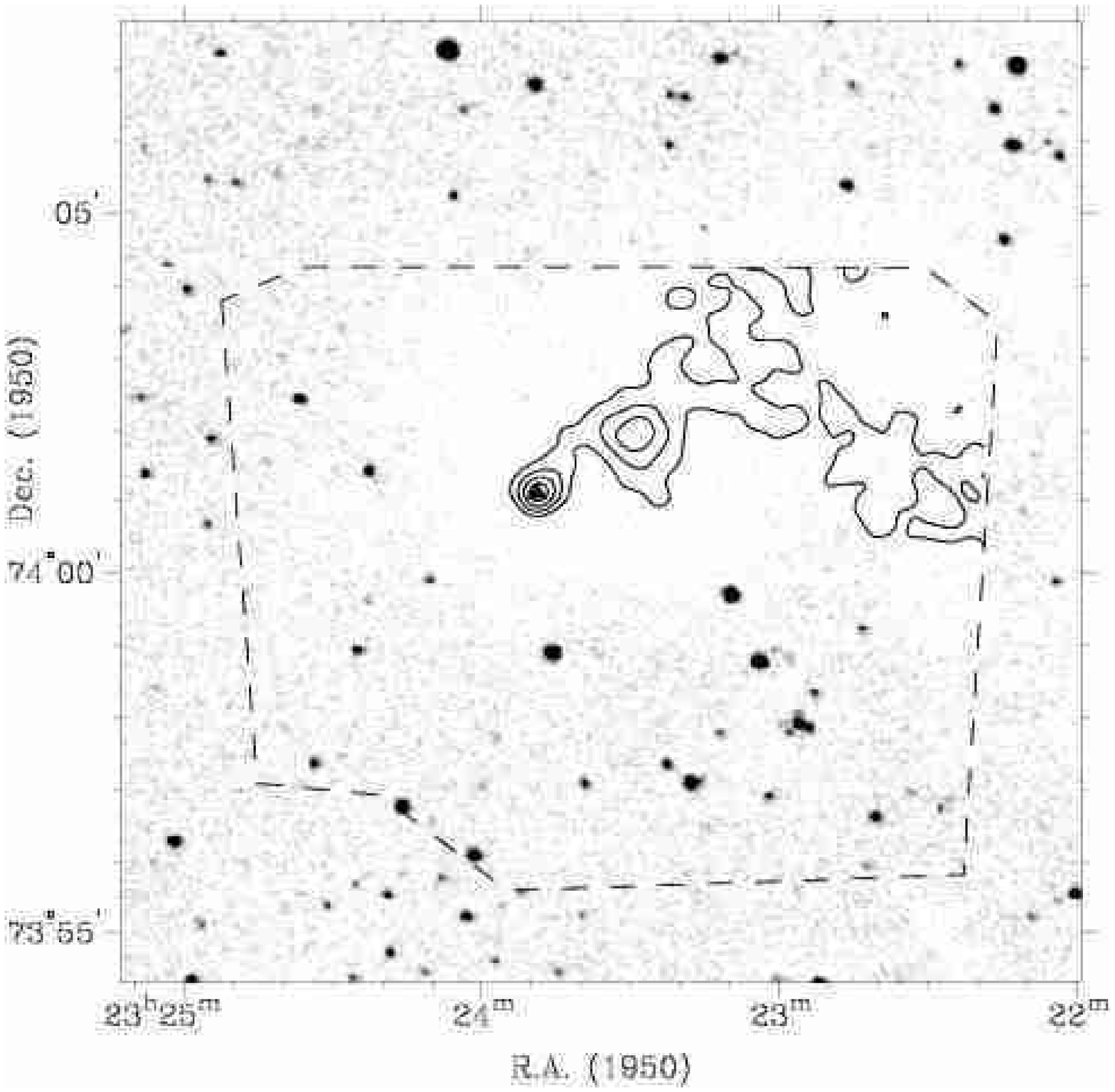,width=5in,angle=0}$$

\figcaption{\label{fig_L1262}L1262 (CB244) contains a protostar,
IRAS~23238+7401, which is clearly detected at 850~$\mu$m (triangle).
The contour levels are at 3$\sigma$, 5$\sigma$, 7$\sigma$, and 9$\sigma$
($\sigma = 30$~mJy~beam$^{-1}$).}

$$\psfig{file=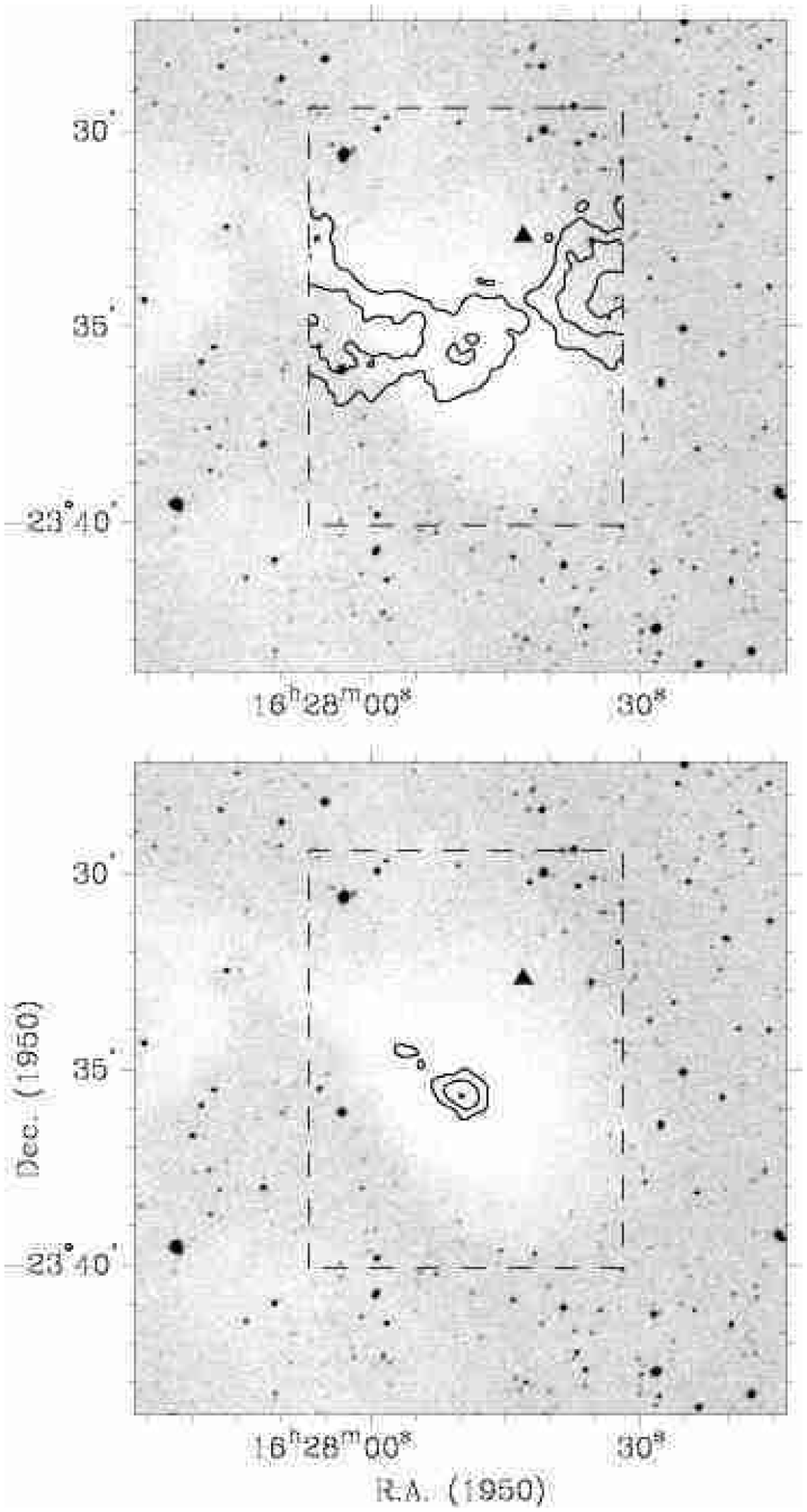,height=7in,angle=0}$$

\figcaption{\label{fig_L1704}L1704 is associated with a continuum source
most clearly detected at 850~$\mu$m (bottom) but also seen at 450~$\mu$m
(top).  IRAS~16277$-$2332 is covered by this map, but is undetected.
Both the 450~$\mu$m and the 850~$\mu$m contours are at 1$\sigma$,
2$\sigma$, and 3$\sigma$ ($\sigma_{450} = 134$~mJy~beam$^{-1}$,
$\sigma_{850} = 18$~mJy~beam$^{-1}$).}

$$\psfig{file=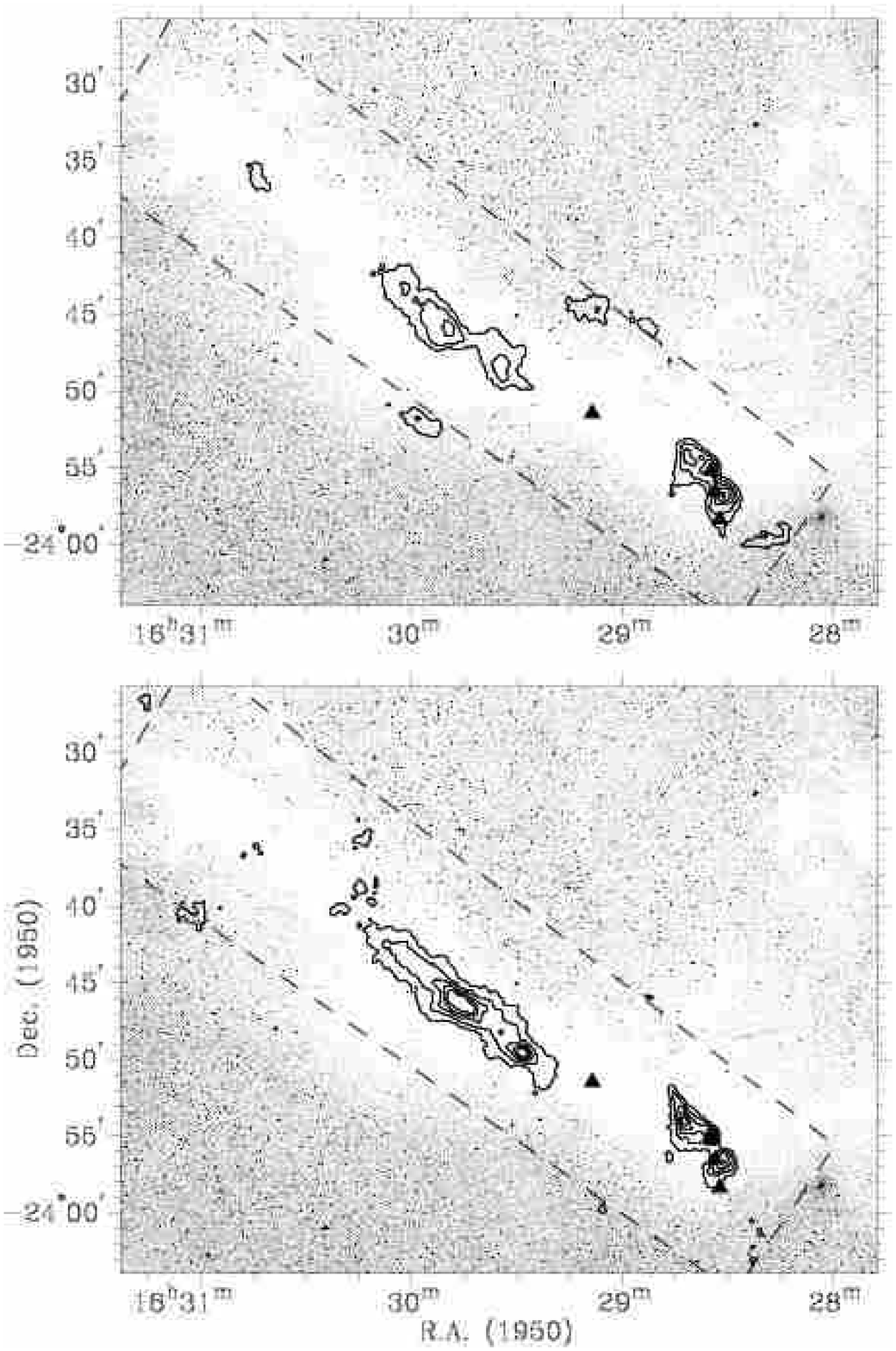,height=6.6in,angle=0}$$

\figcaption{\label{fig_L1709}Extended continuum emission and compact
sources are detected from L1709 at both 450~$\mu$m (top) and 850~$\mu$m
(bottom).  There are four IRAS sources in this cloud (triangles).  The
protostar IRAS~16285$-$2355 can be clearly associated with a submillimeter
continuum source.  South of this protostar, IRAS~16285$-$2356 is very
close to another compact continuum source.  The other two IRAS sources
are not associated with submillimeter emission.  The 450~$\mu$m contour
levels are at 4$\sigma$, 6$\sigma$, 8$\sigma$, 10$\sigma$, and 12$\sigma$
($\sigma = 134$~mJy~beam$^{-1}$), and the 850~$\mu$m contours are at
2$\sigma$, 4$\sigma$, 6$\sigma$, 8$\sigma$, 10$\sigma$, 14$\sigma$,
18$\sigma$, and 22$\sigma$ with $\sigma = 20$~mJy~beam$^{-1}$.}

$$\psfig{file=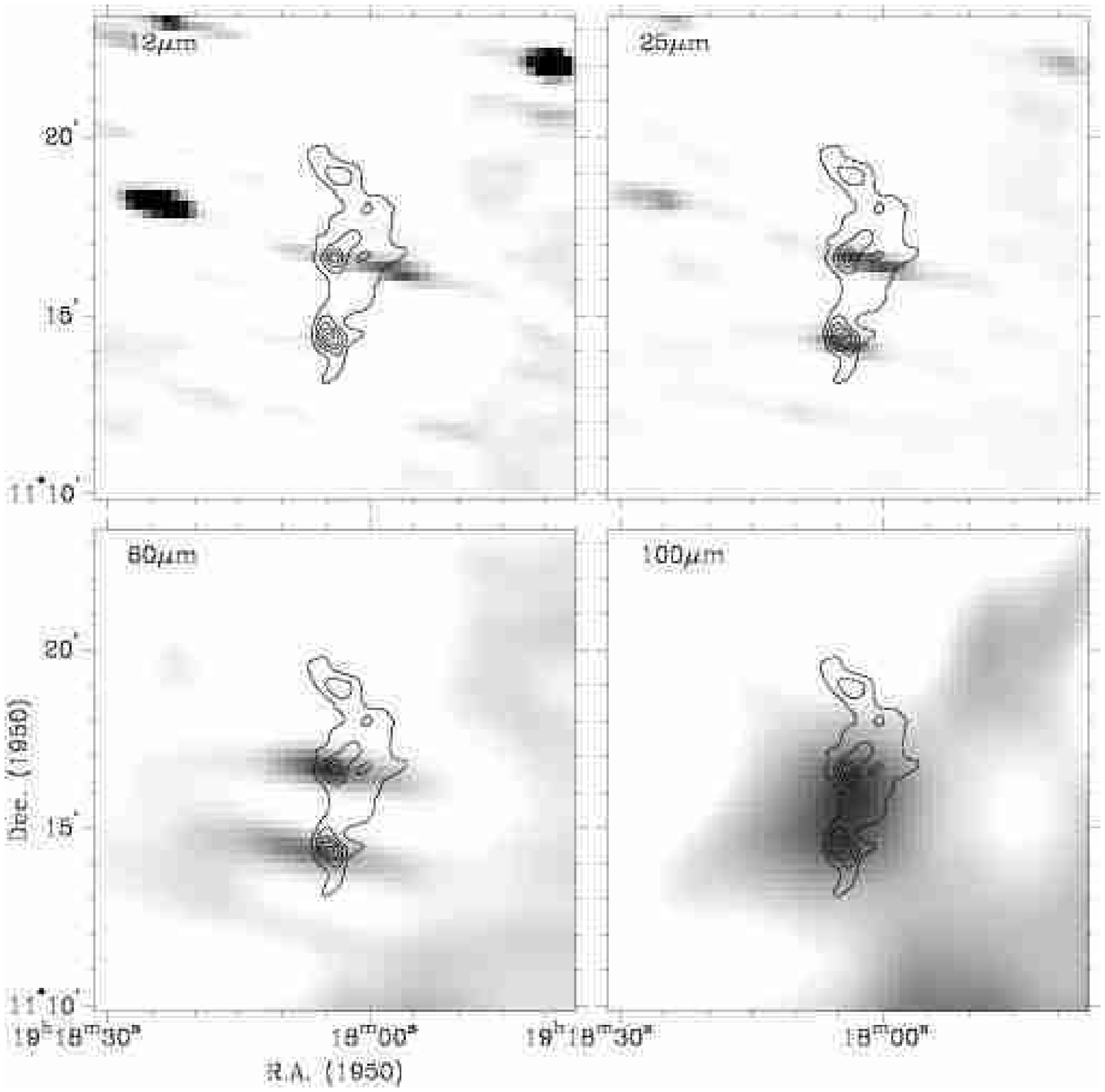,width=6.5in,angle=0}$$

\figcaption{\label{L673_hires}Contours of the 850~$\mu$m emission
from L673 overlaid on HIRES-processed IRAS images in greyscale.}

$$\psfig{file=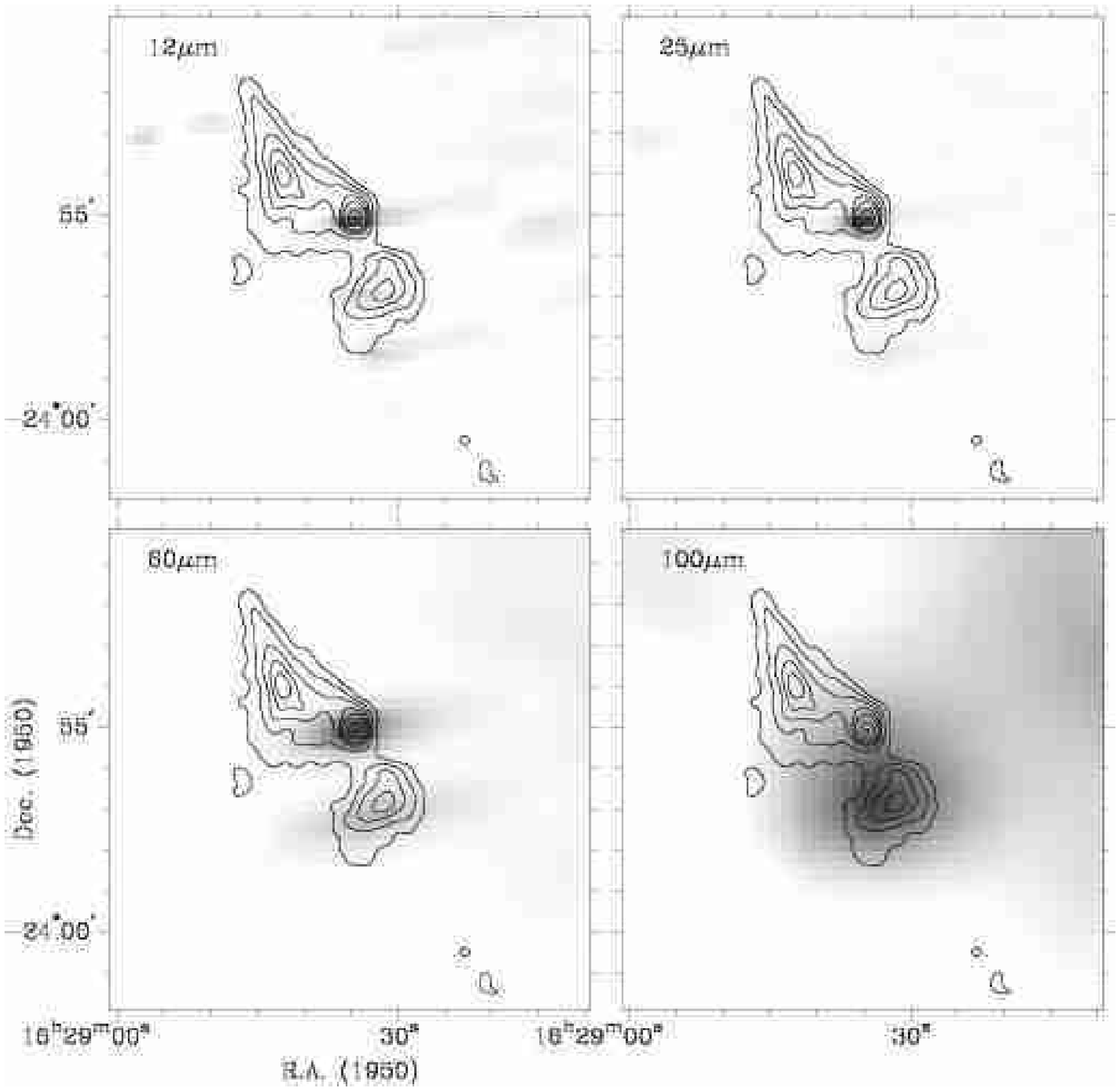,width=6.5in,angle=0}$$

\figcaption{\label{L1709_hires}Contours of the 850~$\mu$m emission from
L1709 overlaid on HIRES-processed IRAS images in greyscale.  The most
southern IRAS source visible in the 12~$\mu$m and 25~$\mu$m bands is
16285$-$2358, a known star (Ichikawa \& Nishida 1989).}

$$\psfig{file=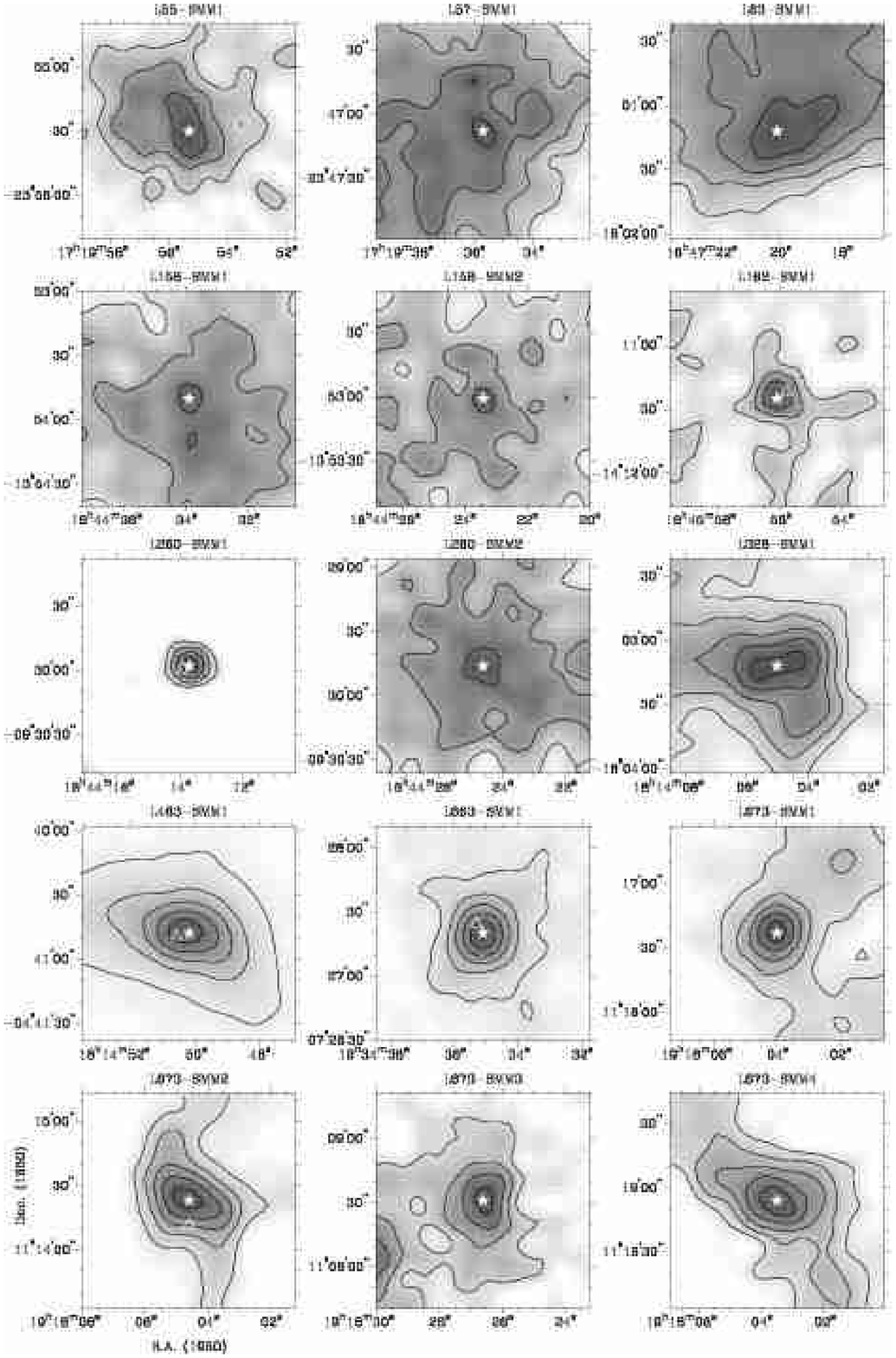,height=7in,angle=0}$$

\figcaption{\label{fig_cores}Images of all 40 submillimeter cores
identified in the sample of Lynds dark clouds.  Contours show the
850~$\mu$m emission, the positions in Table~\ref{tab_cores} are
represented by stars, and triangles represent IRAS sources from the PSC.}

$$\psfig{file=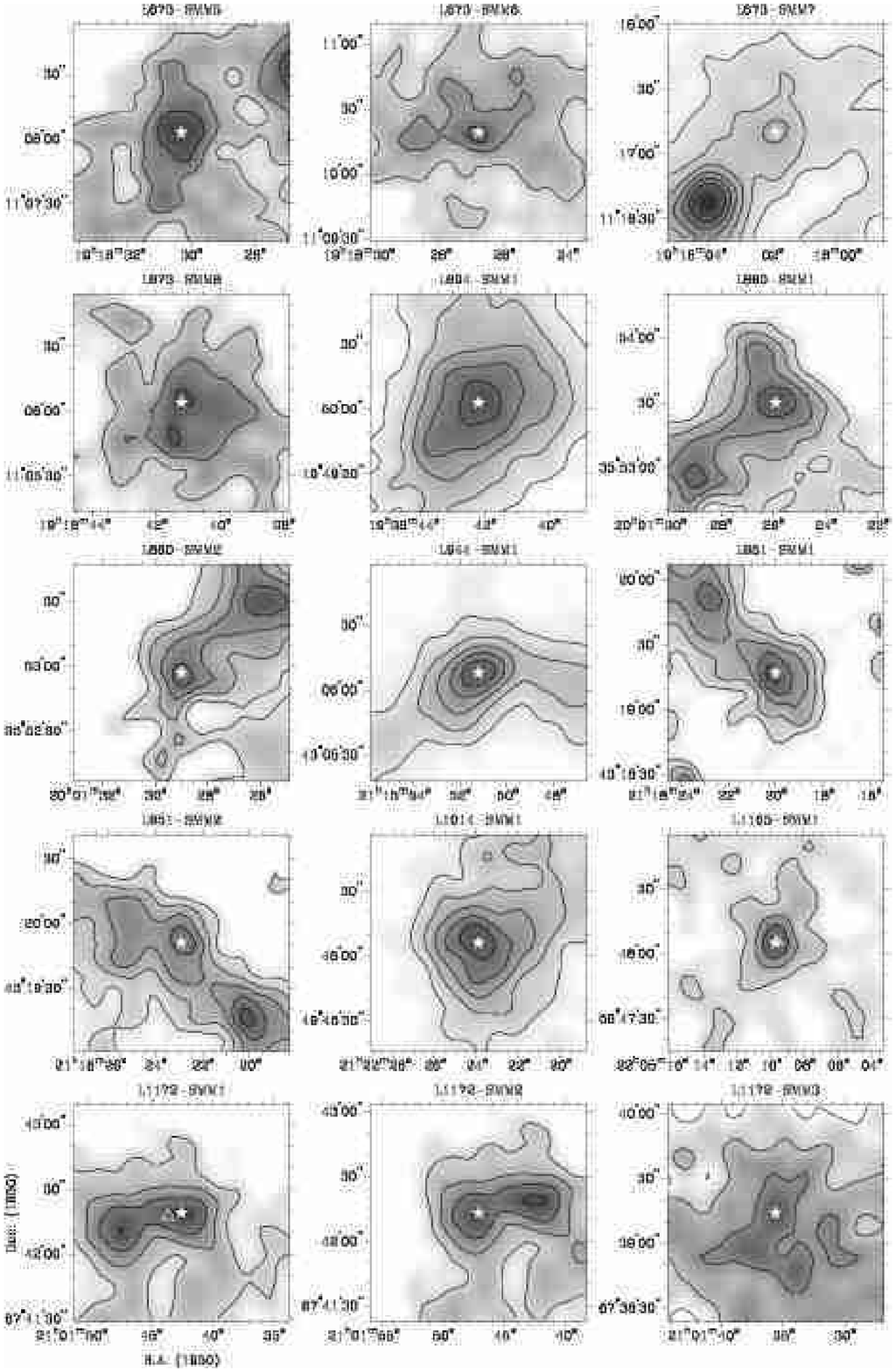,height=7in,angle=0}$$

Fig. \ref{fig_cores}.--- Continued.

$$\psfig{file=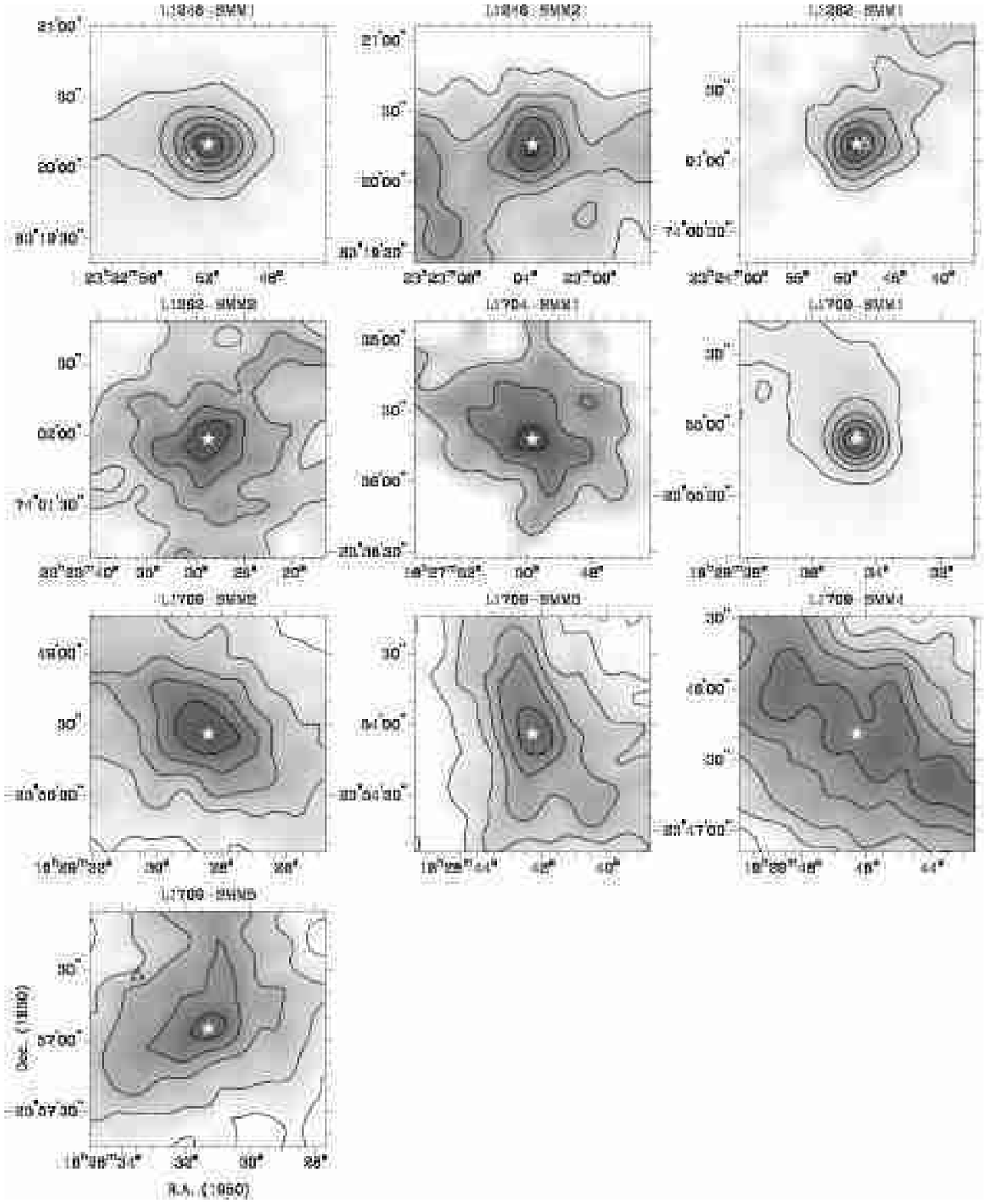,height=7in,angle=0}$$

Fig. \ref{fig_cores}.--- Continued.

$$\psfig{file=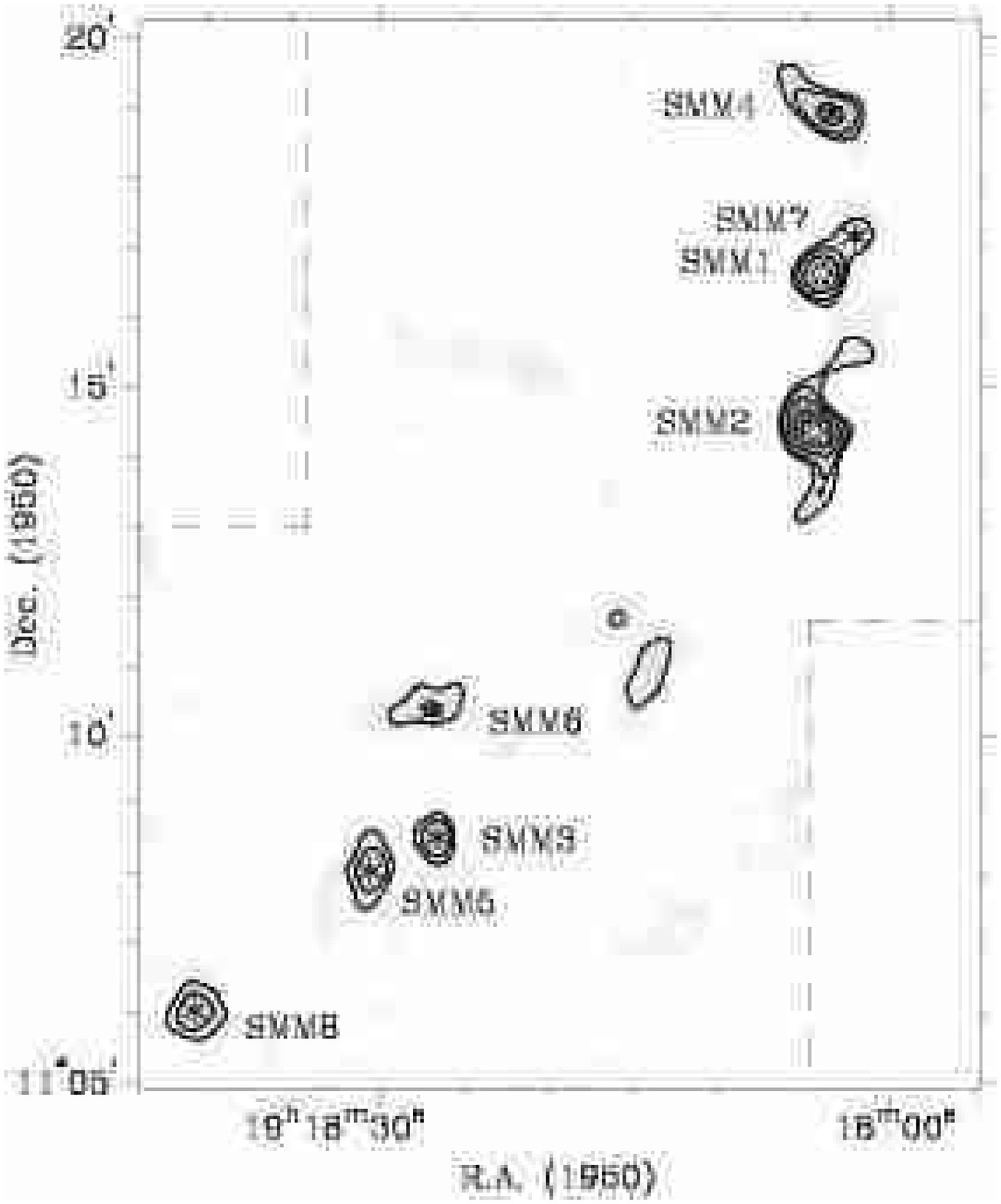,height=3.5in,angle=0}$$

\figcaption{\label{fig673_cores}Positions of the compact submillimeter
cores in L673 from Table~\ref{tab_cores} (asterisks) marked on the
850-$\mu$m SCUBA image.}

$$\psfig{file=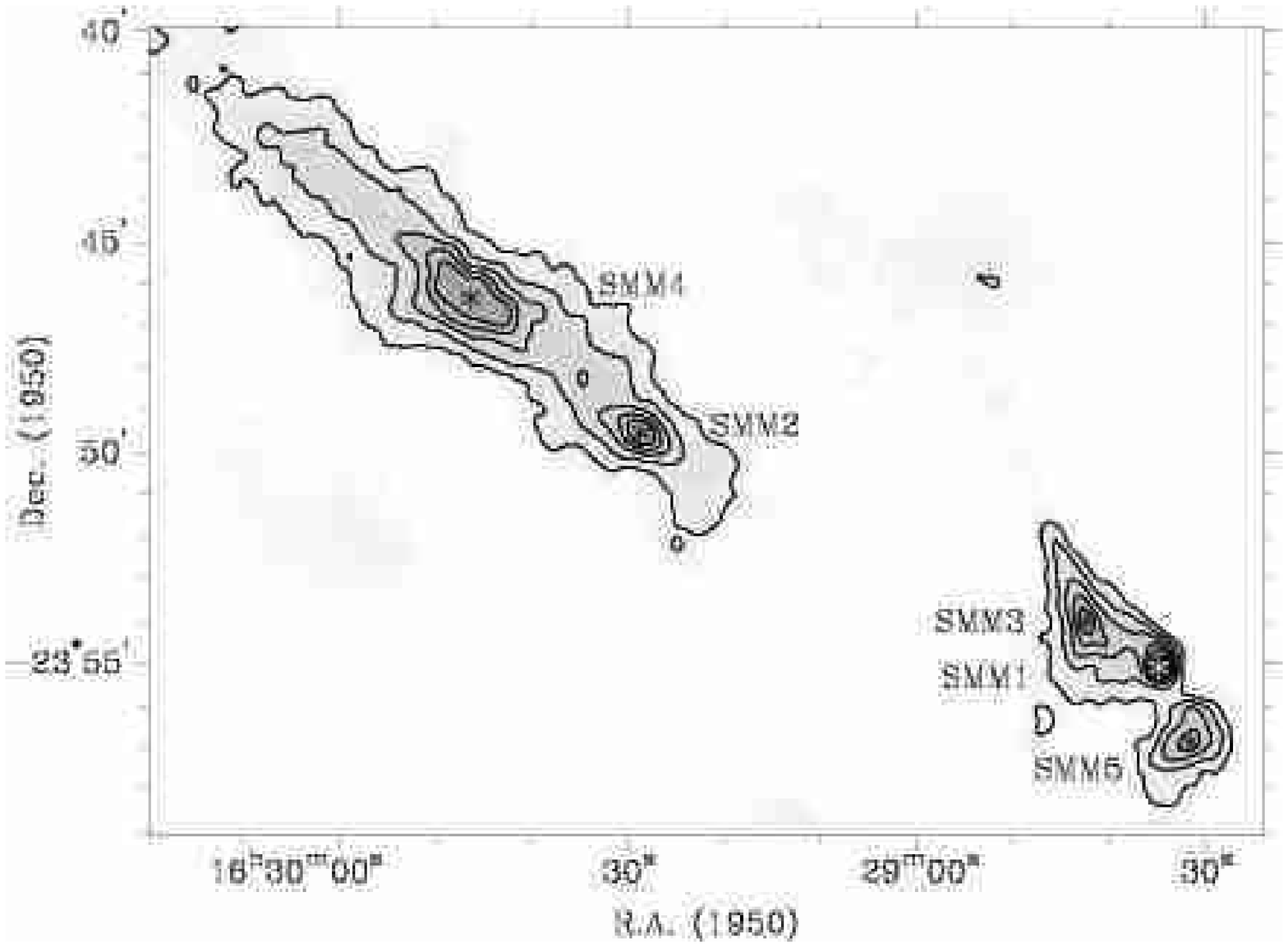,width=4in,angle=0}$$

\figcaption{\label{fig1709_cores}Positions of the compact submillimeter
cores in L1709 from Table~\ref{tab_cores} (asterisks) marked on the
850-$\mu$m SCUBA image.}

$$\psfig{file=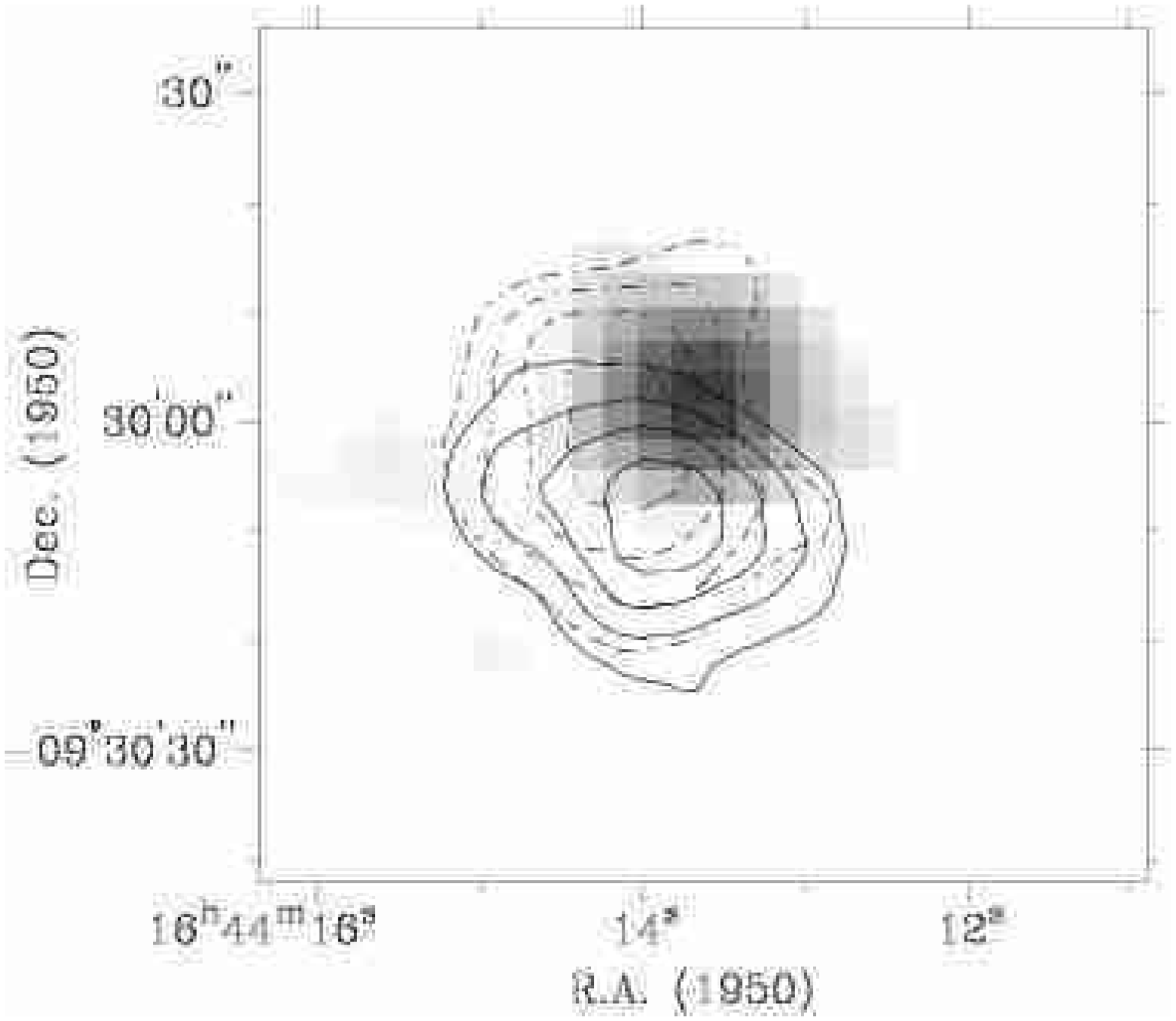,width=3in,angle=0}$$

\figcaption{\label{fig_260outflow}Outflow in L260: red-shifted gas is
integrated over the LSR velocity range 4.5 to 7.5~km~s$^{-1}$ with
contours of $\int T_A^* dV$ at 0.5 + 0.2$n$~K~km~s$^{-1}$, $n$ = 1, 2,
3,..., and the blue-shifted gas is integrated over the range $-$0.5 to
2.5~km~s$^{-1}$ with contours of $\int T_A^* dV$ at 0.4 +
0.2$n$~K~km~s$^{-1}$.  The contours are overlaid on a greyscale of the
SCUBA 850-$\mu$m image.}

$$\psfig{file=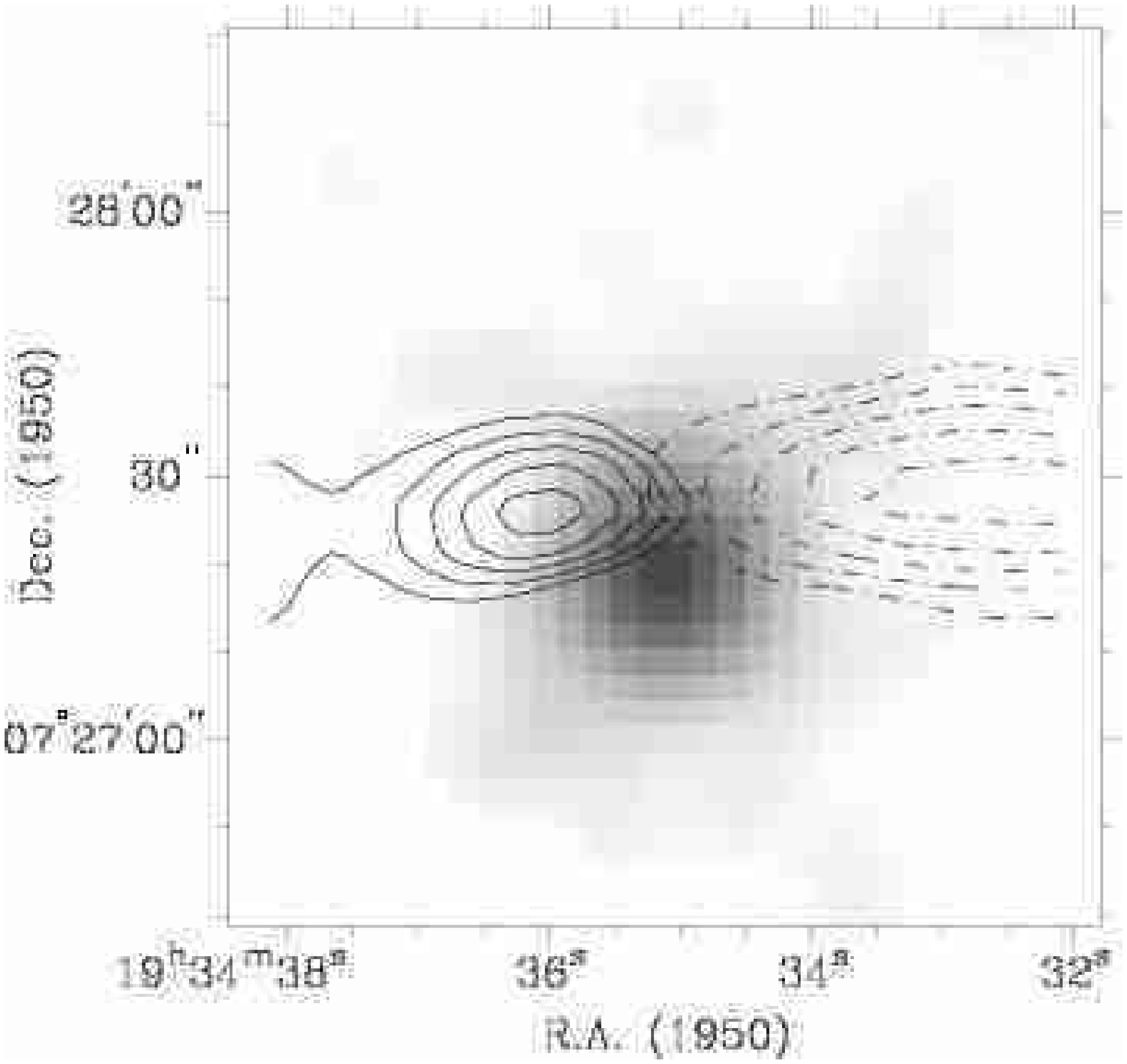,width=3in,angle=0}$$

\figcaption{\label{fig_663outflow}Outflow in L663: red-shifted gas
is integrated over the LSR velocity range 11.5 to 22.0~km~s$^{-1}$
with contours of $\int T_A^* dV$ at 2.0 + 1.0$n$~K~km~s$^{-1}$, $n$
= 1, 2, 3,..., and the blue-shifted gas is integrated from $-$5.0 to
5.5~km~s$^{-1}$ with contours at 3.0 + 0.5$n$~K~km~s$^{-1}$.  The contours
are overlaid on the SCUBA 850-$\mu$m image.}

$$\psfig{file=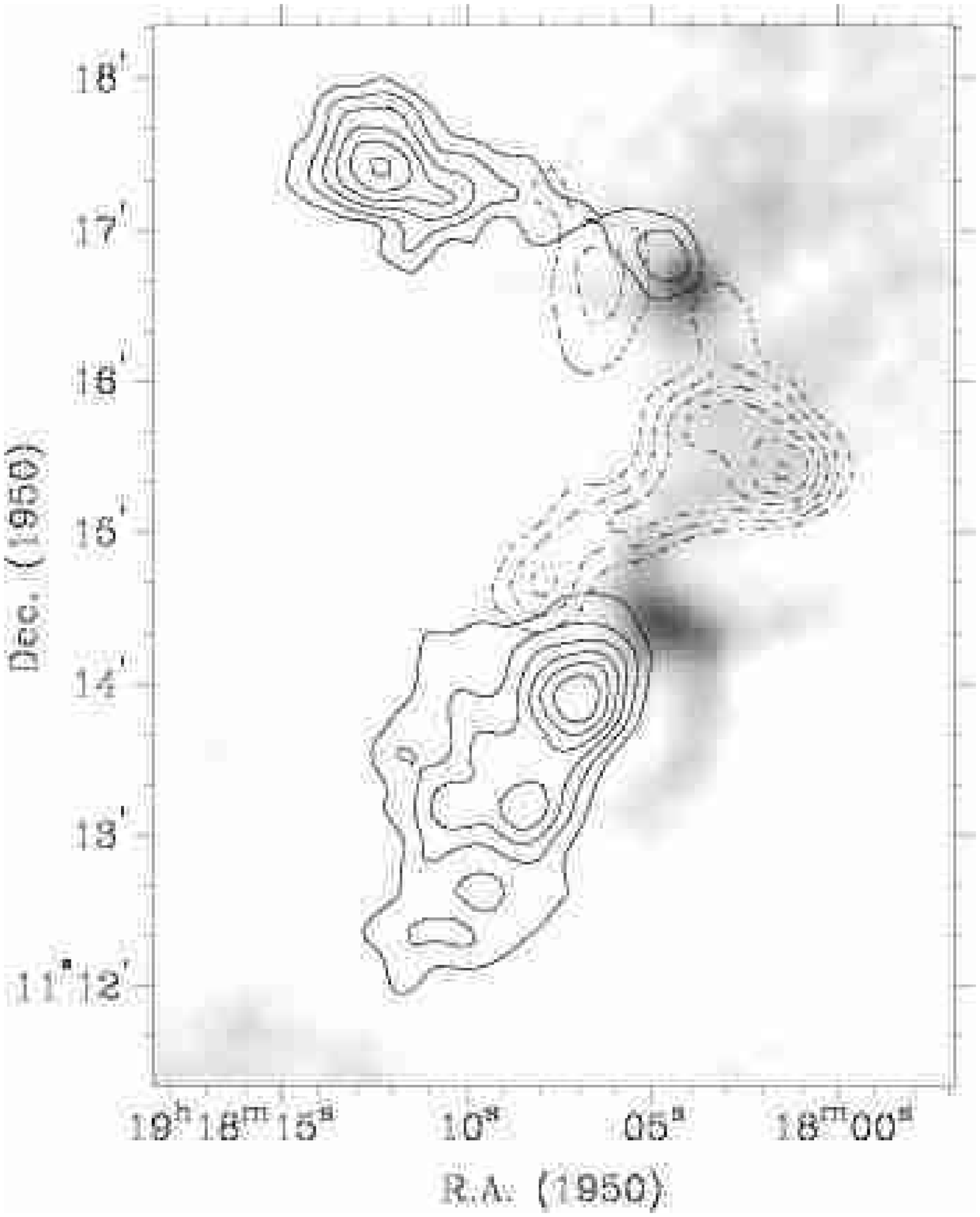,height=5in,angle=0}$$

\figcaption{\label{fig_673outflow}Blue-shifted outflows are clearly
detected from both submillimeter cores in L673, the outflow from
L673$-$SMM1 in the north-east, and the outflow from L673$-$SMM2 directed
toward the south-east.  It is less clear which source is responsible
for the red-shifted flow.  The blue-shifted gas is integrated between
$V_{\rm LSR} = -2.0$ to 3.0~km~s$^{-1}$.  Contours of the blueshifted
flow from L673$-$SMM1 (top) are at 0.7 + 0.4$n$~K~km~s$^{-1}$, and
for L673$-$SMM2 (bottom) they are at 2.0 + 1.0$n$~K~km~s$^{-1}$.
The red-shifted gas is integrated from 9.0 to 15.0~km~s$^{-1}$ with
contours at 7.0 + 2.0$n$~K~km~s$^{-1}$.  The contours are overlaid on
the SCUBA 850-$\mu$m image.}

$$\psfig{file=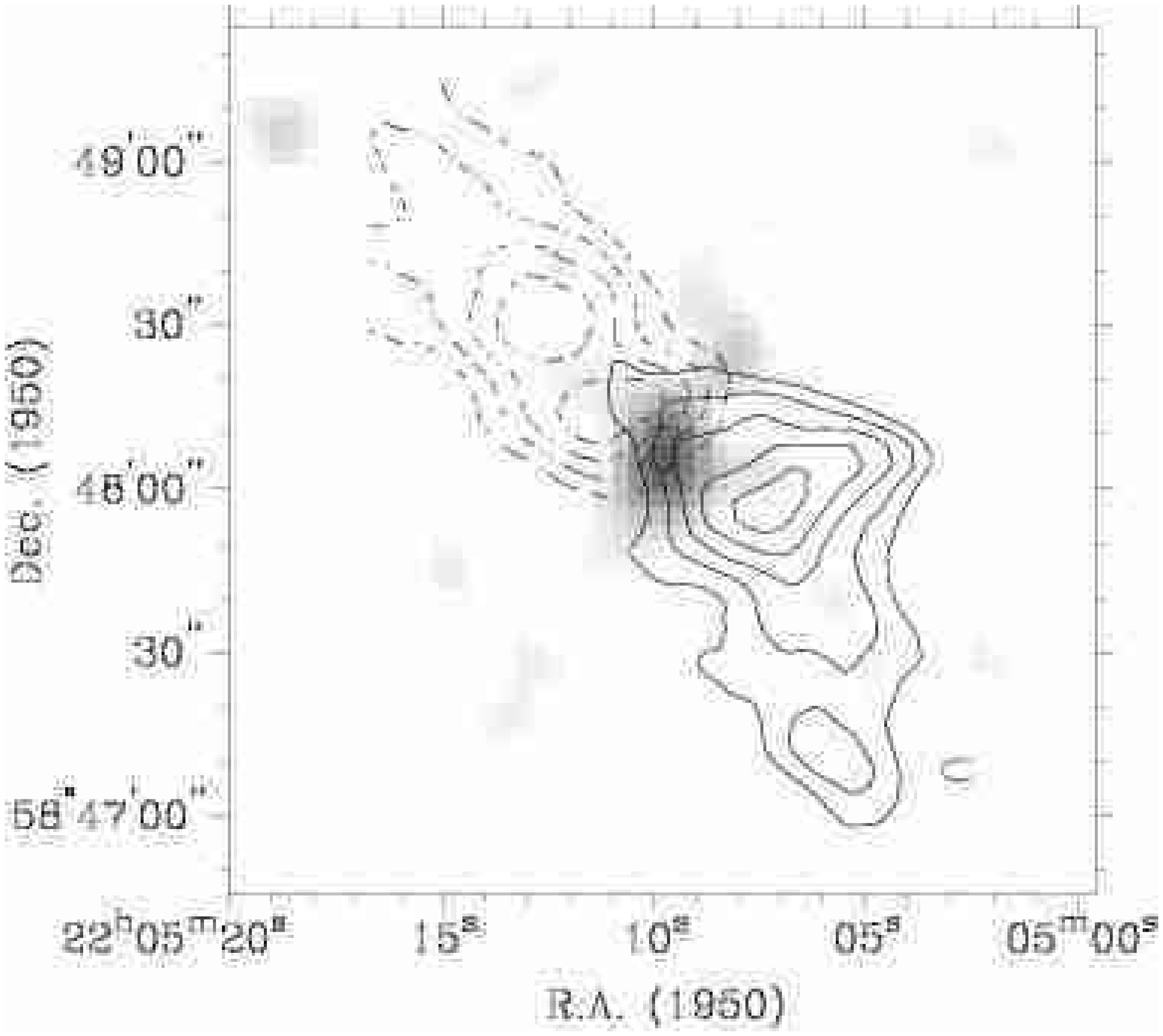,width=3.3in,angle=0}$$

\figcaption{\label{fig_1165outflow}Outflow in L1165: red-shifted gas is
integrated over the LSR velocity range $-$1.0 to 2.7~km~s$^{-1}$ with
contours at 2.6 + 0.7$n$~K~km~s$^{-1}$, and the blue-shifted gas is
integrated from $-$6.0 to $-$3.2~km~s$^{-1}$ with contours at 1.5 +
0.3$n$~K~km~s$^{-1}$, $n$ = 1, 2, 3,...  The contours are overlaid on
the SCUBA 850-$\mu$m image.}

$$\psfig{file=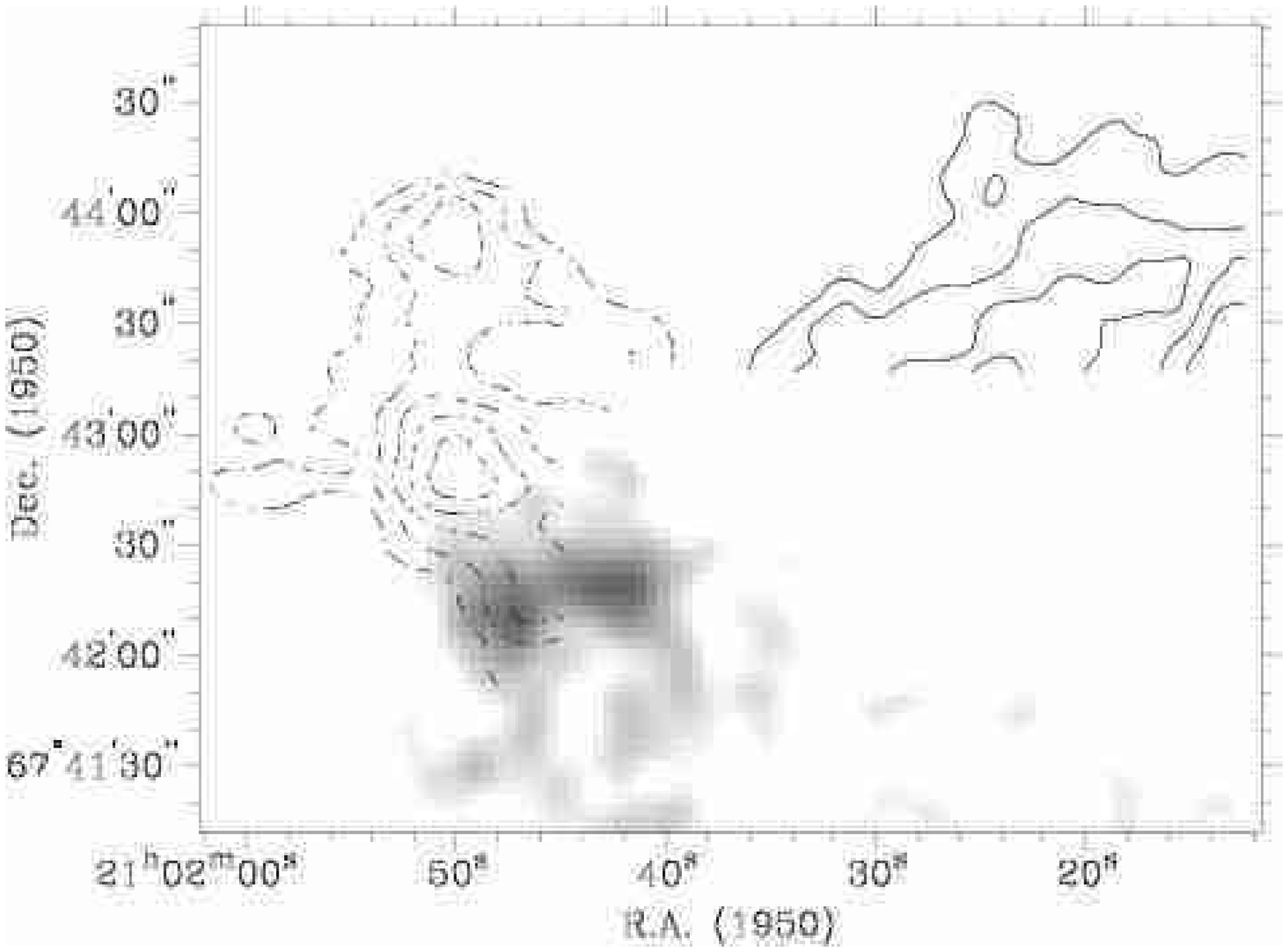,width=4.5in,angle=0}$$

\figcaption{\label{fig_1172outflow}The outflow in L1172 was only partly
mapped: red-shifted gas is integrated over the range 4.0 to
10~km~s$^{-1}$ with contours at 8.0 + 2.0$n$~K~km~s$^{-1}$, and
blue-shifted gas is integrated between $-$5.0 and 1.0~km~s$^{-1}$ with
contours at 4.0 + 1.0$n$~K~km~s$^{-1}$.  The contours are overlaid on
the SCUBA 850-$\mu$m image.}

$$\psfig{file=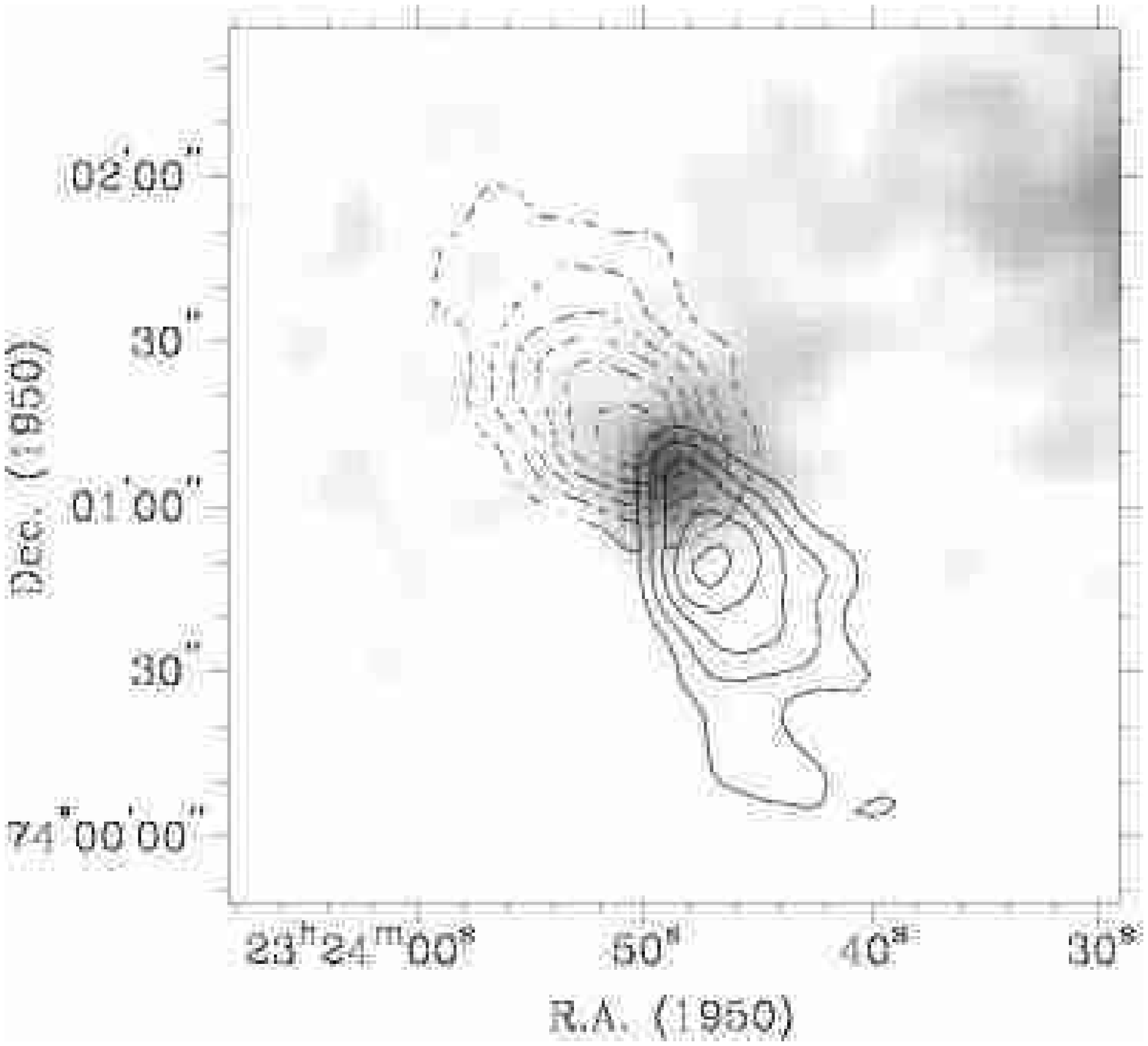,width=3.2in,angle=0}$$

\figcaption{\label{fig_1262outflow}Outflow in L1262: red-shifted gas is
integrated over the range 6.0 to 10.5~km~s$^{-1}$ with contours at 1.0
+ 1.0$n$~K~km~s$^{-1}$, and the blue-shifted gas is integrated from
$-$3.0 to 2.0~km~s$^{-1}$ with contours at 0.5 + 0.5$n$~K~km~s$^{-1}$.
The contours are overlaid on the SCUBA 850-$\mu$m image.}

$$\psfig{file=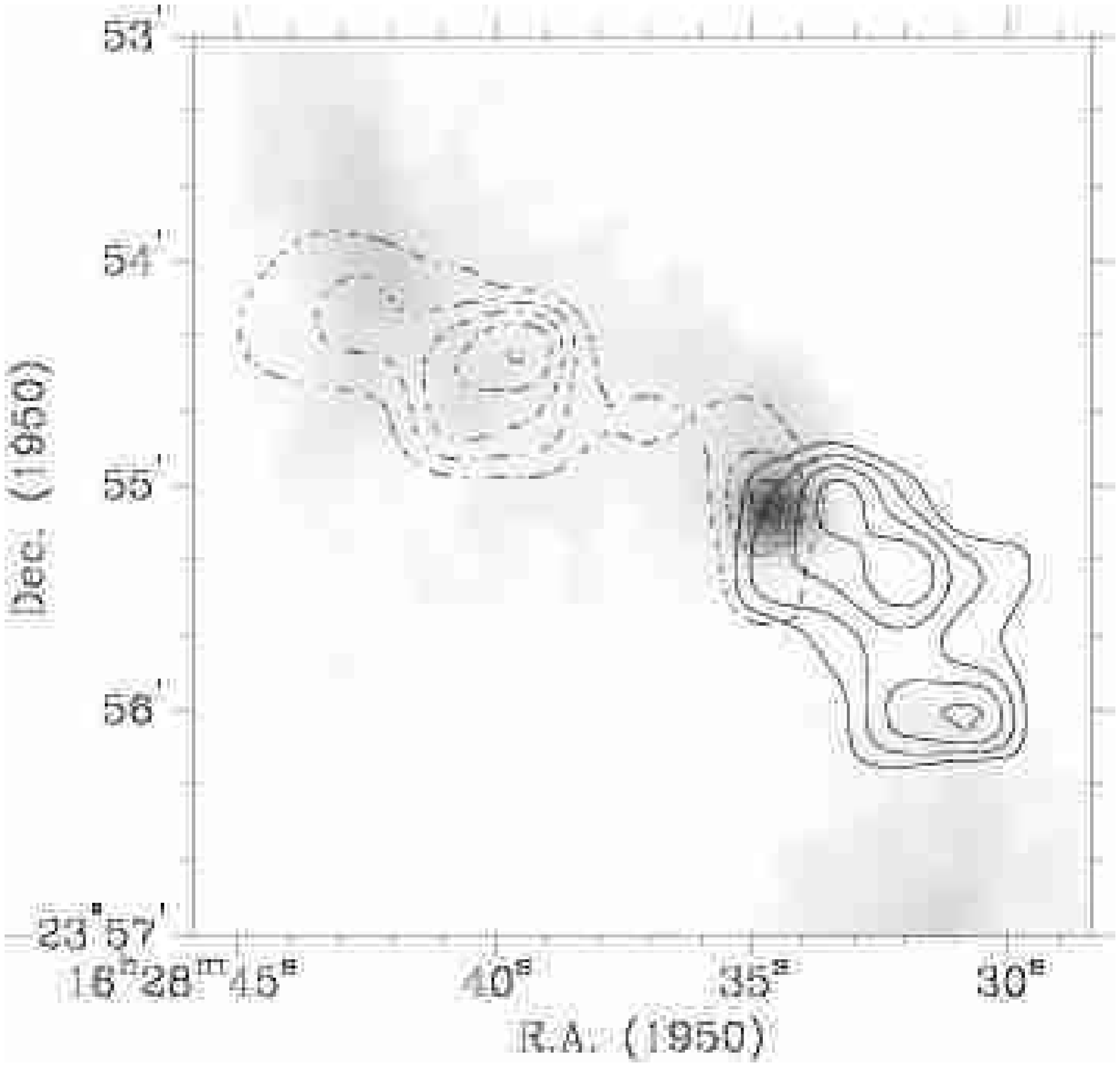,width=3.5in,angle=0}$$

\figcaption{\label{fig_1709smm1outflow}Outflow from L1709$-$SMM1 with
red-shifted gas integrated over the range 4.0 to 9.0~km~s$^{-1}$ and
contours at 0.3 + 0.2$n$~K~km~s$^{-1}$, and blue-shifted gas integrated
from $-$2.0 to 1.0~km~s$^{-1}$ with contours at 0.8 +
0.2$n$~K~km~s$^{-1}$.  The contours are overlaid on the SCUBA
850-$\mu$m image.}

$$\psfig{file=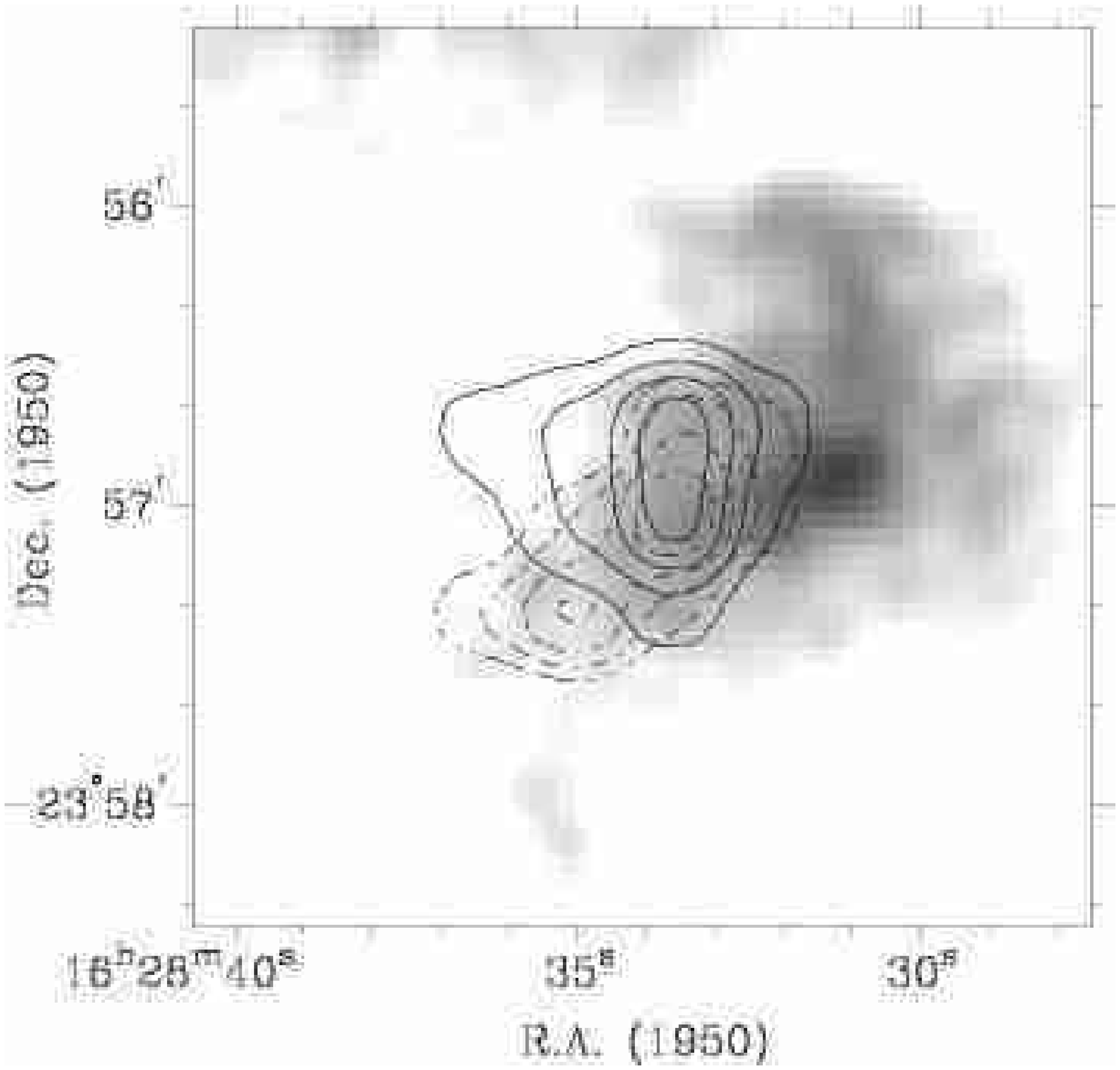,width=4in,angle=0}$$

\figcaption{\label{fig_1709smm5outflow}Outflow from L1709$-$SMM5 with
red-shifted gas integrated over the LSR velocity range 4.0 to
9.0~km~s$^{-1}$ and contours of $\int T_A^* dV$ at 1.0 +
1.0$n$~K~km~s$^{-1}$, $n$ = 1, 2, 3,..., and blue-shifted gas
integrated from $-$2.0 to 1.0~km~s$^{-1}$ with contours at 3.2 +
1.5$n$~K~km~s$^{-1}$.  The contours are overlaid on a greyscale of the
SCUBA 850-$\mu$m image.}

$$\psfig{file=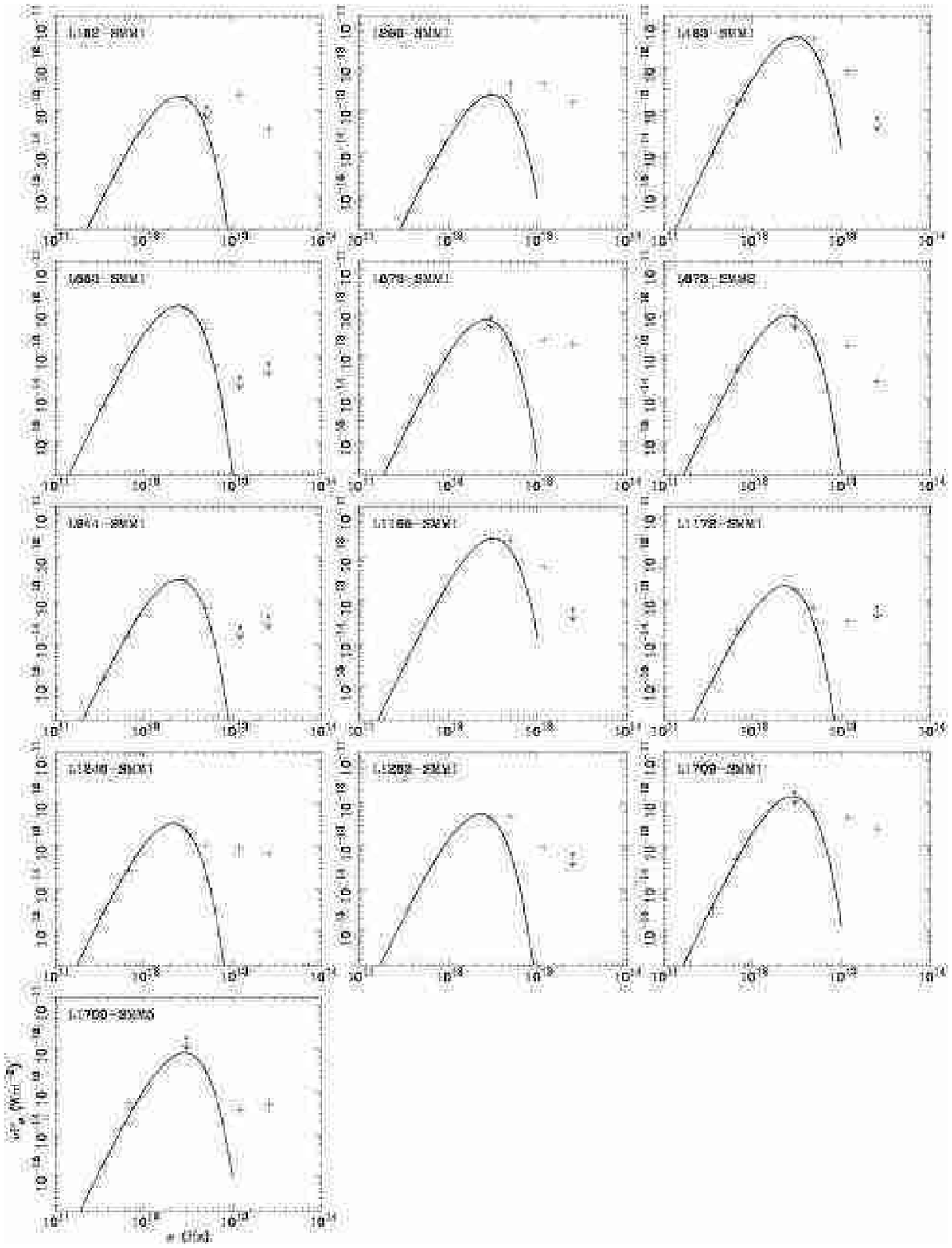,height=7in,angle=0}$$

\figcaption{\label{fig_seds}Spectral energy distributions of all
the protostars in the survey.  The lines are greybody fits to the
submillimeter and far-infrared data, as described in the text.}

$$\psfig{file=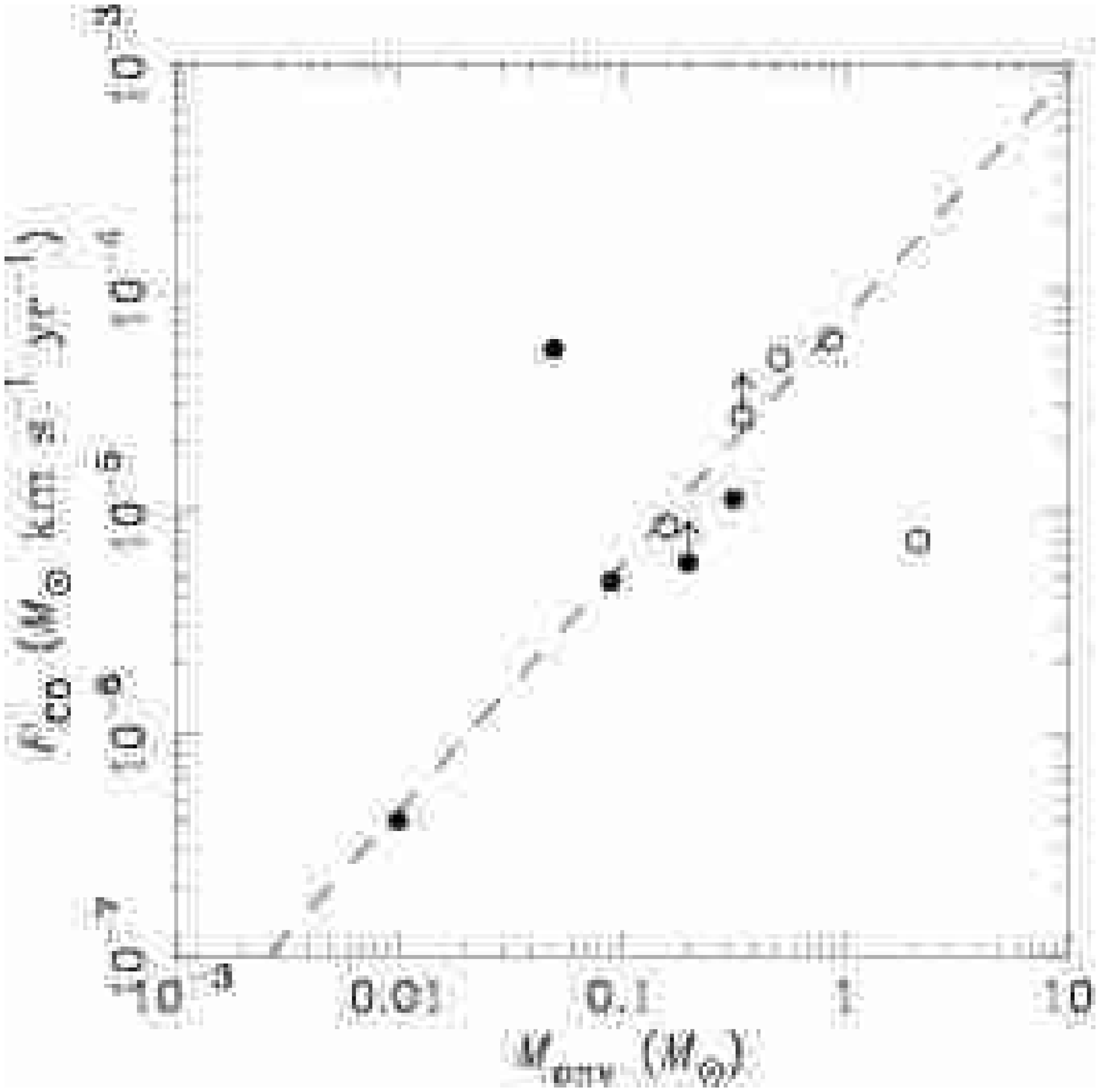,height=4in,angle=0}$$

\figcaption{\label{fig_correlatie}Outflow momentum flux
(Table~\ref{tab_outflows}) versus envelope mass.  Open circles are the
Class~0 sources, filled circles are Class~I sources.  The dashed line
is the best-fit found by Bontemps et al.\ (1996).}

$$\psfig{file=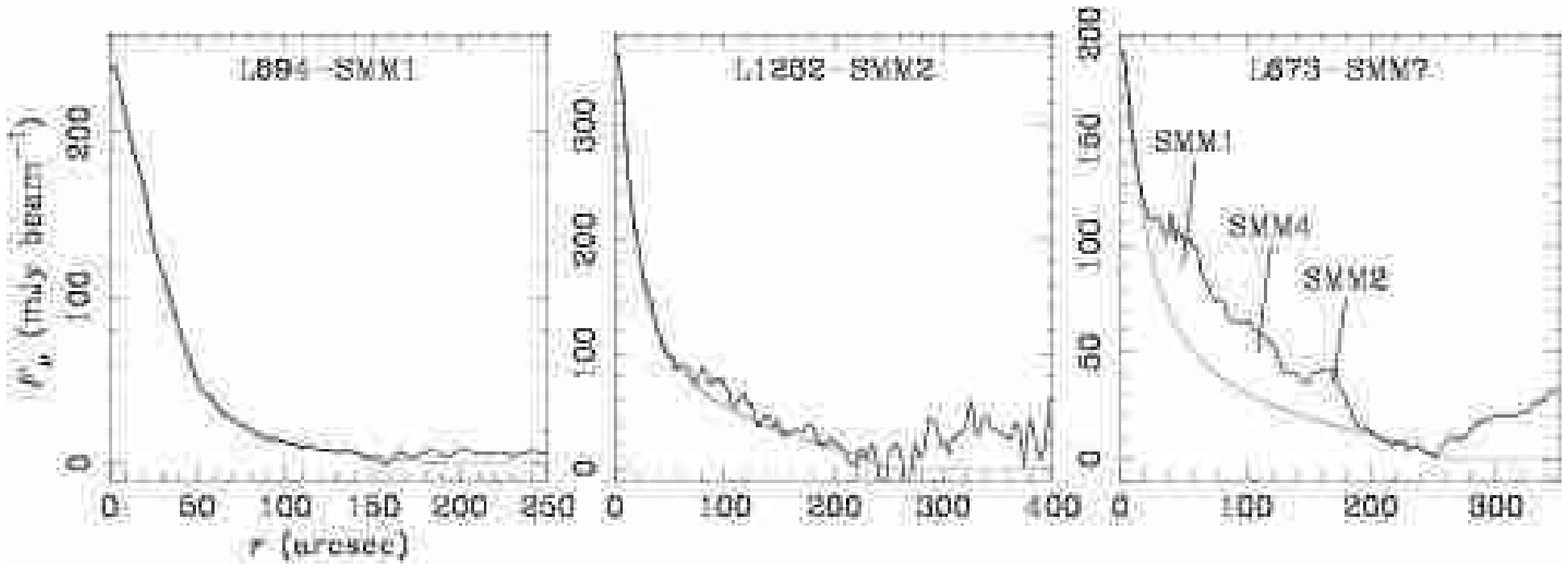,width=6.5in,angle=0}$$

\figcaption{\label{fig_profiles}The 850-$\mu$m intensity profiles of three
starless cores overlaid with best-fit isothermal models (grey lines).
The three have been chosen to illustrate a good fit, L694$-$SMM1 (left),
an average fit, L1262$-$SMM2 (center), and a very poor fit where adjacent
cores can be seen in the radial profile, L673$-$SMM7 (right).}

\newpage

\begin{deluxetable}{lcclrccl}
\tabletypesize{\scriptsize}
\tablecaption{The sample of Lynds opacity class 6 clouds.
\label{tab_sample}}
\tablewidth{0pt}
\tablehead{
\colhead{Source} & \colhead{$\alpha(1950)$\tablenotemark{a}} &
\colhead{$\delta(1950)$} & \colhead{Other names\tablenotemark{b}} &
\colhead{$V_{\rm LSR}$} & \colhead{$D$} & \colhead{References} &
\colhead{Associated complex}\\
\colhead{} & \colhead{(${\rm^h~~^m~~^s}$)~~~} &
\colhead{($^\circ$~~~$'$~~~$''$)} & \colhead{} &
\colhead{km~s$^{-1}$} & \colhead{pc} & \colhead{} &
\colhead{} \\
}
\startdata
L31 & 16 47 31 & $-$19 02 15 & CB67 & 4.8 & 160 & 2,3 & Ophiuchus \\
L53 & 17 20 26 & $-$23 59 45 & \nodata & ? & 160 & \nodata & Ophiuchus \\
L55 & 17 19 55 & $-$23 54 30 & B69  & 4.6 & 160 & \nodata & Ophiuchus \\
L57 & 17 19 36 & $-$23 47 15 & B68, CB82 & 3.5 & 160 & 1,3 & Ophiuchus \\
L63 & 16 47 14 & $-$17 59 00 & \nodata & 5.8 & 160 & 1,2 & Ophiuchus \\
L129 & 16 52 15 & $-$16 17 15 & \nodata & ? & 160 & 2 & Ophiuchus \\
L158 & 16 44 30 & $-$13 52 45 & \nodata & 4.3 & 160 & 1,2 & Ophiuchus \\
L162 & 16 46 13 & $-$14 05 30 & \nodata & ? & 160 & 1,2 & Ophiuchus \\
L222 & 17 38 24 & $-$19 46 00 & CB94 & 10.6 & 160 & \nodata & Ophiuchus \\
L223 & 17 37 53 & $-$19 42 00 & CB93 & 10.3 & 160 & \nodata & Ophiuchus \\
L226 & 17 37 45 & $-$19 38 00 & CB92 & 10.5 & 160 & 3 & Ophiuchus \\
L229 & 17 37 26 & $-$19 31 15 & \nodata & ? & 160 & \nodata & Ophiuchus \\
L231 & 17 37 31 & $-$19 32 45 & \nodata & ? & 160 & \nodata & Ophiuchus \\
L233 & 17 42 18 & $-$20 00 00 & B83A, CB95 & 11.1 & 160 & 3 & Ophiuchus \\
L255 & 16 45 02 & $-$09 49 30 & \nodata & ? & 160 & 1,2 & Ophiuchus \\
L260 & 16 44 33 & $-$09 31 45 & \nodata & 3.5 & 160 & 1,2 & Ophiuchus \\
L328 & 18 14 05 & $-$18 03 00 & B93, CB131 & 6.6 & 190 & 3,9 & \nodata \\
L483 & 18 14 56 & $-$04 41 00 & \nodata & 5.0 & 200 & 1,6 & Aquila Rift \\
L543 & 19 04 11 & $-$06 18 15 & B134, CB181 & 11.3 & 400 & 1,10 & \nodata \\
L663 & 19 34 32 & $+$07 27 30 & B335, CB199 & 8.3 & 250 & 1,7 & \nodata \\
L673 & 19 18 25 & $+$11 07 45 & \nodata & 5.0 & 300 & 1,6 & Cloud B \\
L675 & 19 21 31 & $+$11 01 45 & CB193 & 7.5 & 300 & 6 & Cloud B \\
L694 & 19 38 45 & $+$10 47 45 & B143, CB200 & 9.6 & 250 & 7 & \nodata \\
L709 & 19 11 39 & $+$16 21 45 & CB184 & 6.3 & 300 & 3,6 & Cloud B \\
L771 & 19 18 43 & $+$23 24 15 & CB190 & 11.0 & 400 & 6 & Vul Rift \\
L860 & 20 01 26 & $+$35 53 15 & B146 & ? & 700 & 4 & Cygnus Rift \\
L917 & 20 38 11 & $+$43 58 00 & \nodata & ? & 700 & 4,5 & Cygnus Rift \\
L944 & 21 15 48 & $+$43 05 30 & \nodata & 5.5 & 700 & 4 & Cygnus Rift \\
L951 & 21 18 20 & $+$43 18 45 & \nodata & 5.2 & 700 & 4 & Cygnus Rift \\
L953 & 21 19 32 & $+$43 08 15 & \nodata & 6.3 & 700 & 4 & Cygnus Rift \\
L1014 & 21 22 18 & $+$49 46 30 & \nodata & 4.7 & 200 & 1,8 & \nodata \\
L1021 & 21 19 58 & $+$50 44 30 & \nodata & ? & 200 & 9 & \nodata \\
L1103 & 21 40 34 & $+$56 30 00 & \nodata & ? & 150 & 1,10 & \nodata \\
L1111 & 21 38 54 & $+$57 34 30 & B161, CB233 & $-$0.9 & 150 & 10 & \nodata \\
L1165 & 22 05 17 & $+$58 52 15 & \nodata & $-$1.1 & 300 & 5 & Cloud 157 \\
L1166 & 22 03 47 & $+$59 19 00 & B173, CB236 & $-$2.9 & 300 & 5 & \nodata \\
L1172 & 21 01 19 & $+$67 33 15 & \nodata & 4.5 & 440 & 1,11 & Cepheus \\
L1185 & 22 27 34 & $+$58 54 00 & \nodata & ? & 300 & 11 & \nodata \\
L1246 & 23 23 00 & $+$63 20 00 & CB243 & $-$11.1 & 730 & 3,11 & Cepheus OB3 \\
L1262 & 23 24 19 & $+$74 01 45 & CB244 & 3.9 & 200 & 1,3,11 & \nodata \\
L1704 & 16 27 50 & $-$23 35 30 & CB65 & 2.4 & 160 & \nodata & Ophiuchus \\
L1709 & 16 29 36 & $-$23 47 15 & \nodata & 2.5 & 160 & 1 & Ophiuchus \\
\enddata

\tablenotetext{a}{Positions are from Parker (1988).}

\tablenotetext{b}{B denotes the Barnard (1927) catalogue, and CB denotes
the Clemens \& Barvainis (1988) catalogue.}

\tablerefs{(1) Hilton \& Lahulla 1995; (2) Nozawa et al.\ 1991; (3)
Launhardt \& Henning 1997; (4) Dame \& Thaddeus 1985; (5) Dobashi et
al.\ 1994; (6) Dame et al.\ 1987; (7) Tomita, Saito, \& Ohtani 1979;
(8) Robert \& Pagani 1993; (9) Lee \& Myers 1999; (10) Leung, Kutner,
\& Mead 1982; (11) Yonekura et al.\ 1997.}

\end{deluxetable}

\clearpage

\begin{deluxetable}{llrrrrl}
\tabletypesize{\scriptsize}
\tablecaption{IRAS PSC associations with the Lynds dark clouds.
\label{tab_iras}}
\tablewidth{0pt}
\tablehead{
\colhead{Source} & \colhead{IRAS name(s)} & 
\colhead{$F_\nu$(12~$\mu$m)\tablenotemark{d}} &
\colhead{$F_\nu$(25~$\mu$m)} &
\colhead{$F_\nu$(60~$\mu$m)} &
\colhead{$F_\nu$(100~$\mu$m)} &
\colhead{Identification} \\
\colhead{} & \colhead{} & \colhead{Jy} & \colhead{Jy} & \colhead{Jy} &
\colhead{Jy} & \colhead{} \\
}
\startdata
L53  & 17205$-$2359 & .26L & .48L & .70L & 12.23\phantom{L} & \nodata \\
L158 & 16442$-$1351 & .31L & .52L & 1.70\phantom{L} & 33.80L & \nodata \\
     & 16445$-$1352 & .27L & .33L & 2.22\phantom{L} & 32.62\phantom{L} &
       cirrus \\
     & 16439$-$1353 & .32L & .45L & .78\phantom{L} & 27.56L & \nodata \\
L162 & 16455$-$1405\tablenotemark{b}\tablenotemark{c} & .63\phantom{L} & 
       .90\phantom{L} & 1.23\phantom{L} & 13.29L & T~Tauri \\
     & 16464$-$1407 & .60L & .36L & .63\phantom{L} & 6.48\phantom{L} & 
       \nodata \\
     & 16457$-$1409 & .33L & .34L & 2.33L & 5.26\phantom{L} & \nodata \\
     & 16459$-$1411 & 1.41\phantom{L} & 1.90\phantom{L} & 2.33L & 
       6.47\phantom{L} & T~Tauri \\
L226 & 17375$-$1936 & .64\phantom{L} & .40L & .55L & 19.61L & \nodata \\ 
L255 & 16451$-$0953 & .25L & .32L & .60\phantom{L} & 12.88\phantom{L} & 
       \nodata \\
L260 & 16446$-$0924 & .32L & .92L & .40L & 5.30\phantom{L} & \nodata \\ 
     & 16446$-$0925 & .25L & .69L & .40L & 6.30\phantom{L} & \nodata \\
     & 16442$-$0930\tablenotemark{c} & .57\phantom{L} & 3.31\phantom{L} & 
       7.83\phantom{L} & 7.54\phantom{L} & protostar \\ 
     & 16450$-$0926\tablenotemark{a} & .25L & .34L & .36\phantom{L} & 
       6.02\phantom{L} & \nodata \\
L483 & 18148$-$0440\tablenotemark{c} & .25L & 6.91\phantom{L} & 
       89.05\phantom{L} & 165.50\phantom{L} & protostar \\
L663 & 19343$+$0727 & .32\phantom{L} & .25L & .40L & 2.77L & \nodata \\ 
     & 19345$+$0727 & .25L & .25L & 8.30\phantom{L} & 41.96\phantom{L} & 
       protostar \\  
L673 & 19180$+$1116 & 1.09\phantom{L} & 2.22\phantom{L} & 3.51L & 87.21L & 
       \nodata \\ 
     & 19180$+$1114 & .41L & 1.60\phantom{L} & 2.55L & 105.70L & protostar \\ 
     & 19181$+$1056\tablenotemark{c} & 1.71\phantom{L} & 2.81\phantom{L} & 
       3.69\phantom{L} & 87.16L & T~Tauri \\ 
     & 19181$+$1112 & .38L & .43\phantom{L} & 5.59L & 106.70L & \nodata \\ 
     & 19181$+$1059 & 3.74\phantom{L} & 3.55\phantom{L} & 5.18L & 91.15L & 
       \nodata \\ 
     & 19183$+$1123\tablenotemark{a} & .96\phantom{L} & .50\phantom{L} & 
       5.53L & 82.50L & \nodata \\
     & 19184$+$1055 & 3.74\phantom{L} & 5.46\phantom{L} & 3.69L & 
       9.69\phantom{L} & OH/IR star \\ 
     & 19184$+$1118 & 2.95\phantom{L} & 1.07\phantom{L} & 5.45L & 71.62L & 
       \nodata \\ 
     & 19190$+$1105\tablenotemark{b} & 1.32\phantom{L} & 1.01\phantom{L} & 
       24.52L & 86.05L & \nodata \\ 
L694 & 19389$+$1048 & .25L & .25L & .40L & 7.91\phantom{L} & \nodata \\ 
     & 19380$+$1045 & .54\phantom{L} & .20\phantom{L} & .40L & 12.91L & 
       \nodata \\ 
L709 & 19116$+$1623 & .97\phantom{L} & .57\phantom{L} & .98L & 8.93L & 
       \nodata \\ 
L771 & 19186$+$2325\tablenotemark{a} & .25L & .25L & 1.03\phantom{L} & 
       6.61\phantom{L} & \nodata \\
L951 & 21186$+$4320 & .25L & .25L & 1.84L & 6.84\phantom{L} & \nodata \\ 
L1165 & 22051$+$5849 & 2.16\phantom{L} & .48\phantom{L} & .53L & 105.90L & 
       \nodata \\ 
     & 22051$+$5848\tablenotemark{c} & .25L & 5.22\phantom{L} & 
       51.59\phantom{L} & 94.00\phantom{L} & protostar \\ 
L1172 & 21017$+$6742\tablenotemark{a} & .28L & .26\phantom{L} & 
       1.33\phantom{L} & 5.83\phantom{L} & protostar \\
L1246 & 23228$+$6320\tablenotemark{c} & .26\phantom{L} & .74\phantom{L} & 
       2.06\phantom{L} & 7.98\phantom{L} & protostar \\
L1262 & 23238$+$7401\tablenotemark{c} & .25L & .78\phantom{L} & 
       9.60\phantom{L} & 15.20\phantom{L} & protostar \\ 
L1704 & 16277$-$2332\tablenotemark{a} & .74L & .54L & .92\phantom{L} & 
       20.02L & \nodata \\
L1709 & 16291$-$2351 & .25L & .39L & .86L & 3.59\phantom{L} & \nodata \\ 
     & 16285$-$2355\tablenotemark{c} & 1.19\phantom{L} & 3.73\phantom{L} & 
       9.44\phantom{L} & 60.86L & protostar \\ 
     & 16285$-$2356 & .31L & .42L & 1.01L & 60.86\phantom{L} & \nodata \\ 
     & 16285$-$2358 & .35L & .81\phantom{L} & .97L & 60.86L & star \\
\enddata

\tablenotetext{a}{IRAS source not associated with the cloud by
Parker (1988).}

\tablenotetext{b}{IRAS source not covered by our SCUBA map.}

\tablenotetext{c}{Source studied by Parker (1991).}

\tablenotetext{d}{Flux density in the IRAS bands.  L indicates an upper
limit.}

\end{deluxetable}

\clearpage

\begin{deluxetable}{lcccclcccccc}
\rotate
\tabletypesize{\scriptsize}
\tablecaption{Flux densities and cloud properties of the sample.
\label{tab_fluxes}}
\tablewidth{0pt}
\tablehead{
\colhead{Source} & \multicolumn{2}{c}{Peak $F_\nu$
(Jy~beam$^{-1}$)\tablenotemark{a}} &
\multicolumn{2}{c}{Integrated $F_\nu$ (Jy)} & \colhead{obs.} & 
\colhead{Peak mass} & \colhead{Peak $N({\rm H}_2)$} &
\colhead{Peak $n({\rm H}_2)$} & \colhead{Peak $A_V$} & 
\colhead{Total mass} & \colhead{Mean $N({\rm H}_2)$} \\
\colhead{} & \colhead{450~$\mu$m} & \colhead{850~$\mu$m} &
\colhead{450~$\mu$m} & \colhead{850~$\mu$m} & \colhead{mode} &
\colhead{$M_\odot$} & \colhead{$\times 10^{26}$~m$^{-2}$} &
\colhead{$\times 10^{11}$~m$^{-3}$} & \colhead{mag} &
\colhead{$M_\odot$} & \colhead{$\times 10^{25}$~m$^{-2}$} \\
}
\startdata
L31  & \nodata & 0.11$\pm$0.02 & \nodata & \phn4.6 & jiggle &
     0.06 & \phn1.5 & \phn3.9 & \phn15 & \phn2.2 & \phn5.5 \\
L53  & \nodata & $<$0.12 & \nodata & \nodata & jiggle &
     $<$0.05 & $<$1.5 & $<$4.0 & $<$16 & \nodata & \nodata \\
L55  & \nodata & 0.10$\pm$0.02 & \nodata & \phn1.1 & jiggle &
     0.05 & \phn1.3 & \phn3.6 & \phn14 & \phn0.5 & \phn4.7 \\
L57  & 1.1$\pm$0.19 & 0.14$\pm$0.02 & \phn45 & \phn2.8 & scan &
     0.07 & \phn1.8 & \phn4.9 & \phn19 & \phn1.4 & \phn1.6 \\
L63  & 2.6$\pm$0.23 & 0.31$\pm$0.03 & 945 & 52.3 & scan &
     0.15 & \phn4.1 & 11.1 & \phn43 & 25.4 & \phn6.3 \\
L129 & \nodata & $<$0.09 & \nodata & \nodata & scan &
     $<$0.05 & $<$1.3 & $<$3.6 & $<$14 & \nodata & \nodata \\
L158 & 1.9$\pm$0.42 & 0.22$\pm$0.03 & 415 & 19.6 & scan &
     0.11 & \phn3.0 & \phn8.0 & \phn31 & \phn9.5 & \phn2.2 \\
L162 & \nodata & 0.38$\pm$0.05 & \nodata & 51.9 & scan &
     0.18 & \phn5.1 & 13.7 & \phn53 & 25.2 & \phn4.2 \\
L222 & \nodata & 0.12$\pm$0.02 & \nodata & \phn0.3 & jiggle &
     0.06 & \phn1.5 & \phn4.1 & \phn16 & \phn0.2 & \phn5.5 \\
L223 & \nodata & 0.12$\pm$0.02 & \nodata & \phn0.4 & jiggle &
     0.06 & \phn1.5 & \phn4.1 & \phn16 & \phn0.2 & \phn5.8 \\
L226 & \nodata & $<$0.09 & \nodata & \nodata & jiggle &
     $<$0.05 & $<$1.3 & $<$3.5 & $<$13 & \nodata & \nodata \\
L229 & \nodata & $<$0.09 & \nodata & \nodata & jiggle &
     $<$0.04 & $<$1.1 & $<$3.0 & $<$12 & \nodata & \nodata \\
L231 & \nodata & $<$0.09 & \nodata & \nodata & jiggle &
     $<$0.04 & $<$1.0 & $<$2.8 & $<$11 & \nodata & \nodata \\
L233 & \nodata & $<$0.06 & \nodata & \nodata & jiggle &
     $<$0.03 & $<$0.9 & $<$2.5 & $<$10 & \nodata & \nodata \\
L255 & \nodata & $<$0.09 & \nodata & \nodata & scan &
     $<$0.04 & $<$1.0 & $<$2.8 & $<$11 & \nodata & \nodata \\
L260 & 1.4$\pm$0.24 & 0.21$\pm$0.02 & \phn16 & \phn1.1 & scan &
     0.10 & \phn2.8 & \phn7.6 & \phn30 & \phn0.5 & \phn0.7 \\
L328 & 0.7$\pm$0.12 & 0.21$\pm$0.02 & \phn\phn8 & \phn1.7 & jiggle &
     0.14 & \phn2.8 & \phn6.4 & \phn30 & \phn1.2 & 10.8 \\
L483 & 6.5$\pm$0.23 & 1.30$\pm$0.03 & 400 & 13.1 & scan &
     0.99 & 17.3 & 37.2 & 182 & 10.0 & \phn1.0 \\
L543 & \nodata & $<$0.12 & \nodata & \nodata & scan &
     $<$0.33 & $<$1.4 & $<$1.6 & $<$15 & \nodata & \nodata \\
L663 & \nodata & 1.16$\pm$0.10 & \nodata & \phn7.3 & scan &
     1.38 & 15.4 & 26.6 & 162 & \phn8.7 & 20.9 \\
L673 & 1.5$\pm$0.21 & 0.42$\pm$0.03 & \phn55 & 51.1 & scan &
     0.71 & \phn5.5 & \phn7.9 & \phn58 & 87.4 & \phn3.4 \\
L675 & \nodata & 0.07$\pm$0.02 & \nodata & \phn0.1 & jiggle &
     0.11 & \phn0.9 & \phn1.3 & \phn\phn9 & \phn0.2 & \phn3.4 \\
L694 & 1.0$\pm$0.18 & 0.27$\pm$0.03 & 430 & 21.9 & scan &
     0.32 & \phn3.6 & \phn6.1 & \phn37 & 26.0 & \phn1.0 \\
L709 & \nodata & 0.06$\pm$0.01 & \nodata & \phn0.6 & jiggle &
     0.10 & \phn0.8 & \phn1.1 & \phn\phn8 & \phn1.0 & \phn3.1 \\
L771 & 1.5$\pm$0.25 & 0.14$\pm$0.02 & 105 & \phn9.4 & scan &
     0.43 & \phn1.9 & \phn2.0 & \phn20 & 28.6 & \phn3.9 \\
L860 & \nodata & 0.09$\pm$0.01 & \nodata & \phn1.0 & jiggle &
     0.87 & \phn1.2 & \phn0.8 & \phn13 & \phn9.4 & \phn3.6 \\
L917 & \nodata & 0.09$\pm$0.02 & \nodata & \phn1.1 & jiggle &
     0.85 & \phn1.2 & \phn0.7 & \phn13 & 10.4 & \phn4.4 \\
L944 & 0.8$\pm$0.12 & 0.21$\pm$0.02 & \phn\phn5 & \phn1.4 & jiggle &
     1.99 & \phn2.9 & \phn1.8 & \phn30 & 13.0 & \phn6.0 \\
L951 & \nodata & 0.07$\pm$0.02 & \nodata & \phn0.5 & jiggle &
     0.66 & \phn0.9 & \phn0.6 & \phn10 & \phn4.7 & \phn4.0 \\
L953 & \nodata & 0.06$\pm$0.02 & \nodata & \phn0.2 & jiggle &
     0.59 & \phn0.8 & \phn0.5 & \phn\phn9 & \phn2.2 & \phn3.6 \\
L1014 & 0.8$\pm$0.12 & 0.15$\pm$0.02 & \phn\phn5 & \phn0.9 & jiggle &
     0.11 & \phn1.9 & \phn4.2 & \phn20 & \phn0.7 & \phn6.0 \\
L1021 & \nodata & 0.09$\pm$0.03 & \nodata & \phn0.7 & jiggle &
     0.07 & \phn1.2 & \phn2.7 & \phn13 & \phn0.5 & \phn4.8 \\
L1103 & \nodata & 0.11$\pm$0.03 & \nodata & \phn1.4 & jiggle &
     0.05 & \phn1.4 & \phn4.1 & \phn15 & \phn0.6 & \phn5.6 \\
L1111 & \nodata & 0.10$\pm$0.03 & \nodata & \phn0.7 & jiggle &
     0.04 & \phn1.3 & \phn3.7 & \phn13 & \phn0.3 & \phn5.0 \\
L1165 & \nodata & 0.99$\pm$0.12 & \nodata & 17.2 & scan &
     1.69 & 13.1 & 18.9 & 138 & 29.4 & 25.1 \\
L1166 & \nodata & 0.10$\pm$0.02 & \nodata & \phn0.7 & jiggle &
     0.17 & \phn1.4 & \phn2.0 & \phn14 & \phn1.2 & \phn4.9 \\
L1172 & 2.6$\pm$0.30 & 0.20$\pm$0.03 & 770 & 12.3 & scan &
     0.75 & \phn2.7 & \phn2.6 & \phn28 & 45.2 & \phn1.7 \\
L1185 & \nodata & 0.09$\pm$0.02 & \nodata & \phn0.9 & jiggle &
     0.15 & \phn1.2 & \phn1.7 & \phn13 & \phn1.5 & \phn4.5 \\
L1246 & 1.8$\pm$0.15 & 0.40$\pm$0.02 & \phn13 & \phn4.0 & jiggle &
     4.07 & \phn5.4 & \phn3.2 & \phn56 & 40.1 & \phn8.9 \\
L1262 & \nodata & 0.69$\pm$0.06 & \nodata & 10.5 & scan &
     0.53 & \phn9.2 & 19.8 & \phn97 & \phn8.0 & 14.8 \\
L1704 & 0.8$\pm$0.19 & 0.10$\pm$0.02 & \nodata & \phn0.5 & scan &
     0.05 & \phn1.3 & \phn3.4 & \phn13 & \phn0.2 & \phn1.9 \\
L1709 & 4.9$\pm$0.37 & 1.03$\pm$0.03 & 690 & 64.5 & scan &
     0.50 & 13.7 & 36.8 & 144 & 31.4 & \phn8.4 \\
\enddata

\tablenotetext{a}{Uncertainties are statistical only; upper limits are
3$\sigma$.}

\end{deluxetable}

\clearpage

\begin{deluxetable}{lcclccl}
\tabletypesize{\footnotesize}
\tablecaption{Submillimeter cores identified in the SCUBA maps.
\label{tab_cores}}
\tablewidth{0pt}
\tablehead{
\colhead{Core} & \colhead{R.A. (1950)} & \colhead{Dec.\ (1950)} &
\colhead{IRAS} & \colhead{Searched for} & \colhead{Outflow} &
\colhead{Identification}\\
\colhead{} & \colhead{h~~~m~~~s} & \colhead{~~~$^\circ~~~'~~~''$} &
\colhead{association\tablenotemark{a}} & \colhead{outflow?} &
\colhead{detected?} & \colhead{} \\
}
\startdata
L55$-$SMM1 & 17 19 55.3 & $-$23 55 30 & \nodata & yes & no & starless core \\
L57$-$SMM1 & 17 19 35.7 & $-$23 47 08 & \nodata & no & \nodata & 
            probably starless \\
L63$-$SMM1 & 16 47 20.1 & $-$18 01 12 & \nodata & yes & no & starless core \\
L158$-$SMM1 & 16 44 33.9 & $-$13 53 50 & \nodata & yes & no & starless core \\
L158$-$SMM2 & 16 44 23.4 & $-$13 53 01 & \nodata & no & \nodata & 
            probably starless \\
L162$-$SMM1 & 16 45 56.1 & $-$14 11 25 & 16459$-$1411 & no & \nodata &
            known T~Tauri star \\
L260$-$SMM1 & 16 44 13.7 & $-$09 29 58 & 16442$-$0930 & yes & yes &
            known protostar \\
L260$-$SMM2 & 16 44 24.6 & $-$09 29 46 & \nodata & no & \nodata & 
            probably starless \\
L328$-$SMM1 & 18 14 05.0 & $-$18 03 12 & \nodata & yes & no & starless core \\
L483$-$SMM1 & 18 14 50.2 & $-$04 40 48 & 18148$-$0440 & no & 
            yes\tablenotemark{b} & known protostar \\
L663$-$SMM1 & 19 34 35.1 & $+$07 27 20 & 19345$+$0727 & yes & yes &
            known protostar \\
L673$-$SMM1 & 19 18 04.0 & $+$11 16 37 & 19180$+$1116? & yes & yes &
            new protostar \\
L673$-$SMM2 & 19 18 04.6 & $+$11 14 23 & 19180$+$1114 & yes & yes &
            known protostar \\
L673$-$SMM3 & 19 18 26.6 & $+$11 08 31 & \nodata & yes & no & 
            candidate protostar \\
L673$-$SMM4 & 19 18 03.5 & $+$11 18 54 & \nodata & yes & no & starless core \\
L673$-$SMM5 & 19 18 30.3 & $+$11 08 03 & \nodata & yes & no & 
            probably starless \\
L673$-$SMM6 & 19 18 26.8 & $+$11 10 19 & \nodata & yes & no & starless core \\
L673$-$SMM7 & 19 18 01.9 & $+$11 17 10 & \nodata & yes & no & starless core \\
L673$-$SMM8 & 19 18 41.2 & $+$11 06 04 & \nodata & yes & no & starless core \\
L694$-$SMM1 & 19 38 42.2 & $+$10 50 03 & \nodata & yes & no  & starless core \\
L860$-$SMM1 & 20 01 25.9 & $+$35 53 30 & \nodata & no & \nodata & 
            probably starless \\
L860$-$SMM2 & 20 01 29.0 & $+$35 52 57 & \nodata & no & \nodata & 
            probably starless \\
L944$-$SMM1 & 21 15 51.2 & $+$43 06 08 & \nodata & yes & yes & new protostar \\
L951$-$SMM1 & 21 18 20.0 & $+$43 19 17 & \nodata & yes & no & starless core \\
L951$-$SMM2 & 21 18 22.9 & $+$43 19 51 & \nodata & yes & no & starless core \\
L1014$-$SMM1 & 21 22 23.9 & $+$49 46 06 & \nodata & yes & no & starless core \\
L1165$-$SMM1 & 22 05 09.7 & $+$58 48 05 & 22051$+$5848 & yes & yes & 
            known protostar \\
L1172$-$SMM1 & 21 01 42.5 & $+$67 42 19 & 21017$+$6742 & yes & yes & 
            known protostar \\
L1172$-$SMM2 & 21 01 47.1 & $+$67 42 13 & \nodata & yes & no & starless core \\
L1172$-$SMM3 & 21 01 35.6 & $+$67 39 14 & \nodata & no & \nodata & 
            probably starless \\
L1246$-$SMM1 & 23 22 51.9 & $+$63 20 10 & 23228$+$6320 & yes & yes & 
            known protostar \\
L1246$-$SMM2 & 23 23 03.5 & $+$63 20 16 & \nodata & yes & no & starless core \\
L1262$-$SMM1 & 23 23 48.8 & $+$74 01 07 & 23238$+$7401 & yes & yes & 
            known protostar \\
L1262$-$SMM2 & 23 23 28.7 & $+$74 01 58 & \nodata & yes & no & starless core \\
L1704$-$SMM1 & 16 27 49.8 & $-$23 35 42 & \nodata & no & \nodata & 
            probably starless \\
L1709$-$SMM1 & 16 28 34.6 & $-$23 55 06 & 16285$-$2355 & yes & yes & 
            known protostar \\
L1709$-$SMM2 & 16 29 28.4 & $-$23 49 34 & \nodata & yes & no & starless core \\
L1709$-$SMM3 & 16 28 42.3 & $-$23 54 04 & \nodata & yes & no & starless core \\
L1709$-$SMM4 & 16 29 46.3 & $-$23 46 19 & \nodata & no & \nodata & 
            ridge, probably starless \\
L1709$-$SMM5 & 16 28 31.3 & $-$23 56 55 & 16285$-$2356? & yes & yes & 
            new protostar \\*
\enddata

\tablenotetext{a}{The two IRAS sources indicated with a ? lie within
40$''$ of the submillimeter cores.}
\tablenotetext{b}{L483$-$SMM1 was not searched for high-velocity gas,
but its outflow has been extensively studied by other authors (see
Section~\ref{sec_l483outflow}).}

\end{deluxetable}

\clearpage

\begin{deluxetable}{lcrrcrr}
\tabletypesize{\footnotesize}
\tablecaption{Outflow properties.
\label{tab_outflows}}
\tablewidth{0pt}
\tablehead{
\colhead{Protostar} & \colhead{Mass} & \colhead{Energy} & 
\colhead{Extent} & \colhead{$\Delta V$} & \colhead{$\tau_{\rm d}$} & 
\colhead{Momentum Flux} \\
\colhead{} & \colhead{$M_\odot$} & \colhead{J} & \colhead{AU} & 
\colhead{km~s$^{-1}$} & \colhead{yr} & 
\colhead{$M_\odot$~km~s$^{-1}$~yr$^{-1}$} \\
}
\startdata
L260$-$SMM1 & 0.0002 & 4.8$\times$10$^{32}$ & \phn7200 & \phn5.5 & 
            0.6$\times$10$^{4}$ & 0.4$\times$10$^{-6}$ \\
L663$-$SMM1 & 0.0062 & $>$1.2$\times$10$^{35}$ & $>$23750 & 23.0 & 
            $>$0.5$\times$10$^{4}$ & $\sim$4.7$\times$10$^{-5}$ \\
L673$-$SMM1\tablenotemark{a} & 0.0052 & 7.4$\times$10$^{34}$ & 48000 & 
            \phn7.0 & 3.3$\times$10$^{4}$ & 5.7$\times$10$^{-6}$ \\
L673$-$SMM2\tablenotemark{a} & 0.0260 & 4.0$\times$10$^{35}$ & 54000 & 
            \phn7.0 & 3.7$\times$10$^{4}$ & 2.6$\times$10$^{-5}$ \\
L944$-$SMM1 & 0.0590 & 6.4$\times$10$^{35}$ & 76400 & 11.0 & 
            3.3$\times$10$^{4}$ & 5.7$\times$10$^{-5}$ \\
L1165$-$SMM1 & 0.0134 & 8.2$\times$10$^{34}$ & 48000 & \phn8.5 & 
            2.7$\times$10$^{4}$ & 1.1$\times$10$^{-5}$ \\
L1246$-$SMM1 & 0.0190 & 5.9$\times$10$^{34}$ & 58400 & \phn6.0 &
            4.6$\times$10$^{4}$ & 7.2$\times$10$^{-6}$ \\
L1262$-$SMM1 & 0.0039 & 2.9$\times$10$^{34}$ & 26000 & 10.0 &
            1.2$\times$10$^{4}$ & 8.3$\times$10$^{-6}$ \\
L1709$-$SMM1 & 0.0032 & 5.0$\times$10$^{35}$ & 40000 & 11.0 &
            1.7$\times$10$^{4}$ & 4.7$\times$10$^{-6}$ \\
L1709$-$SMM5 & 0.0171 & 1.5$\times$10$^{35}$ & 21600 & 11.0 & 
            0.9$\times$10$^{4}$ & 5.2$\times$10$^{-5}$ \\
\enddata

\tablenotetext{a}{The outflow properties of both L673$-$SMM1 and
L673$-$SMM2 are based only on the blue-shifted gas.}

\end{deluxetable}

\clearpage

\begin{deluxetable}{lccccccccccl}
\rotate
\tabletypesize{\scriptsize}
\tablecaption{Properties of the protostars in the sample of Lynds dark
clouds.
\label{tab_proto}}
\tablewidth{0pt}
\tablehead{
\colhead{Protostar} & \multicolumn{2}{c}{$F_\nu$ (Jy)} & 
\colhead{$T_{\rm dust}$} & \colhead{$L_{\rm bol}$} & 
\colhead{$L_{\rm submm}$} & \colhead{$L_{\rm bol}/L_{\rm submm}$} &
\colhead{$T_{\rm bol}$} & \colhead{$M_{\rm env}$} & 
\colhead{$M_{\rm env}/L_{\rm bol}$} & \colhead{Outflow} &
\colhead{Class} \\
\colhead{} & \colhead{450~$\mu$m} & \colhead{850~$\mu$m} & 
\colhead{K} & \colhead{$L_\odot$} & \colhead{$L_\odot$} &
\colhead{} & \colhead{K} & \colhead{$M_\odot$} &
\colhead{$M_\odot/L_\odot$} & \colhead{efficiency} & \colhead{} \\
}
\startdata
L162$-$SMM1 & \nodata & 0.28 & 22 & \phn0.5 & $6\times10^{-3}$ & \phn82 &
              120 & 0.03 & 0.07 & \nodata & II \\
L260$-$SMM1 & \phn0.67 & 0.11 & 28 & \phn0.6 & $3\times10^{-3}$ & 250 &
              104 & 0.01 & 0.02 & \phn30 & I \\
L483$-$SMM1 & 23.62 & 2.67 & 27 & \phn9.0 & 0.10 & \phn91 & 
              \phn55 & 0.33 & 0.04 & \nodata & 0 \\
L663$-$SMM1 & \nodata & 1.93 & 21 & \phn2.8 & 0.10 & \phn29 &
              \phn31 & 0.51 & 0.18 & 820 & 0 \\
L673$-$SMM1 & \phn5.28 & 0.60 & 24 & \phn2.8 & 0.05 & \phn59 &
              \phn70 & 0.20 & 0.07 & 100 & I\tablenotemark{a} \\
L673$-$SMM2 & \phn6.88 & 0.97 & 22 & \phn2.8 & 0.07 & \phn38 &
              \phn49 & 0.35 & 0.12 & 450 & 0\tablenotemark{a} \\
L944$-$SMM1 & \phn2.54 & 0.42 & 21 & \phn4.4 & 0.17 & \phn27 &
              \phn30 & 0.88 & 0.20 & 630 & 0\tablenotemark{a} \\
L1165$-$SMM1 & \nodata & 1.23 & 28 & 11.7 & 0.10 & 115 &
              \phn59 & 0.32 & 0.03 & \phn50 & I \\
L1172$-$SMM1 & \phn3.08 & 0.35 & 20 & \phn1.8 & 0.06 & \phn32 &
              \phn62 & 0.31 & 0.17 & \nodata & 0\tablenotemark{a} \\
L1246$-$SMM1 & \phn4.00 & 0.78 & 19 & \phn8.0 & 0.31 & \phn26 &
              \phn62 & 2.15 & 0.27 & \phn40 & 0 \\
L1262$-$SMM1 & \nodata & 0.89 & 20 & \phn1.1 & 0.03 & \phn40 &
              \phn57 & 0.16 & 0.14 & 370 & 0 \\
L1709$-$SMM1 & \phn7.55 &  1.01 & 26 & \phn1.8 & 0.02 & \phn66 &
              \phn67 & 0.09 & 0.06 & 130 & I \\
L1709$-$SMM5 & \phn8.36 & 0.54 & 25 & \phn0.9 & 0.01 & \phn75 &
              \phn44 & 0.05 & 0.06 & 270 & I\tablenotemark{a} \\
\enddata

\tablenotetext{a}{Classifications ascribed by this work, not based on
near/mid-infrared data.}

\end{deluxetable}

\clearpage

\begin{deluxetable}{lccccc}
\tabletypesize{\footnotesize}
\tablecaption{IRAS flux densities derived from HIRES images for several 
protostars.
\label{tab_hires}}
\tablewidth{0pt}
\tablehead{
\colhead{Protostar} & \colhead{IRAS association} &
\colhead{$F_\nu$(12~$\mu$m)} & \colhead{$F_\nu$(25~$\mu$m)} &
\colhead{$F_\nu$(60~$\mu$m)} & \colhead{$F_\nu$(100~$\mu$m)} \\
\colhead{} & \colhead{} & \colhead{Jy} & \colhead{Jy} & 
\colhead{Jy} & \colhead{Jy} \\
}
\startdata
L673$-$SMM1 & 19180+1116 & 1.0\phantom{L} & 3.9\phantom{L} & 11.9 & 57.8L \\
L673$-$SMM2 & 19180+1114 & 0.2\phantom{L} & 0.3\phantom{L} & \phn6.0 & 57.8L \\
L944$-$SMM1 & \nodata & 0.2L & 0.2L & \phn1.3 & 10.0\phantom{L} \\
L1709$-$SMM1 & 16285$-$2355 & 0.7\phantom{L} & 1.8\phantom{L} & \phn3.8 &
26.1L \\
L1709$-$SMM5 & 16285$-$2356 & 0.1\phantom{L} & 1.4\phantom{L} & \phn3.5 &
26.1L \\
\enddata
\end{deluxetable}

\clearpage

\begin{deluxetable}{lcccccccccc}
\rotate
\tabletypesize{\scriptsize}
\tablecaption{Properties of the starless submillimeter cores.
\label{tab_starless}}
\tablewidth{0pt}
\tablehead{
\colhead{Core} & \colhead{$F_\nu$(850~$\mu$m)} & \colhead{Aperture} &
\colhead{Mass} & \colhead{$N({\rm H}_2)$} & \colhead{$n({\rm H}_2)$} &
\colhead{$n_0$} & \colhead{$r_0$} & \colhead{$r_1$} & \colhead{$\gamma$} &
\colhead{$\delta$} \\
\colhead{} & \colhead{Jy} & \colhead{$''$} & \colhead{$M_\odot$} & 
\colhead{$\times 10^{25}$~m$^{-2}$} & \colhead{$\times 10^{10}$~m$^{-3}$} &
\colhead{$\times 10^{11}$~m$^{-3}$} & \colhead{AU} & \colhead{AU} &
\colhead{} & \colhead{} \\
}
\startdata
L55$-$SMM1 & 0.34 & \phn80 & 0.16 & 2.5 & 1.3 & \phn0.20 & \phn8000 & 
             \phn13000  & 1.3 & 2.0 \\
L57$-$SMM1 & 1.43 & 200 & 0.67 & 1.7 & 0.3 & \phn0.47 & \phn6000 & 
             \phn20000 & 0.7 & 1.6 \\
L63$-$SMM1 & 1.31 & 140 & 0.61 & 3.1 & 0.9 & \phn1.55 & \phn4000 & 
             \phn50000 & 0.7 & 1.8 \\
L158$-$SMM1 & 1.02 & 180 & 0.48 & 1.5 & 0.3 & \phn0.20 & 25000 & 
             \phn45000 & 0.9 & 2.0 \\
L158$-$SMM2 & 0.19 & \phn60 & 0.09 & 2.4 & 1.7 & \nodata & \nodata & 
             \nodata & \nodata & \nodata \\
L260$-$SMM2 & 2.10 & 200 & 0.99 & 2.4 & 0.5 & \phn0.22 & 13000 & 
             \phn25000 & 1.0 & 2.0 \\
L328$-$SMM1 & 0.84 & \phn90 & 0.56 & 4.8 & 1.9 & \phn1.72 & \phn3500 & 
             \phn10000 & 0.6 & 2.1 \\
L673$-$SMM3\tablenotemark{a} & 0.34 & \phn60 & 0.56 & 4.3 & 1.6 & \phn6.88 &
             \phn1500 & \phn47500 & 1.5 & 2.0 \\
L673$-$SMM4 & 0.69 & 100 & 1.14 & 3.2 & 0.7 & \phn2.17 & \phn3500 & 
             \phn67000 & 0.6 & 2.0 \\
L673$-$SMM5 & 0.55 & \phn80 & 0.91 & 4.0 & 1.1 & \phn1.60 & \phn3900 & 
             \phn29000 & 0.6 & 2.2 \\
L673$-$SMM6 & 0.53 & 100 & 0.88 & 2.5 & 0.6 & \phn5.15 & \phn1000 & 
             \phn60000 & 2.6 & 1.8 \\
L673$-$SMM7 & 0.24 & \phn50 & 0.39 & 4.4 & 2.0 & \phn1.65 & \phn3000 & 
             \phn75000 & 0.9 & 1.5 \\
L673$-$SMM8 & 1.00 & 110 & 1.65 & 3.8 & 0.8 & \phn1.25 & \phn3500 & 
             \phn33000 & 0.6 & 1.6 \\
L694$-$SMM1 & 1.38 & 120 & 1.58 & 4.4 & 1.0 & \phn0.84 & \phn8000 & 
             \phn45000 & 0.8 & 2.7 \\
L860$-$SMM1 & 0.27 & \phn80 & 2.41 & 1.9 & 0.2 & \phn0.10 & 15000 & 
             \phn40000 & 1.5 & 1.8 \\
L860$-$SMM2 & 0.15 & \phn50 & 1.33 & 2.7 & 0.5 & \phn0.07 & 15000 & 
             \phn37000 & 1.7 & 1.8 \\
L951$-$SMM1 & 0.14 & \phn60 & 1.22 & 1.7 & 0.3 & \phn0.23 & \phn8000 & 
             \phn20000 & 1.2 & 1.6 \\
L951$-$SMM2 & 0.15 & \phn60 & 1.33 & 1.9 & 0.3 & \phn0.57 & \phn2000 & 
             \phn80000 & 2.4 & 1.2 \\
L1014$-$SMM1 & 0.55 & \phn90 & 0.40 & 3.1 & 1.2 & \phn0.71 & \phn4000 & 
             \phn18000 & 1.4 & 1.8 \\
L1172$-$SMM2 & 0.32 & \phn50 & 1.34 & 5.9 & 1.8 & \phn1.74 & \phn3500 & 
             \phn40000 & 0.6 & 1.7 \\
L1172$-$SMM3 & 1.30 & 180 & 4.61 & 1.8 & 0.2 & \phn0.07 & 50000 &
             110000 & 0.9 & 1.7 \\
L1246$-$SMM2 & 0.33 & \phn50 & 3.19 & 6.0 & 1.1 & \phn1.02 & \phn6000 & 
             \phn33000 & 0.8 & 2.3 \\
L1262$-$SMM2 & 1.31 & 100 & 0.96 & 6.0 & 2.0 & 12.40 & \phn1000 & 
             \phn48000 & 0.9 & 1.5 \\
L1704$-$SMM1 & 0.57 & 110 & 0.27 & 2.2 & 0.8 & \phn0.46 & \phn5000 & 
             \phn16000 & 0.9 & 2.1 \\
L1709$-$SMM2 & 1.42 & 120 & 0.67 & 4.5 & 1.6 & \phn2.47 & \phn3500 & 
             \phn40000 & 0.7 & 1.8 \\ 
L1709$-$SMM3 & 1.55 & 120 & 0.73 & 5.0 & 1.7 & \phn3.76 & \phn2000 & 
             \phn45000 & 0.7 & 1.4 \\
L1709$-$SMM4 & 2.51 & 180 & 1.18 & 3.6 & 0.8 & \nodata & \nodata & 
             \nodata & \nodata & \nodata \\
\enddata

\tablenotetext{a}{L673$-$SMM3 is a candidate protostar (see
Section~\ref{sec_l673outflow}).}

\end{deluxetable}

\end{document}